\documentclass[12pt]{article}
\usepackage{jelall}

\newif\ifALLCOMPILE
\ALLCOMPILEtrue
\gdef\ALLCOMPILE{1}

\renewcommand{\saycurrfile}{}

\begin{document}

\addtocounter{section}{0}

\ifdefined\ALLCOMPILE
\else
\usepackage{jelall}

\begin{document}
\fi

\title{Information Cascades and Social Learning}

\author{Sushil Bikhchandani\\David Hirshleifer\\Omer Tamuz\\Ivo Welch}

\maketitle

\thispagestyle{empty}\addtocounter{page}{-1}

\begin{abstract}
We review the theory of information cascades and social learning.  Our goal is to describe in a relatively  integrated and accessible way the more important themes, insights and applications of the literature as it has developed over the last thirty years. We also highlight open questions and promising directions for further theoretical and empirical exploration. 
\end{abstract}

\vspace*{.1in} \noindent We thank the editor, Steven Durlauf; and three anonymous referees for very helpful comments.



\noindent

\needspace{10\baselineskip}

%%%%%%%%%%%%%%%%%%%%%%%%%%%%%%%%%%%%%%%%%%%%%%%%%%%%%%%%%%%%%%%%%%%%%%%%%%%%%%%%%%%%%%%%%%%%%%%%%%%%%%%%%%%%%%%%%%%%%%%%%%%%%%%%%
\section{Introduction}

A central fact of human life is that we rely heavily on the information of others in forming our opinions and selecting actions. As \cite{henrich:17} puts it, ``... we now often put greater faith in what we learn from our communities than in our own personal experiences or innate intuitions.''  The updating of beliefs based on observation of the actions of others, observation of the consequences of these actions, or conversation with others,  is called {\it social learning}.  

The reason that humans and other animals evolved to learn from others is that others often possess valuable information. A basic implication of social learning is that people who observe others’ choices often start behaving similarly. An empirical literature has documented the importance of social learning and its consequences for social outcomes. 
Some examples give a hint of the scope of social learning effects.

In one early experiment, confederates of the experimenter stared up at absolutely nothing, and drew crowds of observers doing likewise (\citeasnoun{milgram:bickman:berkowitz:1969}). Closer to  the heart of human aspirations, there is extensive evidence of social influence in the  mating decisions of humans and other animals. In one experiment, seeing a rival show interest in a member of the opposite sex caused human observers to rate the
individual as more appealing (\citeasnoun{bowers:skyler:place:penke:asendorpf:2012} and the related evidence of \citeasnoun{LBJDC:08}). 
In another, female guppies switched their mate choices to match the mate choices made by other females (\citeasnoun{dugatkin/godin:92}).
A life-and-death example  is the decision by patients to refuse good kidneys for transplant owing to refusal by previous patients in the queue (\citeasnoun{zhang:10}). 

Social learning is also vital for economic activity. For example, in field experiments on financial professionals at the Chicago Board of Trade, investors follow the incorrect actions of others instead of their own signals  (\citeasnoun{alevy:haigh:list:2007}).   In labor markets, unemployment leads to a form of stigma, as employers view gaps in a resume as indicating rejection by other
employers. 
Field experiments identify duration dependence in hiring owing to negative inferences by hiring firms from unemployment periods (\citeasnoun{oberholzer-gee:08} and \citeasnoun{kroft/lange/notowidigdo:13}).

As a final example from the realm of politics, early primary election wins cause later voters to support the winner. For instance, primary election victories   of John Kerry  in competition with Howard Dean generated political momentum and later wins for Kerry  in the U.S. presidential primary contest  of 2004 (\citeasnoun{knight/schiff:10}).

These examples illustrate two patterns. The first is a  tendency for agents to converge upon the same action---one that  surprisingly often generates low payoffs. 
Behavioral convergence seems to arise spontaneously, even in the absence of sanctions upon deviants, and sometimes  despite the opposing force of negative payoff externalities.
Second is that outcomes are often contingent (i.e., path-dependent), and often fragile in the sense that of being  highly sensitive to informational shocks. Examples of contingency are  the effects of spells of unemployment, of strings of kidney rejections, and of political momentum. Examples of fragility  are the rise and fall of surgical fads and quack medical treatments on the part of physicians who rely on what colleagues have done or are doing (\citeasnoun{taylor:1979}, \citeasnoun{robin:1984}). 

The elemental fact of social learning raises several important questions.
When people make decisions in sequence and have the opportunity to observe each other's actions and/or payoff, or communicate with each other, do they eventually make correct choices?
When will beliefs and actions be uniform versus heterogeneous, and how stable is the configuration of  actions and beliefs that the system  converges upon?  Are there fads?
How quickly does (correct) learning occur?
Does rationality lead to better social outcomes?

As emphasized by \cite{hayek:45}, efficient resource allocation requires  aggregating dispersed information about the consequences of economic choices. This raises several questions. Do the beliefs or information of some subset of individuals  unduly influence the  outcomes for society as a whole? How are social outcomes and welfare  shaped by the network of observation and communication? And how are outcomes shaped by  noise or costs of observation of others, by the costs of acquiring private information, and by the ability to delay decisions? 

Early  papers sought to address these questions by means of the concepts of  ``information cascades'' or ``informational herding'' (\citeasnoun{banerjee:1992},  \citeasnoun{bikhchandani:hirshleifer:welch:1992}, and \citeasnoun{welch:92}).\footnote{We  define information cascades later. Banerjee's term, ``herding,'' has essentially the same meaning as information cascades in \citeasnoun{bikhchandani:hirshleifer:welch:1992}, but ``herding'' has several different meanings in economics, and even  within the social learning literature. We therefore use the term  ``information cascades'' for the concept introduced by these two papers.} 
Information cascades were soon applied  in such fields as anthropology, computer science, economics, law, political science, psychology, sociology, and zoology. 
Over time, a  theoretical literature examined the robustness of the conclusions  of the basic information cascades setting to varying different model assumptions.
There has also been a flowering of social learning modeling in which the assumptions used are more distant from the basic cascades settings. 
More recently, a rapidly growing empirical literature (including experimental research) has tested both rational and imperfectly rational social learning models.

In the models of \citeasnoun{banerjee:1992} and \citeasnoun{bikhchandani:hirshleifer:welch:1992}, agents observe private information signals and make decisions in a linear sequence based upon observation of the actions of predecessors.  A key implication of these models is that, under appropriate conditions, there will always be an   {\it information cascade}: a situation in which an agent or sequence of agents act independently of their private information signals. This happens when the information derived from social observation overwhelms the given agents' signals.  When  in a cascade, an agent's action is uninformative to later agents, so that social learning is blocked, at least for a time. So cascading imposes an adverse information  externality on subsequent agents. The   result is poor information  aggregation, so that decisions are often incorrect. 

Cascades in these settings are  precarious; agents in a cascade  are somewhat close to indifferent between two action alternatives.  So in these models the system has  a systematic tendency to reach a  resting point at which behavior is sensitive to small shocks. Social outcomes are fragile; 
even a small influx of new information, or even the possibility of such a change, 
can dislodge agents from the cascade. 
So the cascades model offers a possible explanation for why  social behaviors are often idiosyncratic and volatile, as in the examples discussed earlier. Surprisingly, this is the case even though all individuals are fully rational, and enough  information to make the correct decision is available to society as a whole. 

Models of social learning, including cascades models, have been helpful for understanding volatile aggregate behaviors and socially dysfunctional outcomes in a range of social and economic domains.
In this survey we provide an overview of this research, with a primary focus on theoretical issues. In addition to presenting the basic idea of information cascades, we  explore the robustness of its conclusions to various technical assumptions about  the nature of the action space and the signal distribution.   

We also consider a number of more fundamental deviations from the basic cascades setting. For example, allowing  agents to choose whether to act immediately or to delay provides new insight about boom and bust dynamics in a  range of  behavioral domains. Payoff externalities between the actions of different agents can either reinforce or oppose the disposition to cascade upon predecessors. Alternatively, allowing agents to choose whether to acquire costly private signals encourages cascading upon the actions of others. 

Another key modeling variation  recognizes that in practice the  observation structure is seldom just a simple linear chain in which each agent observes all predecessors and knows everyone's position. 
The network of social connections is  often much more complex than this. 
Furthermore, there is   often nontrivial structure of  meta-knowledge; an agent may not know who a previous agent is able to observe. One of our purposes here is to review the growing literature on social learning in networks.

We also review models in which people have the opportunity to  act and learn from each other repeatedly within a social network. This can lead to loops of informational interaction in which an agent's action can affect the actions of a chain of observers who are then observed by the agent again. These possibilities lead to  informational issues of a very different nature from traditional economic analyses of signaling, adverse selection or moral hazard.

One of the key directions covered here is the study of of how limited rationality affects social learning. Our main focus is on agents who engage in some kind of quasi-Bayesian updating, and are trying to make good choices (i.e., to optimize), as compared with models of mechanistic agents. Allowing for limited rationality is especially important in  realistic  social networks, in which  inference problems are so complex it is unrealistic to assume  perfect rationality. 

Finally, we  cover  models of  information cascades in a variety of applied domains, including    politics, law,  product markets, financial markets,  and organizational structure.  The social learning approach, by focusing on the role of information externalities in determining social outcomes, offers new perspectives about a wide range of human behaviors.

Formal  economic modeling of social learning with many agents is surprisingly  recent, dating primarily to the early  1990s. 
Before that time, models of belief updating and action choice  typically  focused on social learning via interactions either  between small numbers of agents, or else through impersonal markets. 
The latter includes models of learning from market price in financial markets  (\citeasnoun{grossman/stiglitz:76}),  or from actions in product markets (\cite{akerlof:76}). The former includes the ``agreeing to disagree'' literature (\cite{aumann:76}, \cite{geanakoplos:polemarchakis:1982}) and the cheap talk literature (\cite{crawford/sobel:82}). 

In the 1990s, economists started to focus attention upon social learning in general, rather than just as a by-product of market interactions or pairwise game-theoretic interactions. 
A  newer element in this literature  is the possibility that people observe or talk to others even when the target of observation has no incentive or intent to influence others. Also integral to this literature is the possibility
that many agents act and observe {\it sequentially}. 
This captures the reality that when an  agent learns from another, that influences the agent's action, which in turn often affects what later agents can learn from that agent.  

Other social sciences modelled social influence  processes  well before the 1990s. Interactions via social networks were studied heavily in sociology, as with the influential DeGroot model (\cite{degroot:74}), and  cultural evolution was studied in anthropology and other fields  (\citeasnoun{boyd/richerson:85}).  
However, these models typically used mechanistic assumptions about how agents update their actions, traits, or `opinions,' rather modeling  Bayesian or quasi-Bayesian updating of beliefs and resulting optimal actions.\footnote{The field of econophysics (surveyed by \citeasnoun{slanina:14}) also studies the aggregate consequences of agents who are assigned heuristic rules of social updating.} In contrast, the economic approach to social learning  allows for  agents who have a degree of intelligence in their belief updating and process of optimization.   

 Even without  learning, social interaction can cause behaviors to converge, owing, for example, to payoff externalities or utility interactions (\citeasnoun{arthur:89}, \citeasnoun{scharfstein/stein:90}). Our focus is on behavioral convergence or divergence that derives from social learning and information cascades. However, 
 most actual applications involve the interaction of several possible factors, including information, rewards and punishments, and externalities. 
The integration of social learning and cascades with other considerations has led to a richer palette of theory about the process by which society locks into technologies, ideas, governments, organizational choices, conventions, legal precedents, and market outcomes. 

There have been other  excellent surveys of social learning and outcomes (e.g., \citeasnoun{gale:96},  \citeasnoun{chamley:04b}, \citeasnoun{vives:10}, \citeasnoun{acemoglu/ozdaglar:11}, and \citeasnoun{golub/sadler:16}). Our  review includes coverage of more recent research and an array of topics, especially relating to information cascades, in this rapidly evolving field.
Complementing our theoretical focus are several reviews with a partial or full empirical focus, such as  \citeasnoun{bikhchandani:hirshleifer:welch:1998},  \citeasnoun{hirshleifer/teoh:09a}, \citeasnoun{weizsacker:2010}, and \citeasnoun{sacerdote:2014}, and \citeasnoun{BBDI:11}.  These surveys cover tests of cascades theory  in the experimental laboratory (\citeasnoun{anderson:holt:1996aer}, in field experiments \citeasnoun{alevy:haigh:list:2007}, and with archival data (\citeasnoun{tucker:zhang:zhu:2013}, \citeasnoun{amihud:hauser:kirsh:2003}).

%%%%%%%%%%%%%%%%%%%%%%%%%%%%

\ifdefined\ALLCOMPILE\else\end{document}\fi

 % 7
\ifdefined\ALLCOMPILE
\saycurrfile
\else
\usepackage{jelall}
\renewcommand{\citeasnoun}[1]{\textcolor{red}{#1}}

\begin{document}
\fi

%%%%%%%%%%%%%%%%%%%%%%%%%%%%%%%%%%%%%%%%%%%%%%%%%%%%%%%%%%%%%%%%%%%%%%%%%%%%%%%%%%%%%%%%%%%%%%%%%%%%%%%%%%%%%%%%%%%%%%%%%%%%%%%%% 
\section{The Simple Binary Model:  A Motivating Example} 
\label{sec:binary-model}

We illustrate several key intuitions  in a setting with binary actions, states and signals, which we call {\it the simple binary model} (as in the binary example of \citeasnoun{bikhchandani:hirshleifer:welch:1992}, hereafter BHW).\footnote{The  simple binary model is a special case of the BHW model. 
The  model of \citeasnoun{banerjee:1992} differs in several substantive ways, as  described in \S\ref{sec:ActionSpace}.}  We  refer to the  simple binary model as ``the SBM'' throughout this survey. The SBM illustrates how  information cascades can block social learning,  and  can be extended  to  illustrate many  further concepts. 

%%%%%%%%%%%%%%%%%%%%%%%%%%%%%%%%% 

\subsection{Basic Setup: Binary Actions, Signals, and States }
\label{subsec:binary-actions-signals-states}

% \ivo{This definition (here in the latex) now boldfaces 1.  I like the numeral 1 better than the letter I, because the latter is used for notation of agents (individuals).}

\renewcommand{\indicator}[1]{{\ensuremath{\mathbf{1}_{#1}}}}

Each   agent $I_n$ (individuals $I_1, I_2, I_3, \dots$)   chooses one of two actions, High ($a_n =H$) or Low ($a_n= L$), in sequence. The underlying state $\theta$, which is not observed, takes one of two possible values,  $H$ or $L$.  Agents derive utility from matching action to the state, and have the same utility function  $u(\theta, a)$ which is equal to 1 if $\theta=a$ and to 0 otherwise. Under these assumptions, expected utility maximization is equivalent to maximizing the probability, conditional upon the agent's information, of taking an  action that  matches the state. We will often view the $H$ state as one in which there is high payoff to some initiative and the $L$ state in which there is low payoff to that initiative, in which case we call   action $H$ ``Adopt,'' and action $L$ ``Reject.'' Table 1 summarizes the notation used in this survey. 

\begin{center}
 \begin{table}
  \caption{Notation Guide} \label{tbl:Notation Guide}
  \footnotesize
   \begin{center}
    \begin{tabular}{c l l}
      \addlinespace
      \toprule
        &   Notation &  Base case assumption\\
      \hline
      State  &  $\theta$ & $\theta \in \{ \L,\H \}$  \\
      \addlinespace
      Signals & $s_n$ for agent $I_n$  &   $s_n \in  \{ \l, \h \}$  \\
 %     or Signals & $s^\theta \in  \sL, \sH$.  Abbreviated notation: $\l$, $\h$  \\
 %CHECKED SECTION 4 GENERAL SIGNAL DISTRIBUTION; s^\theta, s^L, s^H not used 
      \addlinespace
      Action &  $a_n$ for agent $I_n$ & $a_n \in \{ \L,\H \}$  \\
      \addlinespace
      Utility & $u(\theta, a)$ &$u(\theta,a) = \begin{cases}1&\text{if }\theta=a\\0 &\text{otherwise}\end{cases}$\\
      \hline
      \end{tabular}
       \end{center}
  \vspace{2\baselineskip}
\end{table}
\end{center}

The two states are equally likely ex ante, and each agent $I_n$ receives a binary symmetric private information signal $s_n = h$ or $s_n=\ell$, with $p \equiv \Pr{s_n=h|\theta = H} > 1/2$ and  $\Pr{s_n=h|\theta = L} = 1 - p$.  Signals are independent conditional on $\theta$.        

By Bayes' rule, the posterior probability $\Pr{\theta = H|s_n=h}$ equals $p$, and $\Pr{\theta = H|s_n=\ell}$ equals $1- p$. By  the symmetry of the model,  $h$ and $\ell$ signals have offsetting effects on posterior beliefs, so $\Pr{\theta = H|s_1=h, s_2=\ell} =  0.5$, and if  an agent sees or infers certain numbers of $h$ and $\ell$ signals, the updated belief depends only on the difference between these numbers.

We refer to {\it social information} as information derived from observing or talking to others,  the {\it social belief} at any step in the sequence as the belief based solely on social information, and an agent's hypothetical {\it private belief} as the belief based solely on an agent's private signal.   Each agent's action choice is of course based on the agent's full information set. 

\needspace{5\baselineskip}

We contrast two regimes for social observation:
\begin{description}
\item[1. The Observable Signals Regime:]%

In this regime, each agent can observe both the signals and the actions of predecessors.\footnote{Actions do not convey any additional information as a predecessor's signal is  a sufficient statistic for her action.} 

\item[2. The Only-Actions-Observable Regime:]%

In this regime, agents can observe the actions but not the private signals of  predecessors. 
\end{description}

\vspace*{.1in}

In the Observable Signals Regime, the pool of social information continually expands, and by the law of large numbers the social belief becomes arbitrarily close to correct. So  people eventually settle on the correct choice (adopt if $\theta = H$, reject if $\theta = L$). It follows that they eventually behave alike. A {\it herd} is  a realization in which all agents  behave alike from some point on.  We say that there is  {\it herding} if a herd eventually starts with probability 1. 

In the Only-Actions-Observable Regime, each agent's action depends on the action history and the agent's  own private signal. The precision of the social belief is weakly increasing, so  it is tempting to conjecture  that highly accurate outcomes  will again be achieved.  But in fact,  the precision of the social belief  hits a finite ceiling, as we will discuss. 
The   choices of a few early predecessors determine the actions of later agents. In consequence, there is again herding,  but often   upon the wrong action---the choice that yields a lower payoff. In \S\ref{sec:observability} we consider other observability regimes.

Throughout this review, our default premise will be the Only-Actions-Observable  regime. More specifically, we only mention assumptions when they deviate from the base set of assumptions of the SBM. 
The SBM, and  modest variations thereof, are rich sources of insight into social learning.\footnote{For the purpose of providing minimal examples, we sometimes employ different tie-breaking conventions for the behavior of indifferent agents. Although pedagogically convenient, such ties could be eliminated, with similar results, by slightly perturbations of the game structure.}   

%%%%%%%%%%%%%%%%%%%%%%%%%%%%%%%%% 

\renewcommand{\baselinestretch}{1.3}\par\footnotesize\mbox{}\normalsize

\needspace{5\baselineskip}

\subsection{Why information stops accumulating: Information cascades}
\label{subsec:why-info-stops}

\begin{figure}

    {
        \centering
        \includegraphics[width=0.8\textwidth]{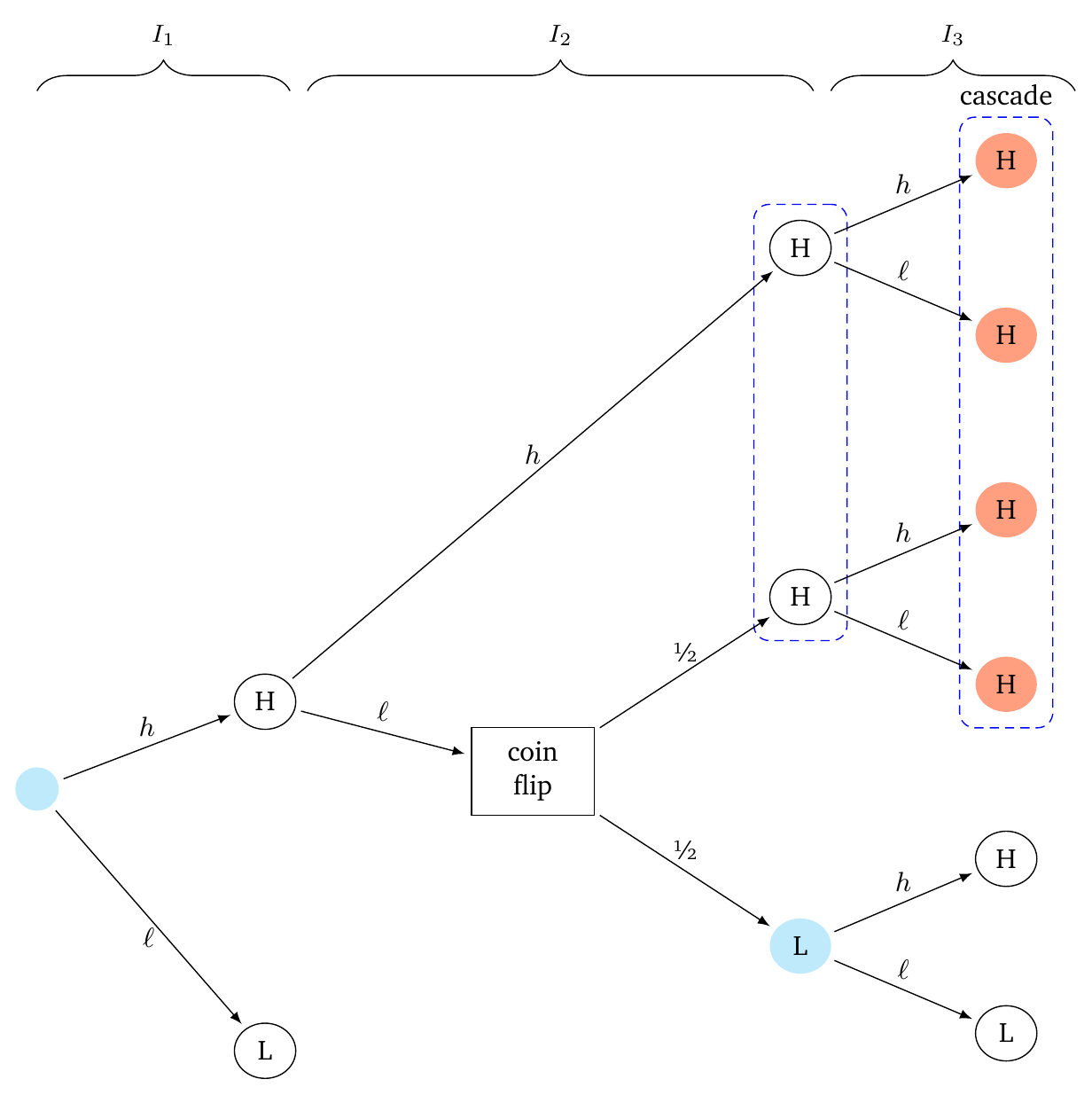}
    }

    \caption{Event Sequence}
    \label{fig:eventchain}
    
    \footnotesize

    This figure shows a partial sequence of events and optimal choices in the SBM. Possible signals are $h$ and~$\ell$, which are indicative of states $H$ and $L$ respectively. The main focus is on events after an initial $h$ signal. Multiple nodes that are in the same information set from the viewpoint of later observers are surrounded by dashed lines.  Action $H$ is correct in state $H$, and  action $L$ is correct in state $L$. The ``L’’ circle for $I_2$  is shaded blue to reflect the fact that when this node is reached, the next player, $I_3$ is informationally in the same position as $I_1$ following the initial blue node. Nodes that are in an information cascade are shaded in peach.
\end{figure}

To see why outcomes are inefficient in the Only-Actions-Observable Regime, consider each  agent in sequence. Figure~\ref{fig:eventchain} shows the possible chains of events.  Agent~$I_1$, Ann, adopts if her private signal is $h$ and rejects if it is $\ell$.  Agent~$I_2$, Bob, and all successors can infer Ann's signal perfectly from her decision.
So if Ann adopts, and Bob's private signal is $h$, he also adopts.  
Bob knows that there were two $h$ signals: he infers one from Ann's actions and has observed one privately. 
If Bob's signal is $\ell$, it exactly offsets Ann's~$h$, making him indifferent between adopting and rejecting.

Suppose that when an agent is  indifferent between the two actions the tie is broken by  tossing  a fair coin.   Agent~$I_3$, Carol, now faces one of three possible situations: (1) Ann and Bob adopted (their actions were $HH$), (2) they  rejected ($LL$), or (3) one adopted and the other rejected ($HL$ or $HL$).  In the $HH$ case, Carol also adopts.  She knows that Ann observed $h$ and that, more likely than not, Bob also observed $h$ (although he may have seen $\ell$ and tossed a coin).  So Carol adopts even if she observes an $\ell$ signal. In other words, Carol's action is independent of her private information signal---she is in an information cascade.

\begin{definition}
An \emph{information cascade} is said to occur if it is optimal for an agent or set of consecutive agents to choose an action which is independent of the agents'   private information.
\end{definition}

Crucially, in the $HH$ case, Carol's action provides no information about her signal to her successors. Her action has not improved the public pool of information;  the social belief remains unchanged.
Everyone after Carol faces the same decision problem that she did.
Once Carol is in an adoption cascade, all her successors also adopt based only on the observed actions of Ann and Bob.
Similarly, in the second case,  two rejections (actions $LL$) put Carol into a reject cascade that never stops.

In the third case,  Ann has adopted and Bob has rejected, or vice versa. 
Carol infers that Ann and Bob observed opposite signals, so the social belief is  that the two states are equally likely.  
Her decision problem is therefore identical to that of Ann, so Carol's decision is based only on her private signal. 
In turn, this makes the decision problem of agent~$I_4$, Dan, isomorphic to Bob's---he need only pay attention to his latest predecessor.

Similarly, agent $I_5$, Eve, is in a position isomorphic to that of Carol's.
Eve knows that Ann's and Bob's signals cancelled, so she can update based only on the actions of the latest two predecessors, Carol and Dan.  So if the latest two predecessors   take the same action, such as Adopt, an adopt cascade starts with Eve.

More generally, if the action sequence starts with an even number of zigzagging reversals between actions Adopt and Reject of any length, the next agent can infer that the private signals alternated, so that this action sequence can be ignored. So the optimal decision rule for
any agent $I_n$ depends on action history only through  $d_n$, defined as the difference between the number of predecessors who adopted and the number who rejected (so $d_1 = 0$).  If $d_n \ge 2$, the agent is in an information cascade and adopts regardless of private signal.
If $d_n = 1$, the agent adopts if the private signal is $h$ and tosses a coin if the signal is $\ell$.  Similarly, if $d_n = -1$ the agent rejects if the private signal is $\ell$ and tosses a coin if the signal is $h$, and if $d_n \le -2$ the agent is in a Reject cascade. If $d_n = 0$, then the agent follows the agent's  private signal. 

With probability 1, the difference $d_n$ eventually hits either the upper threshold of +2, triggering an Adopt cascade, or the lower threshold of $-2$, triggering a Reject cascade.  So cascades start almost surely. (This conclusion holds more generally  for any finite number of possible signal values; BHW Proposition 1.)  Indeed, with  high probability, all but the first few agents end up doing the same thing. In other words, there is herding.

Even in  settings without herding, one or more individuals may be in an information cascade, as can occur with payoff externalities; see \S\ref{sec:externalities}. For example, a long string of adopts may reduce the expected payoff from adopt so much that the next agent rejects.  Furthermore, herding can occur without information cascades, as we discuss in \S\ref{sec:GeneralSignalDistributions}. 
But  in the SBM, there are both cascades and herding  

With many agents, the social outcome in the Only-Actions-Observable Regime is on average much worse than in the benchmark Observable Signals Regime.  This is because once a cascade starts, no new information becomes public; the accuracy of the social belief plateaus. 
An early preponderance of either Adopts or Rejects causes all subsequent agents to rationally ignore their private signals.
These signals thus never join the public pool of knowledge.  Agents disregard their private signals before the social belief becomes very accurate.  
As soon as the social belief is more precise than the signal of just a single agent, the next agent falls into a cascade. We will call this the ``logic of information cascades.''\footnote{In a setting with more than two possible signal values,   two identical successive actions do not necessarily start a cascade. 
The general point is that a fairly low-precision social belief can be enough to trigger a cascade.}

The social outcome depends heavily on the order in which signals arrive.  If signals arrive in the order $hh\ell\ell\dots$, then a cascade starts and everyone adopts.  If, instead, the same set of signals arrives in the order $\ell\ell h h \dots$, everyone rejects.  So cascades and welfare are \emph{path dependent}.

Consistent with path-dependence, cascades are {\it idiosyncratic}, meaning that they are often incorrect. This is because agents tend to fall into a cascade before social information has become very precise.

Cascades tend to form quickly even when private information is very noisy.  For example, if the signal accuracy is $p = 0.5 + \epsilon$, where $\epsilon > 0$ is small, the probability that an adopt or reject cascade forms after the first two agents is about $1-0.5^2=75\%$.  The probability that $I_{10}$ is not in a cascade is only about $0.5^{10}\approx0.001$. This is the probability that $|d_n| < 2$ for each of agents $I_3$ through $I_{10}$. As the only way to avoid a cascade is to have alternating signals over time, a cascade starts eventually with probability one.

Early agents are a small fraction of the large population, so we can think about average accuracy and welfare in terms of late agents. The probability of a correct cascade (Adopt  iff $\theta = H$) is (see equation (3) of BHW) 
\begin{equation}
  \Pr{\text{Correct Cascade}} = \frac{p(p+1)}{2(1 - p + p^2)} ,
  \label{e1}
\end{equation}
where as before, $p$ is signal accuracy.  For a noisy signal as above, the limiting probability that sufficiently late agents end up in a correct cascade  is  $0.5 + 4\epsilon/(3+4\epsilon^2)$, which is close to $0.5$ when $\epsilon$ is small. 

If agents did not have access to  social information, they would choose an action based only on their one private signal, and the probability of a correct choice would be $0.5+\epsilon$.  So the increase in accuracy from being able to observe the actions of predecessors is $\epsilon[(1 - 4\epsilon^2)/(3+4\epsilon^2)] < \epsilon/3$,  which is small when $\epsilon \approx 0$.  Thus, the accuracy improvement from social observation falls far short of  the improvement in the Observable Signals Regime, in which the social belief approaches perfectly accuracy, and late agents almost surely converge upon the right action. 

More generally, let   the relative accuracy improvement from social observation under the Only-Actions-Observable Regime relative to the Observable Signals Regime    be the ratio of two differences. 
The numerator is the improvement in accuracy eventually provided by social observation when only actions are observable minus the accuracy with no social observation, $p$. 
The denominator, $1-p$, measures the improvement relative to this same benchmark, $p$, from achieving certainty of a correct decision, probability 1. Certainty eventually occurs if  agents can observe all past signals. 
So the relative accuracy improvement   is $(\Pr{{\rm Correct} \ {\rm Cascade}} - p)/(1 - p)$, which can be calculated using equation (\ref{e1}).

Using L'H\^{o}pital’s rule, the relative improvement increases from 0 for very noisy signals to 0.5 for very informative signals.  
Remarkably, when only actions are observable, less than half of the potential gains from information aggregation are realized.  
Intuitively, regardless of signal noise,  cascades tends to start early (based on a small preponderance of evidence in the social belief). 
So when  signals are very noisy,  social learning becomes almost useless. 

%%%%%%%%%%%%%%%%%%%%%%%%%%%%%%%%% 
\subsection{Fashion Leaders and Heterogeneous Precision}  
\label{subsec:fashion-leaders-precision}

We call agents who are viewed by others as having more precise private information  {\it fashion leaders}. The actions of fashion leaders can trigger immediate cascades.  
So even a small advantage in signal precision can have a large effect.
For example, if Ann has a signal precision $p' = p + \epsilon$, where $\epsilon > 0$ is small, then Bob will defer to her, as do all later agents.  In this case, the social outcome aggregates the information of only a single signal, which reduces accuracy.

When agents differ in private signal precisions, the order of moves is crucial. If the high-precision agent, Ann, were second instead of first, there would be no immediate cascade.  However, even if agents are ordered in inverse order of precision, cascades  still tend to form quickly---unless later agents have better precision than all preceding agents \emph{combined}.

Social psychologists report that people imitate the actions of experts.  When LeBron James wears a particular brand of athletic shoes, it is an expert product endorsement likely to sway observers.  
Such endorsements may be less compelling for unrelated products. 

The drawback of leading off with the better-informed has not been lost on designers of judicial institutions.
According to the Talmud, judges in the  Sanhedrin (the ancient Hebrew high court) voted on cases in inverse order of seniority.  
Similar voting orders continue in some of today's courts (e.g., those in the U.S. Navy).
Such strategic ordering can reduce the undue influence of older (and presumably wiser) judges. 

%%%%%%%%%%%%%%%%%%%%%%%%%%%%%%%%% 
\subsection{Fragility} 
\label{subsec:fragility-fads}

Cascades can be broken if better-informed agents arrive later in the sequence. Likewise, the arrival of public information can easily dislodge a cascade.  
Each agent knows that any cascade is based upon information that is only slightly more accurate than the agent's own private signal.  
Thus, a key prediction is that even long-standing cascades are fragile with respect to small shocks.

\begin{definition}
 \label{def:fragility}
A cascade is \emph{fragile} if a hypothetical public disclosure of a signal with a distribution that is identical to that of  the private signal possessed by a single agent would, with positive probability,  break the cascade, i.e., causes  the next agent's action to depend on that agent's signal.
 \end{definition} 

In the SBM, once a cascade starts, it remains fragile for all remaining agents. 
For example, a shock to the system in  the form of a public signal that arrives after a cascade has started can easily dislodge the cascade.

\bigskip  %% for now, remove later.  I just wanted to tell you its a break here.

Owing to information cascades, there is a systematic, spontaneous
tendency for the system to move to a position of {\it precarious stability}. 
This is much like the inclination of the hero's car in an action movie chase scene  to  end up teetering at the edge of a cliff.  This tendency is also in the spirit of models of self-organized criticality, wherein certain types of  systems   systematically  evolve  to a critical point in which the system is highly sensitive to small shocks (\citeasnoun{bak/kan:91}). 
In contrast,  equilibrium is much more robust in models in which there are sanctions upon deviants or disutility from nonconformity  (\citeasnoun{kuran:87}).

\subsection{Fads}
\label{subsec:fads}

In reality the best action to take often fluctuates over time, i.e., the state evolves stochastically. 
For example, competing web browsers sometimes leapfrog each other in functionality for users. 
In such a setting, the mere {\it possibility} that a shock to the system (that is, a change in the true value) could occur can be enough to dislodge a cascade, even if the shock does not actually arrive. 

To see this, suppose that just before $I_{101}$'s decision, with probability 0.1 the  state $\theta$  is newly redrawn from its ex ante distribution, and remains fixed thereafter.
Let $\theta'$ denote the state  after $I_{100}$, where $\theta' \neq \theta$ with probability 0.05.  

The possibility of a value shift breaks the cascade; $I_{101}$ optimally follows $I_{101}$'s own signal. 
Eventually the system must settle into a new cascade---one that need not match the old one. 
It is easy to show that the probability of a cascade reversal is a little over  $0.0935 \gg  0.05$ (see BHW). So the probability  that the common behavior  shifts is more than 87\% higher than would be the case  under full information. BHW refer to such  volatile  outcomes as \emph{fads}. 

What if there is a probability {\it each period} that the state shifts?  Then if the probability is not too large, there are still cascades (including incorrect ones).  However, these cascades are temporary; over time the social information in any cascade grows stale and some agent eventually returns to using her own signals (\citeasnoun{moscarini/ottaviani/smith:98}).

%%%%%%%%%%%%%%%%%%%%%%%%%%%%%%%%% 
\subsection{Welfare-Reducing Public Disclosure}
\label{subsec:bad-disclosure}

In purely individual decision making,  an extra signal always makes an agent weakly better off.  However, in a setting with social observation, a public signal can reduce average welfare.\footnote{This can occur in non-cascade contexts, too. For example, in a  context in which  refraining from running on the bank is a public good, disclosure can sometimes trigger collapse (\citeasnoun{teoh:97}, \citeasnoun{morris/shin:00}).} 

In the fashion leader model discussed in \S\ref{subsec:fashion-leaders-precision}, if $I_1$  has a private signal with slightly greater accuracy $p + \epsilon$ than the common precision $p$ of later agents, then $I_2$ is always in a cascade.  All agents other than $I_1$ have lower expected utility than in the basic setting with identical precisions, as all now effectively act based upon just a single signal (that of $I_1$). (It is easy to verify by comparing with equation  (\ref{e1}) that if $\epsilon$ is close to zero, the accuracy  of the cascade is reduced.)  

Suppose next that the more accurate signal is a {\it public} disclosure, and that all private signals have identical precisions. Now agent $I_1$ is the first in the cascade, and all agents have lower expected utility than in the basic setting, as all now effectively act based upon just a single signal (the public disclosure). Of course, a sufficiently  accurate  early signal or public disclosure can improve  the social outcome.  For example, if $I_1$ has perfect information, the cascade is always correct.  

More generally,  a shift in    information regime that might seem to make agents better informed (such as higher signal precision of early agents, greater observability of others, or greater publicly available information) can reduce the average decision  accuracy   in the long run and can reduce average welfare.  
If some variation in the setting directly  makes agents better informed,  they will be closer to the critical information value for a cascade to start, and will tend to disregard their private signals sooner.   
So the direct benefit of better information tends to be opposed by this tendency of agents to disregard their private signals sooner. 
This reduces the information available to later agents. We call this the {\it principle of  countervailing adjustment}.\footnote{However, in general a shift in model structure can have different  types of effects on the quality of information aggregation. For example, a shift in model structure in some cases affects the critical value for a cascade to occur, which can either increase or decrease the ultimate amount of information impounded in the action history. Furthermore, even within the SBM, if each agent's private signal became  more precise,  cascades will tend to be  more informative.} This is reminiscent of Le Chatelier's principle in chemistry, which holds that systems in dynamic equilibrium adjust to oppose the effects of external disturbances. 

%%%%%%%%%%%%%%%%%%%%%%%%%%%%%%%%% 
%%%%%%%%%%%%%%%%%%%%%%%%%%%%%%%%% 

\subsection{Psychological bias can aid or impair social  learning}
\label{subsec:other-model-misspec}

Psychological bias can either aid or hinder social learning. In \citeasnoun{bernardo/welch:01}, occasional overconfident ``entrepreneurs" overestimate the precisions of their own private information signals. This can cause them to make greater use of their own information signals instead of following  predecessors, as in an  information cascade. So  overconfidence can improve  learning and  outcomes for later agents.  

Suppose, for example, that \I2 is overconfident, in the sense that \I2 incorrectly assesses \I2's precision at $p' = p + \epsilon$, where  $\epsilon \approx 0$, and that $I_2$'s true precision of $p$ is known to all except \I2.  
So when \I2 has a signal that conflicts with the action of  \I1, \I2 always follows her own signal instead of flipping a coin.
 It follows that the first two actions perfectly reveal the signals of \I1 and \I2.  
 If both agents adopt, \I3 is in an adopt cascade, and if both reject, \I3 is in a reject cascade. And just as in the SBM, if the first two actions are in opposition, \I3 uses his own signal.  
 What is avoided are  situations where \I2 flipped a coin and chose the same action as \I1, pushing \I3 into a cascade even though there was no justification in the underlying signals. 
 This on average improves all decisions  starting with agent \I3. Algebraic details are in the Online Appendix, \S \ref{onlinesub:binary}. 
 
 %%%%%%%%%%%%%%%%%%%%%%%%%%%%%%
 
 \subsection{Lessons of the Binary Model}
\label{subsec:example-sum-up}

The SBM illustrates a number of key concepts that  also obtain in some more general settings. 

\begin{description}
\item[Conformity:] % Referred to later
  People quickly end up following the behavior of others. 

\item[Idiosyncrasy and Path Dependence:]% Referred to later
 Despite a wealth of private information which, in principle, could be aggregated to achieve correct actions, cascades, and therefore the behavior of most agents, are  often incorrect---idiosyncrasy.
 
In particular, the actions of a few early agents tend to be decisive in determining the actions and success of  a  large numbers of successors. Social outcomes have low predictability. 

This is consistent with  evidence from experiments in which individuals can observe information about the decisions of  predecessors to download songs. When such music marketplaces are run independently,  
the correlation between song popularity rankings across marketplaces is very low, though positive  (\citeasnoun{salganik/dodds/watts:06}). In the field, \citeasnoun{duan:gu:whinston:2009} find that downloads of free software at CNET decline precipitously as the discrete download popularity ranking decreases. They provide evidence suggesting that this is due to   information cascades as compared with several competing hypotheses.

\item[Information Externality:] % Referred to later, especially as inefficient learning
Each agent takes an action that is individually optimal, without consideration of the  informational consequences for  later agents. If early agents were instead to reveal their own signals, they would confer a valuable informational benefit upon later agents.  This loss of private information  blocks efficient  outcomes.
  
\item[Fragility:] % Referred to later
Society  spontaneously wanders to a position that is highly sensitive to small shocks, such as the possible arrival of  noisy  public information.  So small shocks have a good chance of causing  many to change their actions. This contrasts with settings in which social outcomes tend to be insensitive to small shocks, except for very special sets of  parameter values that happen to put the system close to a knife edge (\citeasnoun{kuran:89}).  

\item[Principle of countervailing adjustment:]%
 
Seemingly favorable shifts in information availability do not necessarily improve   average decisions or welfare. The direct positive effect of more information tends to be opposed by the tendency of agents to disregard their private signals sooner, to the detriment of later agents.  An example is   the presence of an agent with slightly better private information (a ``fashion leader''), as discussed earlier. 

There are other settings in which  more information can make agents worse off (\citeasnoun{hirshleifer:71}, \citeasnoun{teoh:97}, \citeasnoun{morris/shin:02}). In the cascades framework, this occurs because of  countervailing adjustment.   

\item[Psychological Biases Can Improve Social Outcomes:]%

Biases that cause agents to make greater use of their private signals can help remedy the informational externality, resulting in more accurate social beliefs. 

\end{description}

The SBM and the cascades model of BHW provide some surprisingly extreme outcomes: complete blockage of learning, and, eventually, identical action choices. However, the broader intuitions that information externality limits learning, and that social information grows relative to private signals, suggests that similar but milder possibilities can occur in more general settings: that learning becomes very slow, and that actions  become very similar. In \S\ref{sec:ActionSpace}-\S\ref{sec:EndogSignalAcquisition} we explore alternative modeling assumptions which  sometimes lead to such broader outcomes.  

\ifdefined\ALLCOMPILE\else\end{document}\fi

%%%%%%
 % 14
\ifdefined\ALLCOMPILE
\saycurrfile
\else
\usepackage{jelall}
\renewcommand{\citeasnoun}[1]{\textcolor{red}{#1}}

\begin{document}
\fi

%%%%%%%%%%%%%%%%%%%%%%%%%%%%%%%%%%%%%%%%%%%%%%%%%%%%%%%%%%%%%%%%%%%%%%%%%%%%%%%%%%%%%%%%%%%%%%%%%%%%%%%%%%%%%%%%%%%%%%%%%%%%%%%%% 

\section{Varying the action space} \label{sec:ActionSpace}

Agents often have only discrete choice sets. Firms may adopt or reject a project. People may attend versus not attend a sports event,  marry one person versus another, or have versus not have children.  Voters must choose one from a set of candidates.  Furthermore, even continuous choice sets are often truncated.  The alcohol consumable at dinner is bounded below at zero.

When the set of available actions is  continuous,  under reasonable assumptions even a small variation in an agent's private signal and belief causes a small  shift in action. If so, agents' actions always depend on their private signals, so that there is no cascade.
In other words, cascades do not occur when actions are {\it responsive} to private signals (\citeasnoun{ali:2018}).

However, a finite action space cannot be responsive.
For example, in the SBM, an agent's decision problem is unresponsive as each of the two actions is optimal for a range of probability beliefs.

Similarly, as \citeasnoun{lee:1993} shows in a setting  with  arbitrary compact (i.e., closed and bounded) action spaces, if the action space is discrete, incorrect cascades always arise, i.e., there is idiosyncrasy. Even if the action space consists of  a finite union of disjoint intervals, incorrect cascades can arise.

Responsiveness is central to whether there is  eventual learning of the true state. If $I_n$'s chosen action is always even slightly responsive to $I_n$'s private signal, subsequent agents can glean information about $I_n$'s  private signal. Based on this, subsequent agents can learn about $I_{n+1}$'s signal from $I_{n+1}$'s action, and so forth.  

This reasoning may seem to suggest that a continuous action space  eliminates cascades, and results in eventual learning of the state. However, even with a continuous action space, an agent's optimal actions could be unresponsive to the signal (\cite{ali:2018}). For example, actions can fail to be responsive if an agent's payoff is a constant function of action in some region of the action space, given the state. This can effectively discretize a continuous action space. 

It is usually simpler to work directly with a discrete action space than to assume a continuous action space and then  effectively discretize it via a discontinuous payoff function. So models with continuous and unbounded action spaces typically have responsiveness, and therefore have  no cascades. 

An exception is the model of \citeasnoun{banerjee:1992}, in which incorrect cascades occur despite a continuous action space.  The unknown state $\theta$ is uniformly distributed on $[0,1]$,  and agents choose an  action $a\in[0,1]$. Each agent obtains a  payoff of 1 if she chooses action $a=\theta$ and a payoff of zero if $a\neq\theta$. Each agent receives either no signal (and is uninformed) or  one signal. An informed agent receives a signal about $\theta$ that is either fully revealing or is pure noise (in which case it is uniform on $[0,1]$).  An informed agent does not know which of these two possibilities  is the case. 

The payoff from matching the state is substantially greater than from even a slight miss. There is a positive probability that any given  predecessor's action $a^*$ matches $\theta$. So when an agent observes that a predecessor chose action $a^*$, the expected payoff from choosing $a = a^*$ is greater  than from choosing $a = a^* + \epsilon$, even when $|\epsilon| >0$ is arbitrarily small. In consequence, the optimal action is not responsive to small changes in an agent's signal.
Thus,  early agents may fix upon an incorrect action---an incorrect cascade.

\newcommand{\Lim}[1]{\raisebox{0.5ex}{\scalebox{0.8}{$\displaystyle \lim_{#1}\;$}}}

\medskip
\noindent
{\sc Enriching the Action Space}

A broader question is whether increasing a finite number of action choices helps prevent incorrect cascades, or at least limits their adverse effects. From a welfare viewpoint, we are now interested not just in whether cascades are incorrect, but in how large the expected mistakes are.

Suppose that the $H$ state of the SBM corresponds to value 1, and the $L$ state to value 0. In other words, $\theta =  0$ or 1. As in the SBM, signals are binary, $h$ or $\ell$. 

Suppose further that there are $M$ possible action choices  that are evenly spaced between 0 and 1 and always include these two values.  (In the SBM, $M=2$.) So the action set is defined to be $ \{ 0, 1/(M-1), \ldots, (M-2)/(M-1), 1 \}$.  The SBM has $M=2$. When $M = 3$, the possible actions are 0, $1/2$ and 1.  As $M$ increases, specific action options can disappear.  For example,  when $M$ increases from 3 to 4, the action choice of 1/2 is removed, and 1/3 and 2/3 become  the two interior choices. 

As in the SBM, each agent $I_n$ receives a binary signal  with precision $p$ about  state~$\theta$.  We assume that each agent's utility  is negatively proportional to the squared distance of action from the state, $u(\theta,a_n) = - (\theta-a_n)^2$.  It follows by direct optimization that each  $I_n$  chooses the action $a_n$ that is closest to her inferred probability that the true state is 1.  For simplicity, we now assume that when an agent is indifferent, she chooses the  action that corresponds to her private signal. 

In this setting, cascades form with probability 1 (as can be shown by extending the proof  for the two-action case as in  BHW).   We can measure the social inefficiency in distance units by the root mean-squared error of the eventual cascade action, i.e., $\Lim{n\rightarrow\infty} \sqrt{E_n[ (\theta - a_n^*)^2 ]}$.

The inefficiency is not in general monotonic in the number of actions. To see this, suppose that $p=0.7$.
When $M = 2$, $I_1$ follows her signal (chooses action 1 if her signal is $h$ and 0 otherwise) for an expected loss of $0.7\cdot 0^2 + 0.3\cdot1^2= 0.3$. Thus, $I_2$ learns $I_1$'s signal.
Adding a third action ($M= 3$) at 1/2  induces $I_1$ to choose $1/2$ regardless of signal, as the more extreme action suggested by her signal leaves her heavily exposed to quadratic disutility of action error.
Her expected loss from this middle action is  $0.7\cdot(1/2)^2 + 0.3\cdot(1/2)^2= 0.25$, which is lower than the expected loss from following her signal ($0.3\cdot1^2$). 
Since $I_1$'s action is  uninformative, all subsequent agents are in a cascade upon the same choice.
More generally, when more actions are introduced, agents may prefer to cascade upon one of the intermediate choices, to the detriment of later agents, rather than using their own information to take a more accurate/extreme choice.  

\begin{figure}[ht]
  \caption{Inefficiency in Relation to Number of Action Choices} \label{fig:actions}
  \begin{center}
     \includegraphics[width=0.9\textwidth]{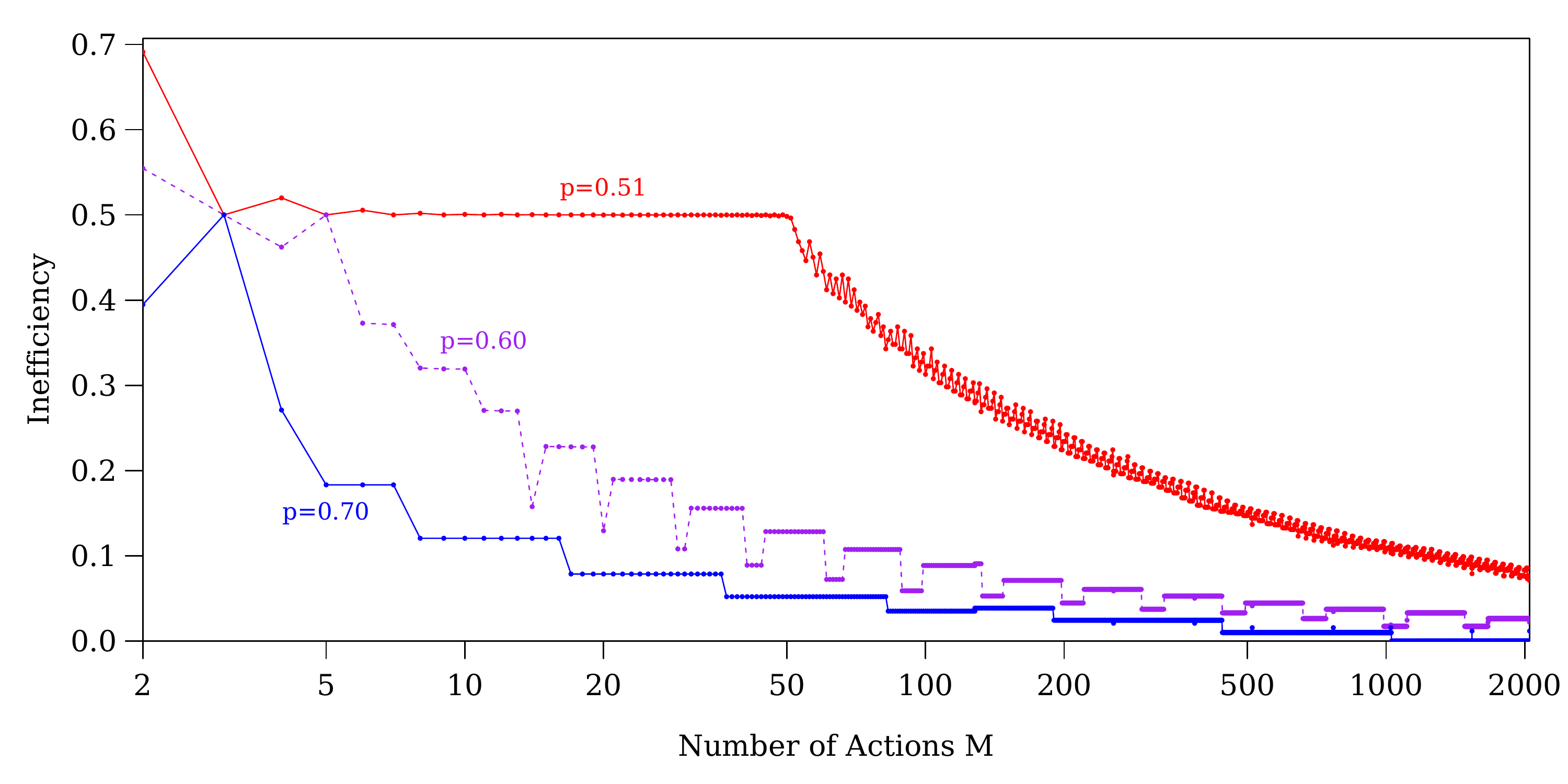}
  \end{center}

   \footnotesize 
   The number of actions $M$ is on the $x$-axis. The actions lie between 0 and 1.  For example, when $M =4$, the actions are $\{0,1/3,2/3,1\}$.  The $y$-axis graphs the social inefficiency in distance units as $\Lim{n\rightarrow\infty} \sqrt{E_n[ (\theta - a_n^*)^2 ]}$.\par

  \noindent\rule{\textwidth}{3pt}
\end{figure}

Figure~\ref{fig:actions} plots the  inefficiency for three different signal precisions, where  the horizontal axis  has   values of $M$ from 2 to 2000  on a logarithmic scale.
As the number of choices becomes large, almost all agents cascade upon an  action that is at or near the correct action (see also the rational agent case of  \citeasnoun{eyster/rabin:10}), and the inefficiency converges to 0.\footnote{The  decline is  nonmonotonic even if the sequence of action sets is chosen so that available options never disappear.} 

Even a few choices can significantly improve the sensitivity of action to signal, thereby improving information aggregation (\citeasnoun{talley:1999}).  
 For example, as seen in Figure~\ref{fig:actions}, when $p=0.7$, moving from $M = 3$ to $M=5$ reduces the inefficiency from 0.5 to less than 0.2. 
 
 It  is striking that when signals are noisy, there can be significant inefficiencies even for surprisingly fine action spaces.  
 If $p = 0.51$, an increase from $M = 3$ to 
the much more refined $M=49$ does not increase efficiency.  Agents continue to cascade immediately on the middle action of 0.5 (or an action near it), which results in low  efficiency of the cascade. Nevertheless, eventually, as $M$ grows large, the social outcome becomes efficient. 
Overall, owing to information cascades, even a modest degree of nonresponsiveness of actions to signals (due to the finite number of action choices) can result in surprisingly large social inefficiencies.

\clearpage

\ifdefined\ALLCOMPILE\else\end{document}\fi

 % 6
\ifdefined\ALLCOMPILE
\saycurrfile
\else
\usepackage{jelall}
\renewcommand{\citeasnoun}[1]{\textcolor{red}{#1}}

\begin{document}
\fi

\def\defeq{\stackrel{\mathrm{def}}{=}}
%%%%%%%%%%%%%%%%%%%%%%%%%%%%%%%%%%%%%%%%%%%%%%%%%%%%%%%%%%%%%%%%%%%%%%%%%%%%%%%%%%%%%%%%%%%%%%%%%%%%%%%%%%%%%%%%%%%%%%%%%%%%%%%%% 
\section{Cascades and Herding with General Signal Distributions}
\label{sec:GeneralSignalDistributions}

\bigskip

We next examine the effects of  general signal distributions, as contrasted with the binary signal distribution of the SBM, on  information externality. In practice people often observe informative signals that have  more than two possible values. For example, a corporate manager might receive  a profitability forecast about a possible investment project that can take one of many possible values. 

The natural questions that arise are: for which signal distributions do information cascades still arise?  When do agents eventually herd  upon identical actions? Under what conditions does society converge to the correct action in the long run? And is long-run behavior fragile to small information shocks?
Understanding these questions requires  technical specifics, so this section is more formal than much  of this survey. 

As before, let $s_n$ be agent $I_n$'s private signal. As in the SBM, private signals are i.i.d.\ conditional on the state $\theta$. But we now consider  a general signal distribution. This setting is very flexible, as we impose no restriction on the set of possible values that signals can take.\footnote{For example, it can capture a situation in which agents have private information about the precision of their own signals. To see this, consider a signal that is comprised of two parts. The first is a symmetric binary signal taking values of either $\l$ or $\h$, as in the SBM. The second is a value $p$ that is random and independent of the state, and which gives the precision of the binary signal.}   
We impose the restriction  that private signals are inconclusive about the state, i.e.,   the {\em private belief} $b_n = \Pr{\theta=H}{s_n}$
 must be in $(0,1)$. 
% As in the SBM, after observing the actions of her predecessors, agent $I_n$ takes action $a_n \in \{H,L\}$, with utility $1$ if the action matches the state, and $0$ otherwise.

Before taking her action, agent $I_n$ observes the actions $a_1,a_2,\ldots,a_{n-1}$ taken by her predececessors , and forms her posterior belief $\Pr{\theta=H}{a_1,\ldots,a_{n-1},s_n}$. She then maximizes the probability of matching the state, i.e., chooses $H$ if this belief is above $1/2$, and $L$ if it is below. 
Let  the {\em social belief} be defined as
\begin{align*} 
  P_n = \Pr{\theta=H}{a_1,\ldots,a_n},
\end{align*} 
the belief held by an outside observer who can observe the agents' actions and has no other information. It is convenient to consider the {\em social log-likelihood ratio}
\begin{align*} 
  R_n =  \log\frac{P_n}{1-P_n}.
\end{align*} 
Likewise, let  the {\em private log-likelihood ratio}\footnote{This definition of the private log-likelihood ratio holds when private signals are discrete. In the case that the conditional private signal distributions have densities, $r_n$ will equal the logarithm of the ratio of the densities.} be
\begin{align*} 
  r_n = \log\frac{\Pr{s_n}{\theta=H}}{\Pr{s_n}{\theta=L}}\cdot
\end{align*} 
By Bayes' rule, agent $I_n$'s {\it posterior log-likelihood ratio} is equal to $R_{n-1}+r_n$:
\begin{align*} \log\frac{\Pr{\theta=H}{a_1,\ldots,a_{n-1},s_n}}{\Pr{\theta=L}{a_1,\ldots,a_{n-1},s_n}} &= \log\frac{\Pr{a_1,\ldots,a_{n-1},s_n}{\theta=H}}{\Pr{a_1,\ldots,a_{n-1},s_n}{\theta=L}}\cdot\frac{\Pr{\theta=H}}{\Pr{\theta=L}}\\ &= \log\frac{\Pr{a_1,\ldots,a_{n-1}}{\theta=H}}{\Pr{a_1,\ldots,a_{n-1}}{\theta=L}}\cdot\frac{\Pr{\theta=H}}{\Pr{\theta=L}}\cdot\frac{\Pr{s_n}{\theta=H}}{\Pr{s_n}{\theta=L}}\\ &=R_{n-1}+r_n.
\end{align*} Since the utility of an action depends only on the state, the expected utility of that action, given the information available to the agent, is a simple monotonic function of the posterior log-likelihood ratio. In consequence,    an  agent will optimally take  action $H$ whenever this ratio is above 0, and  action $L$ otherwise.\footnote{When $R_{n-1}+r_n=0$, the agent is indifferent between the two actions. The tie-breaking convention is not important for our conclusions in this section.}

The social belief $P_n$ (or, equivalently, the public log-likelihood ratio $R_n$)   converges almost surely as the number of agents becomes large. This is a general property of any sequence of beliefs of Bayesian agents who collect more information over time. It follows from the {\em Martingale Convergence Theorem}, and the fact that the sequence $(P_n)_n$ is a {\em bounded martingale}.  

%%%%%%%%%%%%%%%% 
\subsection{Bounded vs.\ unbounded private signals}

The distribution of private signals, and hence of the {\em private beliefs} $b_n = \Pr{\theta=H}{s_n}$, is key to understanding whether incorrect cascades occur, and whether agents eventually learn the state. If agents may receive arbitrarily accurate signals, then it is intuitive that agents will end up making very good decisions. In contrast, bad outcomes are possible when signals are not arbitrarily accurate, as in the SBM.

The concept of arbitrarily accurate signals is formalized with the notion of unbounded signals. We say that private signals are {\em bounded} if the resulting private belief $b_n$ is not arbitrarily extreme, i.e., there is some $\varepsilon >0$ such that the belief is
supported between $\varepsilon$ and $1-\varepsilon$:
\begin{align*} \Pr{b_n>\varepsilon \text{ and } b_n < 1-\varepsilon}=1.
\end{align*}
We say that private signals are {\em unbounded} if, for every $\varepsilon>0$, the probabilities $\Pr{b_n<\varepsilon}$ and $\Pr{b_n>1-\varepsilon}$ are both non-zero.
With an unbounded signal distribution, signal realization sometimes have arbitrarily high informativeness.

Boundedness is a property of  the possible beliefs induced by a signal, rather than the range of the signal values per se. An  example of a bounded signal on   $[0,1]$ is one which has  density  $f_L(s)= 3/2-s$ in state $L$ and $f_H(s)= 1/2+s$ in  state $H$. If, instead, the conditional densities in the two states are $f_L(s)= 2-2s$ and $f_H(s)= 2s$, then the signal is unbounded. The log-likelihood ratio $\log \frac{f_L(s)}{f_H(s)}$ (and therefore the belief) is bounded  in the first case, but can take any value in the real line in the second case.\footnote{A signal can be  neither bounded nor unbounded; for example, it could be the case that the private beliefs can take  values arbitrarily close to 1, but are bounded away from 0.} 

As shown in \citeasnoun{smith:sorensen:2000},  agents  eventually learn the state (in a sense to be made precise later)  if and only if signals are unbounded. If the set of possible signal values is finite, having unbounded signals is equivalent to having some signal values that reveal the state. We have ruled out signals that reveal the state, but in fact their presence has the same implications for eventual  learning of the state as unbounded signals more generally. 

\subsection{Evolution of the Social Belief}
\label{subsec:EvolSocialBelief}

From the viewpoint of an outside observer, the agents' actions form a simple stochastic process. The action process has an elegant stationarity feature: an agent's  action depends on history  only through   the social log-likelihood ratio, which captures all the information contained in past actions  about the probability of the $H$ state. So  
 following \citeasnoun{smith:sorensen:2000}, we denote by \begin{align}
    \label{eq:rho}
    \rho(a,R',\theta') = \Pr{a_n = a}{R_{n-1}=R',\theta=\theta'}
\end{align}
the probability that agent $I_n$ takes action $a$ when the social log-likelihood ratio is $R'$ and the state is $\theta=\theta'$. The stationarity property is manifested in the fact that the function $\rho$ does not depend on $n$.

Using $\rho$, we can define a \emph{deterministic} function $\psi$ that gives the social belief at time $n$, given the social belief at time $n-1$ and the action at time $n$, so that $R_n = \psi(R_{n-1},a_n)$. Since an observer of social information updates her LLR based on the information contained in the latest action, the updated belief satisfies 
\begin{align}
    \label{eq:psi1}
    \psi(R_{n-1},a_n)&=R_{n-1}+\log\frac{\rho(a_n,R_{n-1},H)}{\rho(a_n,R_{n-1},L)}.
\end{align}
This derives from applying  Bayes rule to calculate the updated social belief $P_n$ from $P_{n-1}$ and $a_n$.

\subsection{Asymptotic Learning, Cascades, and Limit Cascades}
\label{section:asymptotic-learning-cascades}

We now put information cascades in a broader social learning perspective, by defining asymptotic learning---a situation in which the social belief  becomes arbitrarily accurate---and consider the conditions under which incorrect cascades versus asymptotic learning occurs. We also define limit cascades, which are a variant of cascades. 

\subsubsection{Asymptotic learning}

We say that there is {\it asymptotic learning} if the social belief $P_n$ almost surely tends to 1 with $n$ when the state is $H$, and to 0 when the state is $L$. Two other possible definitions for asymptotic learning are the following:
\begin{enumerate}
    \item The sequence of actions $a_1,a_2,\ldots$ converges almost surely to $\theta$, i.e., all agents from some $I_n$ on take the action that matches the state.
    \item The probability that agent $I_n$ takes an action that matches the state tends to one with $n$.
\end{enumerate}
It is straightforward  in this setting that both of these are equivalent to asymptotic learning. Intuitively, if beliefs are almost perfectly accurate, so are actions. And for actions to be almost always accurate under the infinite range of possible signal realization sequences, beliefs must also almost always be highly accurate.

\subsubsection{Information Cascades}

Our definition of  information cascade in \S \ref{sec:binary-model}, that the agent takes the same action regardless of her private signal,  can be rephrased as follows. An information cascade occurs when, for some agent $I_n$,  the social log-likelihood ratio $R_{n-1}$ is either so high or so low that no private signal can overturn the cascade, i.e., cause $I_n$'s action to depend on $I_n$'s private signal. This occurs if and only if  
\begin{align}
    \label{eq:psi}
    \psi(R_{n-1},H) = \psi(R_{n-1},L) = R_{n-1},
\end{align}
i.e., whenever the social belief remains unchanged after observing the action. 

\subsubsection{Limit Cascades}

A {\em limit cascade} occurs if the agents' actions converge, the limiting action is sometimes incorrect, and each agent  chooses either  action with positive probability.
As with asymptotic learning, the probability that an agent chooses differently from her predecessor decreases quickly enough that, with probability one, from some point on, all agents choose the same action.
As with cascades, this action is often incorrect, i.e., the social outcome is idiosyncratic.
More formally, in a limit cascade the  social log-likelihood ratio $R_n$ tends to a limit that is not $+\infty$ or $-\infty$, but (unlike cascades proper) does not reach that limit in finite time (\citeasnoun{smith:sorensen:2000});  \citeasnoun{gale:96} provides an example of essentially the same concept. Thus, the outside observer's belief converges to an interior point in $[0,1]$.

If the latest $k$ agents all take the same action, then as $k$ increases,  the next agent  follows the latest action under a wider range of private signal values.
This causes the actions of successive conforming agents to become less and less informative.
But since each agent's  action always depends on her private  signal, 
actions are never completely uninformative. This contrasts with information cascades, in which actions become completely uninformative.
In limit cascades, social learning becomes arbitrarily slow without ever  stopping.

Moreover, the informativeness of conforming agents' action drops so quickly that agents do not, in the limit, learn the state.
Empirically, the implication is essentially identical to that of information cascades: society may fix upon an incorrect action forever.

\subsubsection{Conditions for Different Social Learning Outcomes}

There are three possibilities for the asymptotic outcome of the process: (i) an information cascade, (ii) a limit cascade, and (iii) neither of the two. In case (iii), there is asymptotic learning. This is a consequence of the Martingale Convergence Theorem which implies $R_n$  converges almost surely to some $R_\infty \in [-\infty,+\infty]$.  
\citeasnoun{smith:sorensen:2000} describe the relation between signal structures and these outcomes. 

In the SBM, the only possible outcome is an information cascade. As shown by BHW, this holds more generally when the set of possible signal values is finite.  
When signals are bounded but not finite, either cascades or limit cascades can occur.  
For example,   there are always limit cascades when private signals are distributed uniformly on $[0,1]$ in  state $L$ and have density $f(s) = 1/2+s$ on $[0,1]$ in  state $H$ (\cite{smith:sorensen:2000}). 

When signals are unbounded, there is always is asymptotic learning. In other words,  almost surely $R_n$ converges to $+\infty$ or $-\infty$ and the social belief converges to either 1 or 0, respectively.  
Asymptotic learning fails when cascades or limit cascades occur with positive probability. In this case, the social belief converges to some $P_\infty \in (0,1)$, and the probability that agent $I_n$ chooses correctly tends to $\max\{P_\infty,1-P_\infty\} < 1$.\footnote{For example, when a cascade starts, the social belief reaches $P_\infty$. If $P_\infty > 0.5$, the agent adopts, and this is correct with probability $P_\infty$. Similarly, if $P_\infty < 0.5$,   the agent rejects, and this is correct with  probability $1- P_\infty$.}

Why do bounded signals block asymptotic learning? Signals are bounded if and only if there is some finite $M$ such that the private log-likelihood ratio $r_n$ induced by $I_n$'s private signal is contained in $[-M,M]$. 
Thus, whenever the social log-likelihood ratio $R_{n-1}$ exceeds $M$ in absolute value, the agent disregards her own signal, since $R_{n-1}+r_n$ must have the same sign as $R_{n-1}$. It follows that \eqref{eq:psi} will hold if $|R_{n-1}| > M$, and a cascade  ensues. Thus it is impossible that $\lim_n R_n = \pm\infty$.

Unbounded signals  imply that information cascades and limit cascades are impossible, and asymptotic learning is obtained. To see why, suppose that $R_\infty$, the limit of $R_n$, is finite, so that a limit cascade or a cascade occurs.  Let $q_n$ be the probability that  $I_n$ chooses $H$ conditioned only on social information. This probability depends only on $R_n$, and is strictly between 0 and 1 for any $R_n$, since signals are unbounded. In the long run since $R_n$ converges to some finite $R_\infty$, $q_n$ will approach some probability $0 < q_\infty < 1$.  So an observer who sees the action sequence is essentially receiving an infinite stream of binary signals of approximately constant precision. This observer's beliefs become perfectly accurate, which contradicts the premise that the social log likelihood ratio  is bounded. 

\subsubsection{Herds and Speed of Convergence }
\label{section:speed of convergence}
As defined in \S \ref{sec:binary-model}, we say that a {\em herd} starts from agent $I_n$ if all later agents take the same action, that is, if $a_m=a_n$ for all $m \geq n$. 
BHW show that when private signals are finitely supported, a herd occurs with probability one, i.e., there is herding. \citeasnoun{smith:sorensen:2000} further  show that herding occurs more generally for any bounded or unbounded signal structure.
It follows that when signals are unbounded, with probability one agents eventually take the correct action. This is somewhat surprising; it happens despite the fact that each agent in the herd has a positive probability of taking either action. 

Asymptotic learning may be much slower than the social optimal rate  that would be achieved if private signals were disclosed publicly. In this sense asymptotic learning may be very inefficient. 
The information externality remains;  each agent takes the action that is best for her without regard to the information that her action conveys to others.

So a natural question is: under asymptotic learning, how long does it take for a correct herd to form, and how does the speed of learning depend on the signal distribution? Smith and S{\o}rensen ask whether  the expected time until a correct herd forms (i.e., until the last time the wrong action is taken) is always infinite.  \citeasnoun{hann-caruthers/martynov/tamuz:18} show that this expectation can be either finite or infinite, depending on the tail of the distribution of the private beliefs.

\citeasnoun{rosenberg2019efficiency} also study how quickly  agents converge to the correct action when signals are unbounded. They focus on the expectation of the \emph{first time} that a \emph{correct} action is taken; this is (perhaps not obviously) very closely related to the time at which a correct herd starts. Their very elegant main result is that this is finite conditioned on a given state if and only if $\int \frac{1}{1-F(q)}dq$ is finite, where $F$ is the cumulative distribution function of the private belief in that state. Thus, when private beliefs have very thin tails on the ``correct'' end---i.e., very low probabilities of extremely informative correct-direction signals---the expected time can be long.

Even with  continuous action spaces and unbounded signals, the learning process may still be very slow, as shown by  Vives (\citeyear{vives:1993,vives:1997}). Similarly, \citeasnoun{chamley:04a}, \citeasnoun{ADLO:11} and \citeasnoun{dasaratha2019aggregative} provide  models with slow convergence.  
The review of \citeasnoun{gale:96} points out that very slow asymptotic learning, as occurs in several models, may be observationally indistinguishable from complete learning stoppages as in settings with incorrect cascades.

\subsubsection{Modeling Considerations}

It is  largely a matter  of convenience whether to model signals as unbounded, so that incorrect cascades never occur, or bounded,  so that  either cascades or limit cascades occur. There is no way to empirically distinguish a signal distribution that includes values  that are extremely rare and highly informative, versus one where such values do not exist at all. So for applications, either modeling approach is equally acceptable. (For a similar perspective, see \citeasnoun{gale:96} and \citeasnoun{chamley:04b}.) 

In many applied contexts, the signal space is in fact finite.\footnote{For example, people often obtain information signals through casual conversation with limited nuance. (``Is that movie worth seeing?'' ``Yeah.'') There is  evidence from psychology that there are minimum distinguishable gradations in sensory cues. People often obtain information from experiments with a small number of possible outcomes, such as learning whether a job applicant is or is not a high school graduate.  Course  grades are discrete (such as A through F), as is learning whether  a new acquaintance is married or unmarried, how many children the individual has, or who the individual voted for in the last election.} 
In such contexts, we expect to see cascades rather than limit cascades. 

Regardless of whether the setting implies cascades or limit cascades, if the setting is modified so that acquiring private signals is even a little costly, then, as discussed in \S \ref{sec:EndogSignalAcquisition}, agents eventually stop acquiring private information. So information aggregation is completely blocked regardless of whether  signals are unbounded.

As we saw in \S \ref{sec:binary-model}, cascades are fragile with respect to small shocks.
Since limit cascades, as with cascades proper, do not result in asymptotic learning, they are also fragile, for the same reason that cascades proper are.
Public disclosure of a modestly informative  signal (one that is at least as precise as any single agent's private signal) can break the limit cascade, possibly leading to a later limit cascade  in the opposite direction. 

\subsection{Partial Cascades}

The logic of information cascades is that with a coarse action space, an agent may choose an action independently of her private signal. In consequence,   her action does not add to the social information, and
 asymptotic learning fails. Even when there is asymptotic learning, the improvement of social information tends to be delayed. 

This logic is a special case of a more general point:  that a coarse action space  reduces the informativeness of an agent's action. This effect is in part mechanical. For example, if there were just one possible action, taking that action would convey no information. However, this effect is exacerbated by the information externality that self-interested agents have no incentive to convey information to later agents.\footnote{We mainly  focus on coarseness in conjunction with information externalities.  Coarseness is much less of a problem with altruistic agents, as they could use their actions to convey their private information to others efficiently (see \cite{cheng2018deterministic}).}

To capture this more general point, \citeasnoun{lee:1998} defines a weaker notion of information cascade.    A {\it partial cascade} is a situation where an agent takes the same action for multiple signal values. (This terminology is due to \citeasnoun{brunnermeier:01}; Lee uses the term ``cascades'' for this concept.) An information cascade proper is the special case in which an agent takes one action for all the possible signal values. 
Limit cascades are partial cascades.
Partial cascades occur trivially when there are more possible signal values than actions.  But even when the action space is rich, partial cascades (and cascades proper) can occur.

Lee applies this notion to a model of information blockage and stock market crashes (as discussed in \S \ref{sec:ApplicationsExtensions}). In the model of \citeasnoun{hirshleifer/welch:02}, partial cascades result in either excessive maintenance of early actions (`inertia'')  or excessive action shifts (``impulsiveness''). Overall, there is a wide array of settings in which  partial cascades can occur and hinder information aggregation.

\ifdefined\ALLCOMPILE\else\end{document}\fi % 10

\ifdefined\ALLCOMPILE
\saycurrfile
\else
\usepackage{jelall}
\renewcommand{\citeasnoun}[1]{\textcolor{red}{#1}}

\begin{document}

\fi

%%%%%%%%%%%%%%%%%%%%%%%%%%%%%%%%%%%%%%%%%%%%%%%%%%%%%%%%%%%%%%%%%%%%%%%%%%%%%%%%%%%%%%%%%%%%%%%%%%%%%%%%%%%%%%%%%%%%%%%%%%%%%%%%% 
\section{Endogenous timing of actions} 
 \label{sec:EndogTiming}

When faced with an irreversible decision, people often have an option to either act immediately, or to defer their decision to a time in which more information will be at their disposal. Examples include purchasing a product or making a long-term investment.

Consider a modification of the SBM in which, at any given date, any agent is free to choose among three options:  adopt,  reject, or delay. So there is no exogenous sequencing in the order of moves.   Adopting or rejecting is irrevocable.

The benefit to delay is that an agent can glean information by observing the actions of  others.  Thus, delay generates option value.
The cost of delay could take the form of  deferral of project net benefits, and of ongoing expenditures needed to maintain the option to adopt. 
Since acting early confers a positive information externality upon other agents, in equilibrium there can be   excessive delay.

The incentive to delay in order to observe others plays out most simply when agents have heterogeneous signal precision, as in the next subsection. This  setting  provides insight into a wider set of  models considered in the remainder of this section.

\subsection{Delay with Heterogeneous Signal Precision}

Suppose that agents differ in the precisions of their binary private signals (extending the SBM to agent-specific values of $p$).  At a given point in time, an agent can {\it act} by choosing project $H$ or  project $L$; or alternatively,  can  delay.
As in the SBM, project $H$ is optimal  in state $H$, and  $L$ is optimal  in state $L$.
As discussed in BHW p. 1002, high-precision agents have less to gain from waiting to see the actions of others---in the extreme, a perfectly-informed agent ($p_i = 1$) has nothing to gain from waiting.  So we focus on equilibria in which,  among agents with $h$ signals, those with higher precision adopt earlier than those with lower precision.\footnote{Our focus  on equilibria in which agents choose different timing implicitly requires that time periods be short relative to differences in possible precision. This ensures that an agent with lower precision would prefer to wait one period and learn from a higher-precision agent rather than act simultaneously.  To avoid technicalities, we suppose that agents have precisions drawn independently from discrete distributions such that no two agents have the same precision.}

Suppose first that each agent's signal accuracy is known to all.  The binary signal of the agent with the highest precision dominates the signals of all other agents, as in the fashion leader example discussed in \S\ref{sec:binary-model}. Once the agent with highest precision has acted ($H$ or $L$), all remaining agents are in  a cascade on the selected action. Intuitively, the agent with second-highest precision (the one with the next least gain from delay) acts immediately rather than waiting to learn from others. Since this action is uninformative, for similar reasons, so do all others.

Now suppose instead that precisions are only privately known.  In equilibrium, agents can  infer  the signal accuracy of other agents from time elapsed without action.  In the continuous-time model of \citeasnoun{zhang:97}, this results in an equilibrium in which delay fully reveals precision.   Each agent has a critical maximum delay period, after which, if there are no actions by others, she  becomes the first to act.  The higher the precision, the shorter the critical interval.   Again, all agents wait until the highest-precision agent  acts.  At that point, all other agents immediately act in an  investment cascade.

In this model,  cascades are {\it explosive} in the sense that there is an  initial time period during which all agents delay, and then,  once the highest-precision agent acts, others immediately follow.  Furthermore, since the cascade is based solely on one  agent's signal, actions are also highly idiosyncratic.

In the real options model of  \citeasnoun{grenadier:99}, investors differ in the precisions of their private information about the value of an asset. If two high-precision investors with positive signals invest, all other agents are in an information cascade, and also invest.    Thus the main insight from \citeasnoun{zhang:97}  holds in Grenadier's setting as well---that high precision agents have a stronger incentive to act first, and that once this occurs, low precision agents have an incentive  to mimic,  triggering a flurry of activity.

The fact that low precision agents have a greater benefit to delay suggests that in a setting where there is a decision of whether to acquire information, and where this decision is observable to others, there is a strategic advantage to acquiring less information. 
But even in nonstrategic settings, agents may acquire too little information, since there are  externalities in information acquisition (\S\ref{sec:EndogSignalAcquisition}). 

\subsection{Adopt versus Delay as an Indicator of Degree of Optimism}

In the setting we have discussed, delay is informative about agents' precisions, but not about whether their signals favor project $H$ or $L$.  Suppose that the relevant decision is whether to adopt a given project versus delay, rather than which project to adopt (also with a possibility of delay). Then delay can be an indicator that the project is unattractive. 

The consequences of this are seen most simply in a model in which agents have identical precisions. Typically, the incentive to delay in such a setting rules out symmetric pure strategy  equilibria.  To see why, suppose that there were such an equilibrium in which all agents with $h$ signals adopt immediately.  Then their actions would accurately reveal the state. The value of defecting by waiting one period  would be very high, breaking the equilibrium. 
  Consider instead a proposed  equilibrium in which all agents with $h$ signals  delay one period.  
Then it would pay to defect from the equilibrium by acting immediately rather than acting one period later, since no information is obtained by waiting.

\citeasnoun{wang:17} focuses on asymmetric pure strategy equilibria of such a  model.  At a given point in time, let a {\it bunch}  be a set of agents each of whom  adopts if and only if her private signal is $h$. In the equilibria of interest, there is endogenous bunching. In a given period, some agents are in a cascade of waiting regardless of their private signals.  
The remaining agents are in a bunch. 
A larger number of adopts by members of a bunch is a more favorable indicator of state. As a signal, this number is more informative when the bunch is larger.
So as a bunch becomes larger in a proposed equilibrium, eventually it no longer pays to be a member of the bunch. This endogenously determines the maximum possible bunch size.  

When agents are patient enough, \citeasnoun{wang:17} shows that the bunch size is one agent. The equilibrium play of the bunch agents is much like play in the SBM, until the point in the SBM in which a cascade starts. At that point, in Wang's setting and equilibrium, all  agents who have delayed until then either invest or wait forever.  So the broad insights of the SBM can carry over to a setting in which agents choose the timing of their actions.

Turning to mixed strategy equilibria, \citeasnoun{chamley/gale:94} derive randomization in delay  in a setting in which a subset of agents randomly  receives the option to invest---an option that can be exercised at any time.  More agents receive the option to invest in better states, so receiving the option is a favorable indicator about state. It is the only signal that agents receive. Delay by a greater number  of agents conveys adverse information to others.  In the unique symmetric  perfect Bayesian equilibrium, agents with favorable signals (implicit in receiving the option)  delay with positive probability.\footnote{The idea that information externalities instead result in stochastic delay  is in the spirit of  \citeasnoun{hendricks/kovenock:89}, who examine an experimentation setting in which two firms  with private information decide how soon to drill for oil, where drilling causes the arrival of public information about the payoff outcome.} The information in delays can  cause additional investment to cease. Also, owing to the externality that agents benefit from waiting to observe what other agents do, there tends to be excessive deferral of investment. 

In contrast, if all agents have the option to invest, and each agent receives a direct private signal about the state, there is no bias towards underinvestment relative to full information aggregation,  as shown by \citeasnoun{chamley:04a}. At any date, the number of agents who have already invested is  a positive indicator about the state.  There are  multiple equilibria with different  thresholds for the belief about the state above which agents invest. 
In an equilibrium with a higher threshold, the information conveyed by the decision to invest is stronger. Thus, when  the threshold signal value is higher, the marginal agent has a higher informational benefit of waiting, which compensates for the higher cost of delaying longer.  This supports the equilibrium. 

Overall, these models of timing decision  reveal that inefficient behavior can be manifested in either inefficient delay, or  in a  rush to invest even in unprofitable projects. This suggests that social learning may generate shifts in investment activity that are reminiscent of observed industry-wide booms and busts, or macroeconomic fluctuations.
Such shifts occur within equilibria in the \citeasnoun{wang:17} and  \citeasnoun{chamley/gale:94} models, and might be viewed as occurring occasionally across multiple equilibria in \citeasnoun{chamley:04a}.

Sometimes  agents  may  learn by observing the experienced payoffs of other agents, as in \S \ref{sec:observability}.  Even when there is no private information, if past payoffs are observable, informational externalities can delay social learning from  payoffs (\citeasnoun{caplin/leahy:93}). When agents also have private information and have a timing choice as to when to act, as in the model of \citeasnoun{wagner:18}, in some cases incorrect cascades of delay occur. 

In practice, sometimes firms can repeatedly adopt and terminate projects. \citeasnoun{caplin/leahy:94} analyze information cascades in project decisions when firms can receive multiple private signals over time and can see the actions previously taken by other firms.
After an uneventful period of delay, there can be sudden crashes in which many firms terminate their projects at about the same time.    
Towards the end, with enough signals, firms essentially know the value, and take the correct action. Thus, incorrect cascades do not occur.

\subsection{Clustering in Time and Information Aggregation} 

Suppose that empirically we were to observe a tendency (i) for agents to take similar actions, as in models of information cascades, and (ii) for  actions to be clustered in time, as in models of explosive onset of cascades or avalanches (\citeasnoun{chamley/gale:94} and \citeasnoun{zhang:97}). 
Would this be evidence of incorrect information cascades, or are there non-cascade settings that generate similar patterns?

\citeasnoun{gul/lundholm:95}  offer a model in which there are no information cascades, and in which action and perhaps time clustering may occur. In the model, action is continuous, and therefore fully reveals an  agent's belief. 
 Agents can  choose when to act over continuous time, with  an  exogenous cost of delay. 
 Just as in a continuous action setting without a timing choice (as covered in \S  \ref{sec:ActionSpace}), social learning causes actions to be clustered in the action space even though there are no  cascades.
 
Gul and Lundholm further find that when there are two agents, clustering in {\it time} occurs. However, when there are more than two agents,  agents do not cluster in time (except for the last two agents); timing choices are strategic substitutes. So  models of incorrect cascades provide a more clearcut explanation than the Gul and Lundholm model for  time clustering. 

\ifdefined\ALLCOMPILE\else\end{document}\fi
 % 6
\ifdefined\ALLCOMPILE
\saycurrfile
\else
\usepackage{jelall}

\begin{document}
\fi

\section{Observability Assumptions}
\label{sec:observability}
\label{sec:LimitedObservability}

What an agent learns from others, and whether the agent falls into a cascade, depends on what the agent can observe about past history. What can be observed about  history is, in practice, highly context-dependent. Greater observability can take the form of observation of payoffs, or even private signals, not  just actions.    Observations can  also be  noisy or limited, as with  observation of only a specified subset of agents, a random sample, or a count of adoptions or other aggregate statistic.  Observability can  be asymmetric (as with greater observation of adoptions or of higher payoffs).  Furthermore, there can be meta-uncertainty, wherein agents are not certain whom their predecessors have observed.

Some key questions, under alternative  observability regimes, are whether  incorrect cascades must eventually be dislodged, whether there is asymptotic learning, and whether social outcomes remain fragile. Another key question is whether greater observability increases welfare.  We first discuss these issues in the context of rational settings, and then turn to imperfect rationality. 

\subsection{Rational Models}

There are many different  combinations of observability assumptions in  the literature. Rather  than systematically discussing the proliferation of different possible assumptions, our discussion focuses on some general themes. We organize the discussion in terms of the different general themes rather than by model assumption. 

%% leave these definitions, and then just use \observation{}.  when you want to change formatting later, you can do so by altering the definition once.  moreover, it renumbers automatically.

\newcounter{obscounter}
\newcommand{\observation}[1]{%
  \bigskip\noindent\stepcounter{obscounter}\textbf{Theme~\theobscounter. #1}}

\needspace{10\baselineskip}

Recall that in \S \ref{sec:GeneralSignalDistributions} we  defined  a  herd as starting at agent $I_i$ when all later agents do the same thing as  agent $I_i$.

\observation{When there is sufficient  observation of past actions (and possibly additional social information such as past payoffs), the probability that a  herd eventually starts is 1 in many models. That is, there is herding.}

As a benchmark, if there is no social observation, agents  act based on their own private signals, and there is  no herd.  However, with enough social observation of actions and perhaps payoffs,  more and more information is revealed, at least until agents eventually tilt toward one action. This intuition requires reasonably good observation of past actions. 
An example with insufficient social observation is the model of \citeasnoun{celen:kariv:2004} (which we discuss again in \S \ref{sec:networks}), where each agent observes the action of the immediate predecessor only. Examples with sufficient social observation include the models of \citeasnoun{banerjee:1992}, BHW, \citeasnoun{smith:sorensen:2000}, \citeasnoun{cao:han:hirshleifer:2011}, and models with costly information acquisition discussed in \S  \ref{sec:EndogSignalAcquisition}. 

We discuss the next two themes together.
 
\observation{When there is sufficient observation of past actions or payoffs (and possibly other social information), herds can be incorrect. Specifically, incorrect information cascades can occur, and can last forever with strictly positive probability.  So   in general, asymptotic learning may not occur.  \label{obs-sufficient}} 

\observation{Information cascades can cause insufficient {\it exploration} as well as poor information aggregation.  \label{obs-explore}}

When previous payoffs are observable but are stochastic given the state, there are two types of information externalities.  The first, as in the SBM, is that in choosing an action, an agent does not take into account that her action conveys information  about her private signal to later agents.  This is the problem of {\it  aggregation} of private information into actions.

The second type of information externality is in the {\it generation} of new information. Agents do not take into account the benefit to later agents  of observing the payoffs derived from the chosen action.  This is an externality in  exploration  (see, e.g., \citeasnoun{rob:1991}). Owing to this externality, in a social multi-arm bandit settings with  no private information,  asymptotic learning fails (\citeasnoun{bolton/harris:99}).  An analysis with quasi-Bayesian  agents is provided by \citeasnoun{bala:goyal:1998}.

In principle, with many agents, there are enough private signals for  information aggregation alone to induce asymptotic learning. 
Similarly, exploration alone could generate enough payoff information to pin down the realized state perfectly. 
These facts   raise the hope that when agents can  socially acquire both types of  information there would be asymptotic learning. After all, observation  of predecessors’ payoffs can  sometimes dislodge an incorrect cascade, thereby potentially resulting in new trials and  payoff observations of both choice options. 

Nevertheless, it turns out that even when payoffs are observed, there can be a strictly positive  probability that an incorrect cascade forms and lasts forever, i.e., outcomes are idiosyncratic.   As in the SBM, once a sufficient predominance of evidence from past actions and payoffs favors one action, agents start to take that action even when  their own private signals oppose it. The basic logic of cascades still applies; additional private information is no longer impounded in actions. So this cascade may be incorrect. In contrast with the SBM, such a cascade may be broken by the arrival of payoff information. But asymptotic learning is not assured.

To see this, consider a setting in which all past payoffs as well as actions are observable (\citeasnoun{cao:han:hirshleifer:2011}). Suppose that the payoffs to action $a$ are either $1$ or $-1$, and to action $b$ are either $2$ or $-2$. There are four equally likely states:  $uu, ud, du$ and $dd$, where the first entry indicates a high ($u$)  or low ($d$) payoff to action $a$ and the second  entry indicates a high or low payoff to action $b$. 

Once an action is taken, its payoff is known to all, an assumption that is highly favorable to  effective social learning.
Nevertheless, a problem of inadequate experimentation remains. If private signals and payoff information about action $a$ are initially favorable,  society can lock into $a$, for an expected payoff close to 1, without ever trying  $b$, whose payoff in state $uu$  of $2$ is even higher.
So the ability to socially acquire both types of  information (about  private signals and about payoffs) does not  solve the   information externality problem.
 
More generally, if payoffs are  stochastic even conditional upon the state (or observation of  payoffs is noisy), there is still a strictly positively probability that a given agent will cascade upon an incorrect action.  
Furthermore, agents sometimes  lock into an incorrect action forever even after having tried both alternatives any finite number of times.

A possible interpretation of the payoff signal is that it is an online review posted by an agent who has adopted.  In this  application, \citeasnoun{le:subramanian:berry:2016} show that the probability of an incorrect cascade can increase with the precision of the review. This is another instance of the principle of countervailing adjustment discussed in \S \ref{sec:binary-model}.

\medskip

Alternatively, the additional social information that agents obtain may be about the private signals of predecessors.
If {\it all} past private signals were observed, then trivially agents would converge to the correct action.   
In reality people do sometimes discuss reasons for their actions, but they often do not pass on the full set of reasons that they have acquired from others.
It is not hard to provide an  example with limited communication of private signals in which incorrect cascades form and, with positive probability, last forever. 
In the Online Appendix, \S \ref{subsubsec:RecentSignals}, we provide an extension of the SBM where for all $n > 1$, $I_n$ observes the private signal of $I_{n-1}$, and incorrect cascades still occur and last forever (see \citeasnoun{cao/hirshleifer:97a}).\footnote{An agent might communicate a sufficient statistic (such as the agent's belief) instead of one or a few private signals. In practice, this sometimes occurs, but people  also sometimes seem to convey one or two specific reasons rather than conveying an overall degree of belief.}  

Intuitively, seeing the predecessor’s private signal is much like directly observing an extra private signal oneself, which in turn is much like seeing a more precise signal.
In the SBM, a more precise private signal tends to make decisions more accurate, but  the probability that the long-run cascade is incorrect is still strictly positive. 
So there is reason to expect the same when the predecessor’s signal is observed.

 In particular, the logic of information cascades applies; a point is reached  when the information implicit in past actions overwhelms the bundle of the agent’s own signal and the predecessor’s signal. 
Such a preponderance of evidence is far from conclusive. 

\observation{When each agent observes only a random sample of past actions, incorrect information cascades can occur, and may last forever. 
So asymptotic learning does not necessarily occur.} %  \label{obs-sample}

The conditions for this theme are potentially compatible with those of  Themes 1 and 2, so there are settings where the  conclusions of both hold, i.e., there can be incorrect cascades that last forever, and there can be herding.  To understand this  theme, consider a sequential setting with random sampling of past actions and with no information about the order of past actions (see \citeasnoun{smith/sorensen:20}).   With bounded signals, information aggregation tends to be self-limiting, because more informative  actions tend to encourage  cascading upon the preponderance of actions in the agent's observation sample. Whenever such cascading occurs, the agent's signal is not impounded into the action history. 
 
Suppose, for example, that past agents' actions were to become so accurate that even a single sampled  $H$ action would be sufficient to overwhelm the most extreme possible opposing private signal value. 
Then an observer will sometimes be in a cascade upon the predominant action in the agent's sample (e.g., in a sample of size $k$ in which all observations in the sample are of the same action). 

This reasoning suggests that owing to the possibility of information cascades, learning with random sampling may be quite slow.  Indeed, a stronger claim is true: so long as all agents observe a sample size of at least 1, asymptotic learning fails  (\citeasnoun{smith/sorensen:20}). 

To see this, consider the case of a  sample size of $N = 1$, where private signals are symmetric and binary. The departure from the SBM is that  each agent observes the  action of one randomly selected predecessor instead of all predecessors. Suppose that a point is  reached where for some agent $I_n$ this random observation is more informative than a single private signal. (If this never happens, of course asymptotic learning fails.)
Then agent $I_n$ would be in an information cascade, so  $I_n$'s action would be exactly as informative as a sample of one action from among  $I_n$'s predecessors. In consequence, the sample observed by $I_{n+1}$ is also more informative than $I_{n+1}$'s signal, so $I_{n+1}$ would also be in a cascade. Similarly, for all later agents, so information stops accumulating. Consequently, asymptotic learning does not occur.
This failure is similar to the fashion leader version of the SBM in Subsection \ref{subsec:fashion-leaders-precision}, in which information stops accumulating once an  action is observed with precision greater than an agent's private signal. 

A similar intuition also applies to the random sampling model of \citeasnoun{banerjee:fudenberg:2004}, in which it is possible that past payoffs as well as actions are observed.
\citeasnoun{banerjee:fudenberg:2004} consider  a setting with a continuum of agents, with an arrival rate of new agents that is equal to the death rate. Each  agent takes an action once upon birth, after observation of the actions of a subsample of past agents. The continuum-of-agents approximation has the feature that aggregate outcomes evolve deterministically.\footnote{The continuum of agents assumption disallows  fixation on a mistaken action owing to extinction of alternative action choices (see \citeasnoun{smith/sorensen:20}).  With agents arriving individually in sequence, there is a strictly positive probability that by chance all early agents choose the same action. When this occurs, a random sample by the next agent will only contain one action, which may potentially result in an information cascade wherein no agent ever switches.  In contrast, in \citeasnoun{banerjee:fudenberg:2004} the system is assumed to  start with a positive fraction of both types, so there are always samples containing even the rarer action.  This brings information to bear that might be lost with discrete agents, promoting asymptotic learning.}  
In sequential sampling settings, it is not in  general  obvious whether, with probability 1, agents will converge to the same action.  If there is early diversity in action choice, then there is always a  chance that an agent observes a set of predecessors who did not follow the currently-predominant action (\citeasnoun{smith/sorensen:20}). Similarly, if payoff information is also observed and if an early agent adopted a popular action and experienced low payoffs from doing so, there is always a chance that this agent is later observed.\footnote{\citeasnoun{monzon/rapp:14} consider a sampling setting in which agents also do not know their own positions in the decision queue. In this setting, under a stationarity assumption on  sampling rules, incorrect cascades can last forever.} 

If agents observe samples of payoff outcomes but not of the past actions that led to these outcomes, it is again possible that agents do not converge to the same action. 
The need to simultaneously draw inferences about what actions predecessors have taken and the performance of those actions can confound inferences. \citeasnoun{wolitzky:18} considers a setting  with two actions: action $R$(isky) has a probability of success that depends on state, and action $S$(afe) has a fixed probability of success that is state independent. If action $R$ potentially generates a higher state-contingent probability of success than action $S$, 
then outcomes become close to efficient if the size of samples becomes arbitrarily large.  However, when action $R$ always has lower probability of success (but has lower cost) than  action $S$, then even for very large samples, there is not asymptotic learning (\citeasnoun{wolitzky:18}).  

\observation{Reject cascades can occur even when agents observe the  aggregate number of adopts, but do not observe rejects.}

When only aggregate adoption counts are observed, sequencing information is lost. In general this can induce loss of two types of information.
First, an agent does not know  \emph{the order} in which past actions were taken.  
Second, an agent does not know \emph{how many} predecessors have acted---i.e., agents do not know their own positions in the queue. 
This occurs when an agent does not observe all past actions, one example being when an agent observes only adopts, not rejects. For example, it is not hard to obtain information about how many Tesla 3's have been sold, but we do  not observe how many people considered Teslas but opted not to buy.

In the setting of \citeasnoun{guarino:harmgart:huck:2011},  observation is asymmetric (we will refer to this  as only observing past adopts, not rejects), agents cannot see the order of predecessors’ actions, and they have no direct information about their own positions in the queue. Since all that an agent observes is how many predecessors adopted, there is no way for a cascade on reject to get started. (If it could, then even $I_1$ would reject, since $I_1$ does not see any past adopts nor does $I_1$ know that no agent preceded her. In consequence, all agents would always reject, which is not consistent with equilibrium.) 
So the possibility of cascading is limited to just one action, and indeed, with a large population, such a cascade occurs with probability one regardless of state. 

However, in sharp contrast, when agents do have some idea about their own positions in the queue (based, for example, on observation of own-arrival-time), cascades on either action can occur. 
This is because an agent who is probably late in the queue and who observes few adopts infers that others probably arrived earlier and chose to reject (\citeasnoun{herrera:horner:2013}). 

\observation{Contrarian actions can reveal that an agent has high precision.}  

\citeasnoun{callander:horner:2009} show that with private information about own-precision, agents sometimes act in opposition to the majority of the actions they observe.
Such contrarianism is possible even if only aggregate information about predecessors' actions is observed. 
Consider  the SBM, except that $I_3$ has a conclusive signal, whereas $I_1$, and $I_2$ have very noisy signals. We can think of $I_3$ as a ``local'' who knows whether a restaurant is good.  Even if $I_1$ and $I_2$ (low precision ``tourists'') adopt, if $I_3$ rejects, clearly $I_4$ will imitate $I_3$.  

Furthermore, even if $I_4$ is uncertain about whether $I_3$ is a local, $I_4$ can infer this from the fact that $I_3$ rejected after two adopts.  The very fact that $I_3$ came late and was in the minority (is a ``contrarian'') is an indicator of his high precision.

What if $I_4$ does not know the order of moves, only that one predecessor was in the minority (two adopts and one reject) and that (for simplicity) there was exactly one local?  If $I_1$ were local, then the two tourists, $I_2$ and $I_3$, would have imitated $I_1$ in the hope that she is  local, which would have generated a unanimous choice.  So this possibility is ruled out.  If $I_2$ is the local, there is a chance that $I_3$, a tourist who observed one adopt and one reject, chose an action that matched that of $I_1$ instead of $I_2$, so that the local is still in a minority. If, instead, $I_3$ is local, then $I_2$, being a tourist, would have imitated $I_1$; thus, the two adopts are by $I_1$ and $I_2$ and the reject by $I_3$, the local.    So overall, the preponderance of evidence that two adopts and one reject conveys to $I_4$ is that the local rejected. Being a contrarian can be an indicator of being well-informed, and if $I_4$ is a tourist, $I_4$ optimally follows the minority.  

As noted above, each agent observes only aggregated information about previous actions. In such models, there is no commonly-held social belief. Since  $I_n$ does not know  $I_{n-1}$'s decision, $I_n$ cannot perfectly infer  the history of aggregate decisions observed by $I_{n-1}$ (unless all agents before $I_n$ took the same action). So agents~$I_{n-1}$ and $I_n$ may have differing beliefs about decisions of earlier agents. This is also true of network models discussed in \ref{sec:networks}.

\subsection{Models with Imperfect Rationality}

As discussed in \S\ref{sec:binary-model}, the presence of overconfident agents can break incorrect cascades, improving long-run learning. 
Another important psychological source of bias is limited attention and cognitive processing. Models based on this are discussed next.

\subsubsection{Neglect of Social Observation by Predecessors}
\label{subsubsec:neglect-predecessors}

In a rational setting, an agent should adjust for the fact that the information in an observed action depends on whom the actor in turn is able to observe (see, e.g., \citeasnoun{ADLO:11}).
This requires complex chains of inference. 
Owing to limited attention, in reality people typically do not adjust appropriately. 
\emph{Persuasion bias}, also called \emph{correlation neglect}, is the phenomenon that people sometimes treat information they derive from others as independent even if there is commonality in the sources of this information---a type of double-counting.  
In non-Bayesian and quasi-Bayesian models of social influence this results in greater influence on group outcomes of agents who are more heavily connected in the social network (see \citeasnoun{demarzo/vayanos/zwiebel:03} and the review of \citeasnoun{golub/sadler:16}).

Correlation neglect is captured in a sequential quasi-Bayesian setting in \citeasnoun{bohren:2016}.
In Bohren's model, states are equiprobable, signals are bounded, and there is a given probability that each agent is {\it social} (observes the actions of predecessors) or {\it nonsocial} (does not observe any predecessor actions).  Whether an agent is social is unknown to others.
Agents may either underestimate this probability (a form of correlation neglect) or overestimate it (which could be called correlation overestimation).\footnote{\citeasnoun{guarino:jehiel:2013} offer an alternative  approach to modeling imperfect understanding by agents of the relation between others' private information and their actions.} 

Let $q$ be the probability that any agent observes the actions of predecessors, and let $\hat{q}$ be agents' perception of that probability.  
If the possible values of $\hat{q}$ are in an intermediate interval (an interval which includes $q$), with probability one the social agents eventually make correct choices.
The reason is that even if social agents fall into a mistaken cascade, observation of nonsocial agents tends to dislodge it. So long as there is reasonably accurate understanding of this process (intermediate $\hat{q}$), the beliefs of social agents converge to the true state.\footnote{A rational setting is a special case of Bohren's model with  $\hat q =q$. Continual arrival of nonsocial agents dislodges incorrect cascades in a rational setting as well.}

% Last sentence of BHW 1992
BHW point out a  benefit to  segregating early agents into individual informational quarantines, i.e., making each one nonsocial.  Similarly, \citeasnoun{banerjee:1992} points out that the efficiency of social learning can be increased if some early agents do not observe others.
A similar  effect arises  in the model of cascades with overconfident agents of \citeasnoun{bernardo/welch:01} discussed in \S\ref{sec:binary-model}.
 In either setting, such agents (overconfident, or nonsocial) make heavy  use of their own signals, which helps provide additional social information to later observers.

When $\hat{q}$ is below this intermediate interval, agents view past actions as often being taken independently, and therefore severely overestimate how informative these actions are about private signals. This reasoning is similar to the mechanism of \citeasnoun{eyster/rabin:10}.
When, by chance, a strong enough preponderance of agents favors one of the available actions, the system falls into a correct or incorrect cascade.
So the preponderance of one action tends to grow over time.
Since agents think this growing preponderance is coming largely from independent private signals, social agents grow increasingly confident in the correctness of the latest cascade.  In the limit agents become sure of either the wrong state or the correct one.  

This possibility of strongly held faith in the wrong state provides an interesting contrast with the BHW cascades setting, in which there is failure of asymptotic learning but cascades are fragile. 
It also contrasts with a rational benchmark with continual arrival of nonsocial agents, in which asymptotically the beliefs of the social agents become arbitrarily strong, but converge to the correct state.

When $\hat{q}$ is large (i.e., to the right of the abovementioned interval), beliefs fluctuate forever, so again there is not asymptotic learning.
Even if, at some date, there were a very strong preponderance of action $H$, for example, agents would believe that this derives almost entirely from cascading by predecessors.
So when by chance (as must eventually happen) even a few nonsocial agents take the opposite action, the next social agent will no longer be in a cascade, and will sometimes choose $L$.

As Bohren points out, thinking that $\hat{q} < q$ can cause agents to have excessive faith in a sequence of identical actions relative to expert scientific opinion.  
In the social learning model of \citeasnoun{eyster/rabin:10}, correlation neglect takes a more extreme form---observers think that each predecessor decided independently based \emph{only} upon that agent's private information signal.\footnote{\citeasnoun{hirshleifer/teoh:03} and \citeasnoun{eyster/rabin/vayanos:18} apply such neglect of the signal-dependent behavior of others to financial markets.} 
In their model, state and actions are continuous. Beliefs about others are analogous to $q =1$ and $\hat{q} = 0$ in the \citeasnoun{bohren:2016} model.  
In consequence, the views of early agents are very heavily overweighted by late agents, convergence to the correct belief is blocked (even with sharing of continuous beliefs or actions), and agents become highly confident about their mistaken beliefs.

One lesson that comes from analyses of imperfect rationality and social learning is that biases that cause agents to be more aggressive in using their own signals, such as overconfidence, or such as overestimation of how heavily others have observed their predecessors, tend to promote social information aggregation. 
In contrast, persuasion bias tends to have an  opposite effect, causing agents to defer too much to history. 

Persuasion bias  and overconfidence are just two kinds of psychological bias that can influence social learning.  
Another general  approach to studying the effects of psychological   bias is to make exogenous assumptions about the agent's mapping from observed actions and payoffs into the agent's actions.  
\citeasnoun{ellison/fudenberg:93,ellison/fudenberg:95} provide pioneering analyses using this heuristic agent approach (see the Online Appendix, \S\ref{onlinesubsec:observability} for details).  
In recent years, behavioral economics has moved toward basing assumptions on evidence from human psychology, and endogenizing biased belief formation as part of decisions (e.g., \citeasnoun{daniel/hirshleifer/subrahmanyam:98}, \citeasnoun{rabin/schrag:99}). 
The model of overconfident information processing  in social learning of \citeasnoun{bernardo/welch:01} is an example of this. 

\citeasnoun{bohren/hauser:19} examine a setting that allows for a variety of types of possible psychological biases in social learning, including correlation neglect. In this model, signals are continuous (and may be unbounded).  
They focus on settings in which enough information arrives so that if agents were rational there would be asymptotic learning (via the arrival of either public signals or nonsocial agents).
However, owing to psychological bias, asymptotic learning can fail,  which can take the form of convergence to a mistaken action, permanent disagreement over action, or infinite cycling.  
For example, when agents overreact to private signals, and where there is a positive probability of nonsocial types, there can be  infinite cycling between actions. When agents underreact, there can be fixation upon a mistaken action.
Furthermore, when incorrect herds last forever, beliefs converge almost surely to the incorrect state.
So consistent with \citeasnoun{bohren:2016}, and in contrast with the BHW model, longer herds become increasingly stable.

\ifdefined\ALLCOMPILE\else\end{document}\fi
 % 13
\ifdefined\ALLCOMPILE
\saycurrfile
\else
\usepackage{jelall}
\renewcommand{\citeasnoun}[1]{\textcolor{red}{#1}}

\begin{document}
\fi

\section{Costs of acquiring information (signals or observation of past actions)}
\label{sec:EndogSignalAcquisition}

People often have a choice of whether or not to acquire information.
We next  examine the effects on social learning of costly acquisition of either direct private information signals about the state (Section \ref{subsec:CostlyPrivateSignals}) or about predecessors' actions (Section \ref{subsec:CostlyPreActions}).  

In a  scenario with exogenous private signals and information cascades, such as the SBM, the  signals of late agents do not contribute to social knowledge,  as such signals do not affect actions. 
When agents can acquire private signals, it is unprofitable  to do so if the signal will not (or is unlikely to) affect the agent's action. So in such  settings there is often a similar conclusion, that late agents  mimic their predecessors. 

If private signals are costless, then asymptotic learning occurs when private signals are unbounded (as noted in \S\ref{sec:GeneralSignalDistributions}), and may occur when the action space is continuous (as described in \S\ref{sec:ActionSpace}). A positive cost of observing private signals degrades learning substantially.  A uniform conclusion of several papers to be discussed is that even in settings with unbounded private signals or continuous action spaces, asymptotic learning does not occur if there is even a small positive cost of investigating. 
In practice, costs of gathering or processing information are likely to be positive.
So these results suggest that asymptotic learning will not be achieved. 

On the other hand, if private signals are costless,  introducing a  cost of {\it observing predecessor's actions} can {\it improve} social learning. In such a setting, an agent with a very informative signal realization may choose not to observe others' actions. Thus, her action conveys greater incremental information, which benefits later agents. 

%%%%%%%%%%%%%%%%
\subsection{Costly acquisition of direct private signals about state}
\label{subsec:CostlyPrivateSignals}

Costs of acquiring  private information introduce another information externality of  social learning. In deciding whether to buy a signal, agents do not take into account the indirect benefit that accurate decisions confer upon later observers.

This externality is illustrated by modifying  the SBM of \S\ref{sec:binary-model} so that agents have a choice in   acquiring  private signals about the state. Each agent can choose to pay some cost $c>0$ and observe a binary signal with given precision $p$, or to pay nothing and observe no signal. In this setting,  agents $I_n$, $n > 1$ will not acquire a signal, no matter how small the cost. To see this, suppose that the cost is sufficiently small  that it pays for $I_1$ to acquire a signal. Then $I_2$'s social belief is either $p$ or $1-p$, depending on whether $I_1$ chose $H$ or $L$. Even if $I_2$ were to acquire a private signal, imitating $I_1$'s action remains a weakly optimal action for $I_2$, regardless of the signal realization. Thus, the signal has value zero, so $I_2$ will not purchase it, and instead  imitates $I_1$. Agent~$I_3$ understands this, so by the same reasoning, it is not optimal for $I_3$ nor for any  subsequent agent to  acquire information. All agents imitate $I_1$. So the social outcome impounds even less information than in the SBM.

In more general settings as well, when there are  at least small costs  of acquiring the private signal about state, agents stop acquiring private signals, resulting in complete information blockage. So only a few individuals end up acquiring private signals. This is a version of the ``Law of the Few'' (see 
\citeasnoun{galeotti/goyal:10}).  
As several authors have shown, under appropriate assumptions, asymptotic learning occurs if and only if infinitely many agents have access to unbounded signals at an arbitrarily small cost. 
If agents incur even a small cost of acquiring information, incorrect cascades can arise and therefore the social outcome can be fragile.\footnote{This assertion refers to a slightly generalized definition of cascades: acting irrespective of the value of a potential private signal owing to the fact that the agent chooses not to acquire the private signal.} This point applies in several models of costly information acquisition discussed below.

The main ideas of an early contribution on costly signal acquisition by \citeasnoun{burguet/vives:00} can be seen in a simplified model with unbounded signals and continuous actions and states. Agents seek to choose an action as close as possible to the value of a continuous state  $\theta$.  The common initial prior on $\theta$ is a normal distribution with mean $ \mu_0$ and precision $\rho_0$.  Agent~$I_n$  takes action $a_n$ after observing the actions  of predecessors and a conditionally independent signal $s_n$ which is normally distributed with mean $\theta$ and precision $r_n$.  Suppressing the agent subscript, each agent chooses the precision $r\ge 0$ of her private signal at cost $c(r)$, where $c(r)$ is convex and increasing. Each agent's objective is to minimize the sum of the cost of the signal and the expected quadratic   error.

An agent's unique optimal level of precision is easily inferred by her successors. Given this, and as actions are  continuous, each  agent's action  perfectly reveals the agent's private signal.  So the social information available to $I_n$ is the realization of the $n-1$ conditionally independent normal signals of $I_n$'s predecessors.  Consequently, the social belief of $I_n$ is summarized as a normal random variable $\theta_{n-1}$ with mean $\mu_{n-1}$ and precision $\rho_{n-1}$. Owing to normality, $\rho_{n-1}$ is equal to sum of the initial $\rho_0$ and the precisions of signals of predecessors.

Burguet and Vives show that asymptotic learning is equivalent to the requirement that social precision $\rho_{n-1}$ increases without bound of as $n$ increases. But as $\rho_{n-1}$ increases without bound, $I_n$'s marginal benefit of acquiring additional precision goes to zero. Consequently, asymptotic learning occurs if and only if  $c'(0)=0$, i.e., the marginal cost of additional precision is zero  at $r=0$. For instance, if the smallest available precision is $r_0>0$ at cost  $c_0>0$, so that $c'(0)$ is undefined (infinite), then learning is incomplete. Essentially, public belief becomes exceedingly accurate due to information acquisition by a large number of predecessors. So later individuals find it unprofitable to pay $c_0$ (or more) to acquire a private signal.
Thus, in a setting with unbounded private signals and a continuous set of actions, asymptotic learning occurs if and only if agents can acquire signals, no matter how noisy, at a cost arbitrarily close to zero. 

A similar conclusion obtains in various models with costly information acquisition.   For example, in \citeasnoun{mueller-frank/pai:16}, agents  acquire information about finite samples of predecessors'  actions and payoffs, and act in sequence accordingly. Agents differ in the  realizations of their  costs of sampling, and in their sample outcomes, both of which  are private information.  Asymptotic learning occurs (i.e., probability of taking the best action goes to 1) 
if and only if sampling costs are not bounded away from zero in the sense that costs can be arbitrarily close to zero for an unlimited number of agents.

As discussed in \S \ref{sec:ActionSpace}, \citeasnoun{ali:18} introduces a notion of responsiveness which, loosely speaking, requires that any change in an agent's beliefs changes the optimal action.  Ali shows that even with responsiveness, there may not be asymptotic learning if information is costly to acquire, since the benefit of greater accuracy may not outweigh the cost of information.  Responsiveness implies asymptotic learning if and only if  costs of gathering information are arbitrarily close to zero across agents.

Thus, the consistent message from \citeasnoun{burguet/vives:00}, \citeasnoun{mueller-frank/pai:16}, and \citeasnoun{ali:18} is that asymptotic learning is not robust to introducing costs of acquiring (or processing) private information.

In the SBM, agents  costlessly obtain a private signal and observe the actions of all predecessors. Suppose now that  for a small fixed cost an agent can {\it talk} to a predecessor to find out the rationale behind her action choice. In other words, the agent can learn the predecessor's  belief (which may reflect information that she has acquired in conversations with her predecessors).
Nevertheless, as long as there is even a small cost of such conversations, incorrect cascades  occur  with positive probability. Intuitively, as beliefs become increasingly informative, at some point it pays for an agent to simply follow the action of the agent's immediate predecessor rather than paying to learn the predecessor's belief. So there is not asymptotic learning.
 % see online appendix, \S\ref{subsubsec:CostlyPrePrivateSignals}  

%%%%%%%%%%%%%%%%
\subsection{Costly or noisy observation of past actions}
\label{subsec:CostlyPreActions}
 
It is often costly to observe the actions of others. For instance, in evaluating the decision to invest in a  startup firm, a venture capitalist can choose to devote time and effort  to gathering information about the decisions of earlier potential investors.

Consider a setting in which, for a fixed cost, an agent can observe all predecessors' actions. If the cost is high enough, early agents will not incur it, and therefore will act solely on the basis of their own private signals. However, for this very reason, at some point the action history may become so  informative that an agent  finds it worthwhile to learn the choices of  predecessors.  Once this  point is reached, all subsequent agents will also find  observation worthwhile. So observation costs can turn early agents into sacrificial lambs, to the benefit of many later decision makers.

Based on this,  from the viewpoint of improving the accuracy of decisions (and perhaps welfare as well), typically the observation cost should be  positive but not too large. With a zero cost, as in the SBM, cascades tend to be very inaccurate. With too high a cost of observing predecessors, no one will ever incur it, so that social learning is blocked. 

This insight is developed in the model of  \citeasnoun{song:16}. Each agent~$I_n$ first observes a costless private signal and then decides whether to pay a cost $c$ in order to observe the actions of up to $K(n)\le n-1$ predecessors (see also \citeasnoun{kultti/miettinen:07}). This generalizes the scenario just described  by allowing for selective observation of predecessors. Individual decisions about which predecessors to observe build a (directed) social network of observation links. However, agents do not know the full structure of the network, as each agent's decision about which predecessors to observe is private information.  

Consistent with the intuition above, social learning may improve as $c$ increases. To see why, suppose that  $K(n)=n-1$. That is, each agent can observe all predecessors after paying cost $c$. If $c=0$, it is optimal to observe all predecessors and results from the standard model apply.  With bounded private signals, there is a chance of incorrect cascades. But if   private signals are unbounded and the observation capacity is unlimited (i.e., $\displaystyle{\lim_{n\to\infty}K(n)=\infty}$), then asymptotic learning always occurs.\footnote{The endogenous social network  has expanding observations in the sense of \citeasnoun{ADLO:11}; see \S\ref{subsec:NetworksSeqAct}.}  

For sufficiently large  $c$, agents who receive strong signal realizations will choose not to observe any predecessor. As in the papers summarized in \S\ref{subsec:CostlyPrivateSignals}, there is no
asymptotic learning for such $c$, because  such agents  decide without observing others. On the other hand, such  agents increase the pool of social information,  so  agents who do acquire information (who will exist if  $c$ is not prohibitively large) decide correctly with probability that tends to one.

%%%%%%%%%%%%%%%%%%%%

\subsection{Costly Information Acquisition, Limited Observation and Groupthink}
\label{subsec:inv-costs-and-noisy-observation-actions}

Can social observation lead to decisions that are  even worse   than the decisions that agents would make under informational autarky? This might seem impossible, since any information gleaned by an  agent via social observation  is incremental to  her own private information.  However, \citeasnoun{janis/mann:77} and \citeasnoun{janis:82} emphasize that  ``groupthink'' in  group deliberations causes disastrous decision failures, as if interaction with others were harming instead of improving decisions.  There is also evidence suggesting that observation of others sometimes result in degradation in decision quality (a zoological example  is provided by  \citeasnoun{gibson/bradbury/vehrencamp:91}).
 % p. 175 Solo females made choices more correlated with male phenotypic traits (which should be associated with quality) than female groups. Furthermore, the mating success of males when females were making solo choices was poorly correlated with  mating success when femaled chose as groups, suggesting that the group decisions involved copying of females who had not themselves assessed males directly.

Analytically, when there are investigation costs and noisy observation of past action, agents in groups can come to decisions that are on average worse than if there were no social observation. Owing to free-riding in investigation by agents who are potentially knowledgeable, social observation can actually reduce decision quality.

To see this, first suppose that, as in the SBM, that others are observed without noise, but that there is a small cost of acquiring private signals.  As discussed at the start of \S\ref{subsec:CostlyPrivateSignals}, starting with $I_2$ all agents follow $I_1$, so the social belief reflects only a single signal.  This is no more accurate than if agents decided independently (though welfare is higher as agents save on investigation costs).

Suppose instead that observation of predecessors is noisy, where each agent observes binary signals about the actions of all  predecessors. Suppose further that all agents $I_n$ observe the same binary noisy signal about the action of any given predecessor  $I_m$, where $n > m$. 

If the noise is sufficiently small relative to the cost of the signal, the net gain to $I_2$ of investigating is still negative, so she still does not investigate.  But now, owing to observation noise, her action is less accurate than if she were to decide on her own.  So observation of others reduces decision quality relative to informational autarky. (Nevertheless, $I_2$'s welfare is higher than under autarky, as  observation of others economizes on observation costs.)

What about later agents?  Agent $I_3$  also just follows $I_3$'s signal about $I_1$'s action.\footnote{Agent $I_3$ ignores her signal about $I_2$'s action, because she knows that $I_2$ imitated $I_1$ based on the same signal realization about $I_1$'s action that $I_3$ observes. So if  $I_3$'s signal about $I_2$'s action differs from $I_3$'s signal about $I_1$'s action, $I_3$ knows that this discrepancy {\it must} be caused by error in observation of $I_2$'s action.} 
The same applies to all later agents, so everyone's action is less accurate than if they had decided independently.   In a related setting,  suppose that agents observe only the latest predecessor. In this case noise can compound repeatedly until a point is reached at which an agent again pays to acquire a private signal (\citeasnoun{cao/hirshleifer:97a}). 

An important empirical question in social learning settings is who makes better decisions on average, the  agents who follow the predominant action, or those who deviate.
In the SBM (and in the general BHW cascades model, which allows for  nonbinary signals as well), in any realization, it is the later agents who are in a cascade, so those in a cascade have observed more predecessors than those who precede the cascade. If there are many agents, then such cascading agents predominate. So in expectation, those agents who take the predominant action are better informed than deviants.

Notably, this prediction is reversed when there is modest observation noise and costly investigation.  
Now deviants are more accurate, because they acquire a signal directly, whereas cascading agents copy a garbled version of past actions.  This garbles the information content of the single past action that was needed to trigger the cascade.

\ifdefined\ALLCOMPILE\else\end{document}\fi

 % 8
\ifdefined\ALLCOMPILE
\saycurrfile
\else
\usepackage{jelall}
\renewcommand{\citeasnoun}[1]{\textcolor{red}{#1}}

\begin{document}
\fi

\section{Payoff Externalities}\label{sec:externalities} 

The SBM focuses on information externalities, under which  an agent's action affects the payoff of another agent by influencing the other agent's action. Often, however, there are payoff externalities, wherein an agent's action directly affects the payoff of another agent. 
Externalities or network effects have been analyzed in settings without information asymmetry (see \citeasnoun{katz/shapiro:86}, \citeasnoun{arthur:89}, and in settings in which actions affect agents' reputations \citeasnoun{scharfstein/stein:90}). 
We focus on the interaction between  payoff externalities, information, and social learning. This topic is discussed extensively  in \citeasnoun{chamley:04b}.\footnote{We discuss pecuniary externalities, i.e., changes in the price of adoption due to predecessors' actions,  in \S\ref{subsec:MktprSL}.}

We distinguish between an externality that is  (i) backward looking only, or (ii) both backward and forward looking.  In an  externality of type (i), an agent's payoff depends only on predecessors' actions. An example is  an agent's decision to join one of two queues, where the cost of waiting is increasing with the length of the queue.  In such a situation, agents have no incentive to influence the inferences of later agents. Our primary focus here is  on  such settings.

In an externality of type (ii), an agent's payoff  depends on the actions of both earlier and later agents. An example is the decision of individuals arriving in sequence to line up to get into a restaurant, if there is disutility from dining in a crowded restaurant.  This can result in strategic incentives. 
A literature on sequential games with learning encompasses strategic issues (see, e.g., \citeasnoun{jackson/kalai:97}, and \citeasnoun{dekel/fudenberg/levine:04}).

%%%%%%%%%%%%%%%%%%%%%%%%%%%%%%%%%

Payoff externalities, such as network externalities or congestion, can be captured by modifying the utility function of the SBM.  Recall that, $d_n$ is the difference between the number of predecessors of $I_n$ who have taken action $H$ and action $L$. 
 Consider the following  utility function:
\begin{eqnarray}
u(\theta,H, d) & = \indicator{\theta=H}+\epsilon d \nonumber \\
u(\theta,L,d) & = \indicator{\theta=L}-\epsilon d, \label{eq:payoff-extern} 
\end{eqnarray}
 where setting  $d=d_n$ give the utility of agent $I_n$. 
When   $\epsilon > 0$, there is   complementarity between the actions of different agents. This tends to reinforce cascades and herding. Recall that in the SBM, if $I_2$ sees a private signal contrary to $I_1$'s action, $I_2$  randomizes between the two actions. In contrast, in the  setting here, when  $\epsilon>0$, $I_2$ strictly prefers to imitate $I_1$ regardless of her private signal realization. Thus, all agents follow the action of $I_1$. 

If $\epsilon < 0$, there are  {\it negative} payoff externalities,  such as congestion costs. Now the interplay between social learning and negative payoff externalities is more interesting.  
Under negative externalities and backward looking social learning, and if the externalities ($\epsilon$ above) are not too large in absolute value, then agents, at least for a time,   imitate predecessors. \citeasnoun{veeraraghavan:debo:2011}  model congestion as the cost of waiting in a queue with random service times.  The queue length conveys favorable information about the value of the service provided.  As long as the waiting cost is small relative to the difference in the length of the two queues, agents ignore their private information and join the longer queue in a cascade. But when the difference is sufficiently large, the extra waiting time from a longer queue can outweigh the favorable inference. 

\citeasnoun{EGKR:14} describe how backward-looking negative externalities prevent action fixation and improve social learning without eliminating incorrect information cascades.  We illustrate  with a variation on the SBM.  As before, there is a payoff component of +1 deriving from taking the correct action ($a=\theta$ when the state is $\theta=L$ or $H$). In addition, there is a negative payoff component deriving from congestion costs.  The authors use a generalization of the utility function in (\ref{eq:payoff-extern}). So the utility of agent $I_n$ from taking action $a$ at history of actions $h_{n-1}$ is
\[
u(\theta, a, h_{n-1}) \ = \ \indicator{a=\theta}-k c_a(h_{n-1}),
\]
where $k > 0$ is a  constant and $c_a(h_{n-1})$ is the congestion cost of taking action $a$ at history $h_{n-1}$. We focus on the case where $k$ is small, which provides insight into the robustness of the SBM to the introduction of small negative externalities. 

Consider two illustrative cases:\vspace{-3mm}
\begin{enumerate}
    \item[(i)] {\it Absolute congestion costs:} $c_a(h_{n-1})=n_a(h_{n-1})$, the number of predecessors who have taken  action $a$ at history $h_{n-1}$. Absolute congestion costs increase without bound.\vspace{-3mm}
    \item[(ii)] {\it Proportional congestion costs:} $c_a(h_{n-1})=\rho_a(h_{n-1}) \leq 1$, where $\rho$ is the proportion of predecessors who have taken that action $a$ at history $h_{n-1}$. Proportional congestion costs are bounded above.
\end{enumerate}\vspace{-3mm}

Under absolute congestion costs but not under proportional congestion costs, costs  can in principle grow arbitrarily large. Loosely speaking,  proportional congestion costs  describe  applications in which queues are gradually processed rather than being allowed to grow arbitrarily long. When there are proportional congestion costs of modest magnitude, on the whole the  main conclusions of the SBM about incorrect cascades and herding carry through, because with $k$ small, the informational incentive to imitate outweighs opposing payoff interaction effect. 

For the case of absolute congestion costs, as in the SBM, an $H$ cascade starts the first time that $d_n=2$. (A similar analysis applies starting with an $L$ cascade if $d_n =-2$ is reached first.) However, this cascade is temporary, as the  congestion cost of action $H$ increases over time. Eventually, an agent is reached who finds  it optimal to switch to  $L$ if and only if the agent sees the signal $\ell$.  This agent's action reveals her signal, and an interlude of informative actions continues  until  another cascade starts, this time at some threshold $|d_n|>2$.

Ultimately, permanent cascading, meaning a situation in which all agents starting from agent $I_n$ make choices independently of their private signals,  must  start. Remarkably, at this point agents alternate between actions! The intuition rests upon two observations. First, as the social belief approaches 0 or 1, an agent's belief about the true state becomes less sensitive to the agent's private signal. Second, even though agents become almost certain about a state, they do not herd upon the action corresponding to that state. 

To see the second point, suppose that $k=0.1$. Then even if $I_n$ is almost  certain that $H$ is the true state, she prefers action $L$ if $d_n > 10$,  prefers action $H$ if $d_n< 10$, and is indifferent between the two actions if $d_n =10$.   It follows that if the social belief of $I_n$ is that the state is very likely $H$, $I_n$ prefers $H$ even after seeing signal $\ell$ if $d_n\le 9$, and prefers $L$ even after seeing signal $h$ if $d_n\ge 10$. So  agents cascade in alternation between actions $H$ and $L$ as $d_n$ alternates between 10 and 9 forever. Such cascading starts with probability one, so there is never herding upon a single action.

Also,  for any given $k$, as in the SBM, there is no asymptotic learning. It follows that actions are potentially fragile, i.e., sensitive to the introduction of small informational shocks.  However,   \citeasnoun{EGKR:14} show that when absolute congestion costs $k$ approaches zero, the system becomes arbitrarily close to achieving  asymptotic learning. Of course, in applied settings the magnitude of congestion costs are often non-negligible. 

What can we conclude about how congestion externalities affect social learning and information cascades? 
With bounded congestion costs (as in the case of proportional costs discussed above), the insights of the SBM carry through. The informational pressure to imitate eventually outweighs congestion costs, resulting in  cascades  and idiosyncratic behavior. In contrast, in the case of unbounded congestion costs, as occur in the case of congestion costs that are proportional to the number of adopters of an action, learning outcomes differs qualitatively from the SBM. While cascades occur with probability one, herding does not. Instead, agents alternate between the two actions owing to congestion costs.  Once cascades start all learning ceases, so there is no asymptotic learning. 

\ifdefined\ALLCOMPILE\else\end{document}\fi

 % 4
\ifdefined\ALLCOMPILE
\saycurrfile
\else
\usepackage{jelall}
\renewcommand{\citeasnoun}[1]{\textcolor{red}{#1}}

\begin{document}
\fi

%%%%%%%%%%%%%%%%%%%%%%%%%%%%%%%%% 
\section{Price Determination in Markets with  Social Learning}
\label{sec:Prices}

In a market for a product or security of uncertain value, the decision to buy depends on price, an agent's (buyer's) private information signal, and the decisions of predecessors that the agent can observe.  This raises several questions. Does  the price setting process promote or prevent cascades, including incorrect ones? How should a seller manage the social learning process? How does social learning affect securities markets efficiency? What are the welfare consequences of social learning and cascades?  We first discuss the case of monopoly pricing, in which the seller chooses prices to maximize expected profits. We then turn to competitive price-setting.

%%%%%%%%%%%%%%%%%%%%%%%%%%%%%%%%% 
\subsection{Social Learning under Monopoly}

A monopolist may have an incentive to set price low enough to induce cascades of buying. The dynamics of prices and buying depend on whether the monopolist must commit to a single price or can adjust prices in response to observation of the purchase decisions of  early potential buyers. We first discuss the fixed price case. This applies to products with menu costs (costs of changing prices; \citeasnoun{sheshinski/weiss:77}). It also applies to  the sale of equity shares of a firm in an Initial Public Offering (IPO), since a fixed price per share is mandated by U.S. law.

\subsubsection{Fixed Price Case}

As in \citeasnoun{welch:92}, consider an uninformed monopolist who offers to sell one unit of a product to each agent  in a sequence at a fixed price for all buyers until the monopolist's supply of the product, $n$ units, is exhausted.\footnote{\citeasnoun{caminal/vives:96} and \citeasnoun{caminal/vives:99} study social learning about product quality via market share in a duopoly.} In this setting  the net gain to adopting (buying the product) is endogenously determined.
The monopolist is risk neutral and does not discount the future. As in the SBM, each agent receives a binary private signal about the state, which is the unknown value $\theta$ of the product, i.e., its quality, and can observe the choices of all predecessors.

We assume that when indifferent the customer  buys.  If the price is sufficiently low, all agents buy independent of their private signals, i.e., there is an information cascade of buying. At a somewhat higher product price, an agent's choice depends on her private signal, in which case her choice reveals her private information to subsequent agents.  If the price is high enough, a non-buying cascade  occurs, but such a price  is not optimal for the seller. 

Consider three possible prices ($P = P_\ell, P_0$, and $P_h$), where:
  \begin{itemize}
  \item%
    $ P_\ell  = \E{\theta}{\ell} = 1 - p$. The first agent  starts a buying cascade, yielding the monopolist a  per-buyer expected net revenue of $P_\ell$ for the first  $n$ buyers (and zero thereafter).
  \item%
      $ P_0  = \E{\theta} = \frac 1 2$. A buying cascade starts if  the difference between the number of buys and the number of  sells reaches 1. A non-buying cascade starts if  this difference reaches $-2$.
  \item%
     $ P_h  = \E{\theta}{h} = p$.   A buying cascade starts if and when the buy/sell difference reaches +2.  A non-buying cascade starts if the difference reaches $-1$.
  \end{itemize}
 From the monopolist's perspective, demand is fragile.  Just a few early agents with negative signals would cause buying to collapse.   For a sufficiently low signal precision $p$, the profit-maximizing price is $P_\ell$, so that all agents buy.   
 
Intuitively, with low precision, $P_\ell = 1{-}p$ is only slightly below~0.5, so it is not worth risking  collapse of demand for slightly higher prices $P_0$ or $P_h$. Since $P_\ell < E[\theta]$, the seller  underprices the product. This implication is consistent with the empirical finding of underpricing in  IPO markets in many countries. For higher precision $p$, raising the price to $P_0$ is worth the risk, so that there is not underpricing. 

In a setting with a uniform prior on $\theta$, \citeasnoun{welch:92} shows that when the seller also has a private information signal, a seller whose private signal indicates  higher quality (a high-quality seller, for short) sets a higher price (with higher failure probability) to separate  from a lower-quality seller type.
\citeasnoun{welch:92} focused on  cascades in the IPO market.
Empirically, \href{s/px19veb62rx1kap/amihud-hauser-kirsh-2002.pdf?dl=0}{\citeasnoun{amihud:hauser:kirsh:2003}} find that IPO opportunities for investors tend to be either  heavily oversubscribed or undersubscribed. This is consistent with information cascades, in which there is positive feedback from early  investor decisions to later ones. 

\subsubsection{Flexible Price Case}

In many applied contexts a monopolist can change prices  after observing each agent's buying decision.\footnote{\citeasnoun{newberry:16} studies empirically how fixed versus flexible pricing regimes affects buyer social learning and seller profits.} In  \citeasnoun{bose:orosel:ottaviani:vesterlund:2008}, the seller is risk neutral and uninformed, and can modify price after observing each buyer's decision. 
The relevant prices for the seller to consider (in the SBM as modified above) are a low price that leads to an immediate sale and a higher price that results in a sale only for a high signal. At each stage, the values of the low price and the high price depend on the actions of preceding buyers.

Bose et al. show that the seller starts with a price that induces the first buyer to reveal her private signal. 
Once enough information is revealed, the value of additional information revelation to the seller is low. Eventually, the seller fixes  a low price that induces a  buying cascade thereafter.  As the seller's discount factor increases, the value discovery phase on average becomes longer, and more information is revealed. In the limit as the discount factor goes to one, there is complete value discovery, and the seller achieves a revenue of $\E{\theta}$ per buyer. This is the best conceivable asymptotic outcome for the seller, as rational buyers will never, on average, pay more than their  ex ante expected valuation. 
 
%%%%%%%%%%%%%%%%%%%%%%%%%%%%%%%%% 

\subsection{Social Learning in Competitive Markets}\label{subsec:MktprSL}

We have seen that monopolists sometimes set prices to induce information cascades. In contrast, it is not immediately clear whether cascades of buying or of selling will occur under  competitive price setting in  product or securities markets.  With regard to securities markets,  when an agent buys or sells based on her private information, market prices should incorporate at least some of the agent's private information. This makes it less attractive for an observer to imitate the trade, which opposes the formation of a trading cascade. 

To see the consequences of this effect, consider a setting in which each agent receives a signal about an object with value $\theta = 0$ or 1. The agent can buy one unit at the  market maker's ask price $A$, sell one unit to the market maker at bid price $B \leq A$, or not trade.  If the market maker is uninformed and there are no noise traders, then the market maker sets bid and ask to prevent trade (e.g., $A \geq 1$, $B  \leq 0$), as otherwise the market maker would lose money. A  bid-ask spread that is tight enough to accommodate  trade would allow the trader on average to make money conditional on a trade. But this occurs iff the market maker on average loses money conditional on a trade. 
In equilibrium, there is a cascade on the choice of not trading. 
No trade is a standard conclusion from models of information and securities markets with no noise trading (\citeasnoun{glosten/milgrom:85}).

This cascade is quite different in nature from cascades in the SBM. In the SBM, there is a fixed net value of engaging in each action. In contrast, in this competitive market example, the endogenous price is set such that the net benefit of buying or of selling is nonpositive.

However, in models of securities markets with noise traders (\citeasnoun{glosten/milgrom:85}), market makers can profit at the expense of noise traders, so bid ask spreads are set to accommodate trading. The adjustment of competitive market price to reflect private information discourages the occurrence of information cascades of buying or selling by making it optimal for agents (traders) to use their private information. 
Intuitively, a hypothetical  price that led  all agents to want to buy, or all to want  to sell, would not clear the market.

\citeasnoun{avery:zemsky:1998} point out that in  a basic setting with noise traders, there are no information cascades.
 To see this, it is convenient to modify SBM as follows:   
\begin{itemize}
    \item Each trader (agent) takes one of three (rather than  two) actions: buy one share, sell one share, or hold (do not trade). Each trader $I_n$ trades only once, at date $n$.
    \item Traders buy from or sell to perfectly competitive risk-neutral market makers.
\end{itemize}  
  The value of the asset is $\theta=0$ if the state is $L$ or $\theta=1$ if the state is $H$. With probability $1-\mu$, trader $I_n$ has private information (with signal realizations $\ell$ or $h$ as in the SBM) about the value of the asset and with probability $\mu$,  $I_n$ is an uninformed noise trader.  
 Traders' information types  are independently distributed and privately known. 
  
There are two prices in each period $n$: an ask price $A_n$ at which $I_n$ may buy the stock, and a bid price $B_n$ at which $I_n$ may sell the stock.  An informed $I_n$'s utility from buying is $\theta-B_n$, and from selling is $A_n-\theta$.  
Thus, based on ${\cal S}_n$, the publicly observed history of trades by agents $I_1,I_2,\ldots, I_{n-1}$, and on $s_n$, her own private signal, an informed $I_n$ sells if $\E{\theta}{{\cal S}_n, s_n}< B_n$, buys if $\E{\theta}{{\cal S}_n, s_n}> A_n$ and holds otherwise. Noise traders buy, sell, or hold with probability 1/3 each.  Perfect competition among market makers implies that the bid and ask prices satisfy
\begin{equation}\label{compr:1}
B_n=\E{\theta}{{\cal S}_n, a_n=\mbox{Sell}}\ \le \ \E{\theta}{{\cal S}_n}\ \le \ A_n=\E{\theta}{{\cal S}_n, a_n=\mbox{Buy}},
\end{equation}
where $a_n$ is $I_n$'s action. 

In an information cascade, the action of an informed trader is uninformative as it does not depend on her private information. A noise trader's action is always uninformative. 
Thus, if an information cascade were to start then (\ref{compr:1}) would be satisfied with equality and $A_n=B_n = \E{\theta}{{\cal S}_n}$. But then informed traders would  sell if $s_n=\ell$ and buy if $s_n=h$ because
$$\E{\theta}{{\cal S}_n, \ell} <B_n = \E{\theta}{{\cal S}_n}=A_n < \E{\theta}{{\cal S}_n, h},$$
which is a contradiction.  Hence, information cascades do not form.  Since there is no cascade and $I_n$'s action has information content, the inequalities in (\ref{compr:1}) are strict.
 Moreover, an informed $I_n$ buys if $s_n=h$ and sells if $s_n=\ell$ regardless of the public history ${\cal S}_n$. Over time, bid and ask prices converge to the true value of the stock and volatility of the stock price decreases.

Avery and Zemsky define a concept of herds that is adapted to financial markets.  We will refer to this concept as a  ``momentum herd.'' An informed trader is in a {\it momentum herd} if the trader's optimal action based on her social and private information is contrary to the optimal action the trader would have taken had she moved first (i.e., if she had no social information and  faced  the initial bid and ask prices).\footnote{The adjective `momentum' differentiates this concept from the usual notion of herding in most social learning models. Momentum herd behavior differs in that the action depends on the signal realization. An agent with a negative signal might be in  a buy (momentum) herd after a certain history, but by definition an agent with a positive signal cannot be in a buy herd.}  
Thus, an informed $I_n$ is in a {\it buy (momentum) herd} if 
\begin{eqnarray*}
\mbox{(i) $I_n$ would sell in period 1:} & &  \E{\theta}{s_n}<B_1\ =\ \E{\theta}{a_1=\mbox{Sell}} \\
\mbox{(ii) $I_n$  buys in period $n$: $\quad \ \ \ $}  & & A_n\ =\ \E{\theta}{{\cal S}_n, a_n=\mbox{Buy}}\  <\  \E{\theta}{{\cal S}_n, s_n}.
\end{eqnarray*}
In our modified SBM setting, condition (i) implies that $s_n=\ell$ while condition (ii) implies that $s_n=h$.  Thus, a buy herd is impossible in this example. A similar argument rules out a sell herd in the modified SBM.

Avery and Zemsky present an example with three states and non-monotone signals in which momentum herds are possible.\footnote{The signals are non-monotone in the sense that the posterior distribution of state  conditional upon signal is  not ordered by first-order stochastic dominance. That is, the signal distribution does not satisfy the Monotone Likelihood Ratio Property.}  
A more general treatment is provided in \citeasnoun{park-sabourian:2011}, who provide necessary and sufficient conditions for momentum herds, and show that momentum herds are possible with monotone signals.

Incorrect momentum herds  can arise, as  illustrated in \citeasnoun{cipriani-guarino:2014}.  Cipriani and Guarino modify the Avery-Zemsky stock-market model by dividing time into days, where each day consists of a finite number of trading periods. Each trader acts once in her lifetime, in one trading period on a particular day. 
At the end of each day~$d$, after all trading for the day is over, the value of the asset for that day, $\theta^d$, is revealed. The value of the asset may change from one day to the next; for instance,  $\theta^{d+1}$ is equal to $\theta^d+1$, $\theta^d$, or $\theta^d-1$.
%$\E{\theta^{d+1}}=\theta^d$.  
On each day, investors who are active on that day act in an exogenous sequence.
In addition to knowing $\theta^d$ and previous trades on day $d+1$, a trader acting on day $d+1$ receives a private signal about $\theta^{d+1}$ before trading. 

As in the Avery-Zemsky model, information cascades do not occur, as prices adjust to reflect prior trades. 
A momentum herd is  incorrect if investors in the herd buy on a day when the value of the asset declines  (from its level the previous day) or sell when the value rises. \citeasnoun{cipriani-guarino:2014} provide an example with monotonic signal distributions in which incorrect momentum herds can arise. 
Using NYSE data to estimate their model, \citeasnoun{cipriani-guarino:2014} find  evidence of momentum herds and  informational inefficiencies (due to a failure of prices to  impound private signals).

In contrast with the  above finding of  no cascades in   trading in competitive markets, in some  models there  are cascades of no-trade.
\citeasnoun{romano:07} shows that information cascades occur  if the model of  Avery-Zemsky is modified so that the market maker incurs a transaction cost with each trade.\footnote{A similar result is obtained by \citeasnoun{cipriani/guarino:08a} when the transaction cost is incurred by traders.} Information asymmetry  about the asset value decreases as successive traders buy or sell. Ultimately, a no-trade cascade starts when the value of an informed trader's private information is less than the cost of trading induced by the market maker's transaction cost.

\citeasnoun{romano:07} also examines a variation where the transaction cost is proportional to the stock price. After a price rise (due to more buys than sells), the cost rises, resulting in a  cascade of no-trade. Also owing  to this  higher  cost, this cascade aggregates less information than a cascade after a drop in prices (which is accompanied by lower transaction cost). So a cascade during a boom is more fragile with respect to the arrival of opposing  public information than a cascade during a bust. When a boom reverses,  the transaction cost drops precipitously. Informed agents start trading again and the no-trade cascade is broken.

\citeasnoun{lee:1998} considers a model in which each trader incurs a one-time transaction cost that enables her to  trade repeatedly based upon a single private signal. Temporary information blockage is possible even without exogenous public information arrival. Each agent $I_n$ enters in period $n$, and after entering, can  buy or sell any amount of a risky  asset. 
Owing to the transaction cost, private information can be sidelined during  several periods with no trading.  
This quiescent interval is shattered if a later agent trades upon observing a sufficiently extreme signal.
Since multiple signal values can result in the same action the equilibrium at a given date is an example of a partial cascade as discussed in \S \ref{sec:GeneralSignalDistributions}. When agents suddenly start to trade, there is a sudden drop or jump in price. Lee calls this phenomenon an information avalanche.

There are settings in which information cascades of buying or selling by informed traders occur. 
In the preceding papers, informed traders have a common value for the asset, as is plausible for highly liquid assets (owing to the  opportunity to resell and the availability of close substitutes). 
In contrast, for illiquid assets, agents may differ substantially in their valuations, as modelled by  
\citeasnoun{cipriani-guarino:2008b}. When assets have a private value component, the authors show that information cascades, both incorrect and correct, may occur, with no asymptotic learning. Such heterogeneity can occur, for example, for shareholdings of privately held assets and firms, or when investors place value upon control rights.  Even though bid and ask prices adjust to reflect  previous trades, informed traders with low private valuations sell and those with high valuations buy regardless of their respective signals realizations. A similar result is obtained by \citeasnoun{decamps/lovo:06}, who consider trader heterogeneity in valuations deriving from  differences in risk aversion and initial endowments.

Also in contrast with the no-cascades result in Avery and Zemsky's setting, \citeasnoun{chari/kehoe:04} find that if agents own investment projects and have a choice as to when to trade them, then information cascades can occur.
In \citeasnoun{chari/kehoe:04}, agents decide when to buy or sell one unit of a risky project.
An agent  has the option to wait, but once an agent buys or sells,  she leaves the market (becomes inactive). In each period, one randomly-selected active agent receives a binary private signal about the value of the project. All active agents, including those who have not yet received a private signal, may buy or sell in any period in a market with bid and ask prices set competitively by market makers.
Agents observe the history of buy and sell decisions, as well as  prices. There is a discounting cost of waiting, but early on, uninformed and informed agents may prefer to wait to exploit the arrival of new information.  As public information accumulates, the value of further  information decreases with time, so a point is reached when  all active but uninformed agents take a decision (either all buy from or all sell to  market makers), thereby generating a possibly-incorrect cascade of buying or selling among the uninformed.
\footnote{This is a cascade in the sense of models with a cost of acquiring information as discussed in \S \ref{sec:EndogSignalAcquisition}. So where the model of \citeasnoun{avery/zemsky:98} rules out cascades by informed traders, here there are cascades by uninformed traders.}  

\ifdefined\ALLCOMPILE\else\end{document}\fi
 % 3
\ifdefined\ALLCOMPILE
\saycurrfile
\else
\usepackage{jelall}
\renewcommand{\citeasnoun}[1]{\textcolor{red}{#1}}

\begin{document}
\fi

%%%%%%%%%%%%%%%%%%%%%%%%%%%%%%%%%%%

\section{Heterogeneous preferences}
\label{sec:hetero}

The information that a vegetarian chose a restaurant has a different meaning for another vegetarian from the information that a meat-lover chose it.  If the preference types of past decision makers are common knowledge, then it is straightforward to draw an inference from  a predecessor's action about her information signal. But if instead agents have private information about their preference types, observers need to disentangle past private signals from preferences. We focus on this case of ignorance of others' preferences, under the assumption that preference types are independently distributed. 

In this setting, even when signals are unbounded, there is no guarantee of asymptotic learning, as is seen in the following example. There are two equally likely states, two equally likely preference types, and two actions. An action either matches or mismatches the  state.  The first type wants to match the state, and the second type wants to mismatch it. Each agent draws a conditionally independent signal, possibly unbounded,  from the same distribution. For each signal realization, the two types take opposite actions as their preferences are opposed. 

As the two preference types are equally likely, conditional upon either state, $I_1$ has an equal probability of choosing either action. So $I_1$'s action is uninformative. It follows that $I_2$, and by similar reasoning all later agents, also have equal probability of taking the two actions, and there is no social learning. 

In a model of social learning with heterogeneous preferences, \citeasnoun{smith:sorensen:2000} show that if agents have a finite number of preference types and signals are bounded, then asymptotic learning does not occur. This failure can take the form of  information cascades, limit cascades, or confounded learning as in the example above. The failure of asymptotic learning in this case is not surprising, since asymptotic learning fails with bounded signals even with homogeneous preferences.

In contrast, \citeasnoun{goeree/palfrey/rogers:06} find that if there is a continuum of preferences types, then asymptotic learning is possible. (See also \citeasnoun{AMMO:17} for a model of customer reviews with this feature.)
 An agent's payoff from taking an action is the sum of a common value, which is imperfectly known, and her private value. Agents receive private signals about the common value. When the range of   possible private values is greater than the range  of  possible common values,
an agent's action is always informative, i.e., cascades do not occur, and  asymptotic learning obtains. 

\citeasnoun{zhang:liu:chen:2015} compare a simple version of the heterogeneous preferences model of Goeree et al. (two actions, two states, and uniformly distributed private values)  with the SBM (homogeneous preferences). For sufficiently large $n$, $I_n$ has a higher probability of making a correct inference about the state under heterogeneous preferences than under homogeneous preferences. Zhang et al. argue that the homogeneous preferences case may fit applications such as a social media network of friends, whereas the heterogeneous preference case applies to social observation among strangers. Consequently, Zhang et al. suggest that sellers of low quality products may prefer to advertise on social media  networks of friends while sellers of high quality products may prefer to advertise on social media networks of strangers. 

So far, we have considered settings in which the true distribution of preference types is common knowledge. More generally, \citeasnoun{frick/iijima/ishii:20a} find that misestimation of this distribution (psychological bias) severely hinders social learning.  In their model, a large population of agents chooses binary actions repeatedly in each discrete period.  Action payoffs depend on the continuous state and on the agent's type.  
Initially, each agent observes a single private  signal, which may be unbounded.  In each subsequent period, agents are randomly selected to meet in pairs, with each observing the action that was taken by one other agent in the preceding period.  

As a benchmark case for this setting, if agents correctly understand the type distribution, then there is asymptotic learning.  
In the spirit of  \citeasnoun{goeree/palfrey/rogers:06}, random heterogeneous preferences preserve action diversity. This allows  the information content of agent private signals to be revealed over time.

However, even arbitrarily small amounts of misperception break asymptotic learning.  In the long run, agents approach full confidence in one state, regardless of the actual state.  Intuitively, when the action space is continuous, small misperceptions can repeatedly compound with successive random drawings and observations, so that misperceptions ultimately induce extreme beliefs. This contrasts with \citeasnoun{bohren/hauser:19} (discussed in \S \ref{subsubsec:neglect-predecessors}), who find failures of asymptotic learning only when psychological bias is sufficiently strong. It will be useful for future research to delineate more fully the circumstances under which small bias iterates to eventually generate  very large effects on social outcomes.  

%%%%%%%%%%%%%%%%%%%%%%%%%%%

 As discussed in \S \ref{sec:EndogSignalAcquisition},  in models with homogeneous preferences, even with responsiveness, asymptotic learning occurs only if
 costs of gathering information are arbitrarily close to zero across agents.
 One might think that when preferences are heterogeneous,  incorrect cascades would tend to be occasionally dislodged by the  arrival of  agent with deviant tastes. This suggests that even without responsiveness, if information costs are small, there may be asymptotic learning. However, in a model with heterogeneous preferences and a  cost of acquiring private information, \citeasnoun{hendricks/sorenson/wiseman:12} find that learning can be incomplete.\footnote{Hendricks et al. test the predictions of their model with online music market data generated by \citeasnoun{salganik/dodds/watts:06}.} In  \citeasnoun{bobkova/mass:20}, each agent  can acquire two pieces of   costly information: (i) about a common value and (ii) about the agent's private value. Her payoff is the sum of the two values. An arbitrarily small cost of information acquisition about the common value   blocks asymptotic learning.  
 
\ifdefined\ALLCOMPILE\else\end{document}\fi

 % 3
\ifdefined\ALLCOMPILE
\saycurrfile
\else
\usepackage{jelall}
\renewcommand{\citeasnoun}[1]{\textcolor{red}{#1}}

\begin{document}
\fi

\section{Cascades on social networks}
\label{sec:networks}
%%%%%%%%%%%

Social networks---from word of mouth networks in iron-age villages to modern online social media websites---play an important role in the spread and aggregation of information in human society. 
In general, what an agent learns by observing others depends on the agent's position in the social observation network. The overall structure of the observation network also affects aggregate social outcomes, such as whether there is asymptotic learning. So network structure is a source of empirical implications for behavior, and of insight for  policy interventions to improve welfare.  Accordingly, a large recent literature models social learning in networks.  

Networks increase model complexity. By placing heavy computational demands on  agents in the model,  the rationality assumption becomes less plausible.  It is hard to calculate expected utilities, as such calculation  requires drawing inferences about the state of nature, taking into account the inferences of other agents based upon the structure of the entire social network (\citeasnoun{mossel:sly:tamuz:2015}).  
So for   tractability, network economists often make strong assumptions about the geometry of the network, and for both tractability and realism, often  study non-Bayesian agents. Nevertheless, models of rational  utility maximization provide valuable  benchmarks for evaluating how different  heuristics for updating beliefs and choosing behaviors affect social outcomes. 

A key question  is   how the geometry of the social network affects  learning outcomes. As we will discuss, a  general lesson from both rational and boundedly-rational models is that egalitarian networks---loosely speaking, networks in which there are no agents who observe far more agents than do other agents, and there are no agents who are observed much more heavily than are other agents---tend to facilitate social learning (\citeasnoun{bala:goyal:1998}, \citeasnoun{golub:jackson:10}, \citeasnoun{ADLO:11}, and \citeasnoun{mossel:sly:tamuz:2015}).

We next discuss the spectrum of models of social learning on networks.
In \S \ref{subsec:NetworksSeqAct} we consider network models of rational social learning with sequential actions. In \S\ref{sec:repeated_actions}  we study models with repeated actions, featuring rational agents in \S\ref{subsec:rational-repeated} and heuristic agents in \S\ref{sec:heuristic_network_models}. 

%%%%%%%%%%%%%%%%%%%%%%%%%%%%%%%%% 
\subsection{Model Spectrum}

Models of social learning on networks  vary across a number of  dimensions.  We outline here the model spectrum, and in later subsections discuss the insights provided by  these models. 

\subsubsection{Rationality}

Substantial literatures examine settings with either Bayesian (i.e., rational) agents (\citeasnoun{parikh:krasucki:1990}, \citeasnoun{ADLO:11}, \citeasnoun{mossel:sly:tamuz:2015}, \citeasnoun{arieli:mueller-frank:19}), quasi-Bayesian agents (\citeasnoun{bala:goyal:1998},  \citeasnoun{molavi:tahbaz-salehi:jadbabaie:18}), and  agents who use (non-Bayesian)  heuristics (e.g., \citeasnoun{golub:jackson:10}).

We will say that agents are {\it quasi-Bayesian} if they use Bayesian updating to incorporate some of the information available to them, and either ignore the rest (as in \citeasnoun{bala:goyal:1998}) or use non-Bayesian updating to incorporate the rest, as in \citeasnoun{molavi:tahbaz-salehi:jadbabaie:18}.

Heuristic agents are those whose action choices are far from following any expected utility maximizing decision process or whose beliefs are not driven by Bayes' Law. For example, in the early and influential model of \citeasnoun{degroot:74}, agents repeatedly update their beliefs to equal the average of their social network neighbors' previous period beliefs.

In settings with repeated moves, rational  agents  either use exponential discounting or are myopic. Myopic agents maximize expected utility in each period, but completely discount future utility, and thus do not take into account the effects that their actions have on the actions of others. In contrast, forward-looking agents may wish to take actions that in the short run yield lower utility, in order to influence their peers to reveal more information in the future (see, e.g., \citeasnoun{mossel:sly:tamuz:2015}). The myopia assumption (see, e.g., \citeasnoun{parikh:krasucki:1990}, \citeasnoun{mossel:sly:tamuz:2012}, \citeasnoun{HMST:21}) is often made for tractability, to facilitate the study of rational agents. In some papers, such as \citeasnoun{gale:kariv:03}, agents are effectively made  myopic by assuming  that an agent's action has no effect on the behavior of others.\footnote{In the Gale and Kariv model, this occurs because there is a continuum of agents in each node of the network.}

\subsubsection{Sequential Single Actions vs.\ Repeated Actions}

Early social learning models assumed that each agent acts only once in an exogenously determined order (\citeasnoun{banerjee:1992} and BHW). Most of the network literature follows suit (notably \citeasnoun{ADLO:11}); other  models consider  agents who act repeatedly.  An early example of a quasi-Bayesian network model with repeated actions is that of \citeasnoun{bala:goyal:1998}. An early repeated action model with rational myopic agents on a social network is that of \citeasnoun{parikh:krasucki:1990}, who generalize the two agent model of \citeasnoun{geanakoplos:polemarchakis:1982} to a network setting. Later repeated action myopic models include \citeasnoun{rosenberg:solan:vieille:2009} and \citeasnoun{mossel:sly:tamuz:2012}. Forward-looking agents acting repeatedly on social networks were studied by \citeasnoun{mossel:sly:tamuz:2015}.

%%%%%%%%%%%%%%%%%%%%%%%%%%%%%%%%% 
\subsubsection{Signal Structure, Action Space, and State Space}

Most papers in the networks literature deviate in other ways from the SBM.  Some  allow for unbounded signals or non-atomic signals, as in \citeasnoun{smith:sorensen:2000}.  Unless otherwise mentioned, the results we discuss in this section apply to general signals, under the assumptions of a  binary state and binary actions. 

%%%%%%%%%%%%%%%%%%%%%%%%%%%%%
\subsubsection{Information about Who Others Observe}

In the basic cascades setting, each agent knows exactly who her predecessors have observed.  More generally, an agent's neighborhood of observation can be private information of that agent. In the model of \citeasnoun{ADLO:11}, agents know the distribution from which the neighborhoods of other agents are drawn, but not the realizations. In the imperfectly rational model of  \citeasnoun{bohren:2016} discussed in \S \ref{sec:ActionSpace}, there is a chance that any agent has an empty neighborhood. Misestimation of this probability by others leads to failures of asymptotic learning. 

%%%%%%%%%%%%%%%%%%%%%%%%%%%%%%%%% 
\subsection{Sequential Actions}\label{subsec:NetworksSeqAct}

In this section we consider models of rational social learning with sequential actions on networks. \citeasnoun{banerjee:1992} and BHW assume that every agent observes the actions of all predecessors in the queue. This is a simple network structure wherein agents can be identified in order of moves with the positive integers, and each agent $I_n$ observes the actions of all of her predecessors $I_m$, where $m < n$. 

A subsequent literature retains the exogenous ordering of actions, but relaxes the complete observation structure, so that each agent observes only a subset of her predecessors. We  define $I_n$'s {\em neighborhood}, $N_n$, as the set of agents whose actions agent $I_n$ observes before acting.

\citeasnoun{ccelen2004observational} study a model in which $N_n = \{I_{n-1}\}$: each agent observes her immediate predecessor. In their model, the state is equal to the sum of the agents' private signals, rather than being binary as in the SBM.  With this state and network structure, neither herding nor information cascades arises, but the probability that later agents mimic their immediate predecessors  tends to one.

\citeasnoun{ADLO:11} introduce a general network structure: the neighborhood $N_n$ of agent $I_n$ can be any subset of $\{I_1,\ldots,I_{n-1}\}$, and, moreover, can be chosen at random, exogenously and independently. They study  asymptotic learning (as defined in \S \ref{section:asymptotic-learning-cascades}):  under what conditions does the probability that agent $I_n$ takes the correct action tend to 1 as $n$ becomes large? Part of the answer is that asymptotic learning  never occurs when agents can observe all predecessors and have bounded signals (BHW, \citeasnoun{smith:sorensen:2000}), owing to incorrect information cascades (or limit cascades). Nevertheless, \citeasnoun{ADLO:11} show that asymptotic learning is possible even with bounded signals in some networks with incomplete observation structures. 

For example, the presence of {\it sacrificial lambs}---agents who are unable to observe others---can induce asymptotic learning. To see this, suppose that $N_n = \{\}$ with probability $1/n$, and that  $N_n = \{I_1,\ldots,I_{n-1}\}$ with the remaining probability $(n-1)/n$.  Sacrificial lambs act according to their private signals only.  The rest observe all their predecessors.  The sacrificial lambs choose the wrong action with a constant probability that does not tend to zero with $n$. But these mistaken actions become exceedingly rare, as the  frequency of sacrificial lambs tends to zero. Furthermore, their actions reveal independent pieces of information to their successors.  Because there are infinitely many sacrificial lambs, the rest eventually choose correctly with probability one.\footnote{The beneficial effect of sacrificial lambs is similar to an effect  in \citeasnoun{bernardo/welch:01} and \citeasnoun{bohren:2016}, wherein imperfectly rational agents act based on their private signals instead of cascading.}
\citeasnoun{ADLO:11} provide a more general condition that ensures asymptotic learning. As in the sacrificial lambs example, this condition applies to stochastic networks only; indeed, deterministically placed sacrificial lambs cannot produce asymptotic learning. 

Acemoglu et al.\ also show, conversely, that  asymptotic learning cannot hold when some agents play too important a role in the topology of the network. This happens when there is a set of important agents $\{I_1,\ldots,I_M\}$ that constitute the only social information for an infinite group of agents. When their signals happen to indicate the wrong action---which occurs with some positive probability---infinitely many agents follow suit. 
Along these lines, Acemoglu et al.\ say that a network  has \emph{non-expanding observations} if there is some $M$ and $\varepsilon>0$ such that, for infinitely many agents $I_n$, the probability that $N_n$ is contained in $\{I_1,\ldots,I_M\}$ is at least $\varepsilon$. In this case, asymptotic learning does not occur.   The lesson that important agents impede the aggregation of information is one that---as we shall see---recurs frequently across a wide spectrum of models.

Turning to short-term dynamics,  \citeasnoun{ADLO:11}  show that social learning can sometimes induce contrarian beliefs and anti-imitation (see also \citeasnoun{eyster/rabin:14} discussed in \S~\ref{sec:observability}).  Intuitively, suppose that both $I_3$ and $I_2$ observe $I_1$ only, and that $I_4$ observes $I_1$, $I_2$ and $I_3$. Then $I_4$ should place positive weight on $I_3$ and $I_2$, and negative weight on $I_1$ to offset double-counting. So, within a broad stream of imitation, there can be eddies of contrarian behavior.\footnote{For a formal example, see ``Nonmonotonicity of Social Beliefs'' in \citeasnoun[Appendix B]{ADLO:11}. Such contrarian behavior does not occur in networks in which each agent can only observe one predecessor (a {\it tree network}).}

Overall, these papers   show that sometimes smart observers do not just follow the herd. Sometimes, smart agents may put much greater weight on the actions of fewer agents (resulting in following the minority, as in \citeasnoun{callander:horner:2009}, also discussed in \S\ref{sec:observability}), or even put negative weight on some (as in Acemoglu et al.). Although such effects are far from universal properties of social learning models, such findings provide a useful caveat to the intuitive notion that agents put positive and similar weights on the actions of predecessors. 

Until now we have considered network models in which the neighborhoods $N_n$ are either deterministic, or drawn independently.  Less is known about the case in which  neighborhoods $N_n$ are not independent (as analyzed by \citeasnoun{lobel:sadler:2013}). For example, a given agent may have a chance of being observed by either everyone or by no one.

%In their model, there is an underlying network of \emph{possible} ties between agents. Before the start of the game, a  sub-network is realized at random. In particular, 
In \citeasnoun{arieli:mueller-frank:19}, agents are placed on a two (or higher) dimensional grid, and the timing of their actions is given by their distance from the origin. The observation structure is chosen at random according to a parameter $p$; each agent is independently, with probability $p$, {\em connected} to each of her grid neighbors who are closer to the origin. An agent $I_n$ observes an agent $I_m$ if there is a path of  connected agents starting from $I_n$ and ending in  $I_m$. This is a special case of  \citeasnoun{lobel:sadler:2013}, but not  of the \citeasnoun{ADLO:11} setting, since the realized neighborhoods are not independent. Specifically, if $I_n$ is not observed by any neighbor, then she  would not be observed by any other agent, and so these events cannot be independent. 

After the observation structure is realized, the agents take actions sequentially, according to  their distance from the origin.
\citeasnoun{arieli:mueller-frank:19} study what they call {\em $\alpha$-proportional learning}, meaning roughly that at least an $\alpha$ fraction of the (infinitely many) agents choose the correct action. More accurately, this obtains whenever an $\alpha$-fraction or more of the agents in the ball of radius $r$ choose the correct action with probability that tends to one as $r$ tends to infinity.

Fixing $\alpha<1$, \citeasnoun{arieli:mueller-frank:19} ask for which values of $p$ is $\alpha$-proportional learning obtained. When $p=1$, each agent observes all agents who lie between that agent and the origin. In this case cascades form as in the SBM, so  $\alpha$-proportional learning does not hold. Their main result is that, nevertheless,  for all sufficiently large $p < 1$, $\alpha$-proportional learning \emph{does} hold.

As in  \cite{ADLO:11}, this conclusion is  based upon sacrificial lambs. For any $p < 1$ there is a constant fraction of agents who observe no other action. The actions of these agents provide independent information to observers. As $p \rightarrow 1$ these agents become more rare, but the network becomes more connected, delivering this information to a larger and larger fraction of the population. Thus, learning is achieved as long as there are some sacrificial lambs, regardless of how small their fraction is. 

\citeasnoun{ADLO:11}, \citeasnoun{lobel:sadler:2013} and \citeasnoun{arieli:mueller-frank:19} leave open an interesting  question: does there exist a {\em deterministic} network structure  in which asymptotic learning is attained for some bounded private signal distribution? The sacrificial lamb mechanism by which asymptotic learning is achieved in \emph{stochastic} networks is simple, but does not seem to have an obvious analogue in deterministic networks.\footnote{In a deterministic network, if there is an infinite number of sacrificial lambs, then the probability that $I_n$ makes the correct choice cannot tend to one, since  the sacrificial lambs have a fixed positive probability of taking the wrong action.} Thus, learning in deterministic networks would have to result from a different mechanism; it is interesting to understand whether such mechanisms exist.\footnote{An alternative possible definition of asymptotic learning is that the  probability that the majority of  the first $K$ individuals takes the correct action approaches one.  Under this definition, deterministic sacrificial lambs can  induce asymptotic learning if they are sparse enough, e.g., at positions $1$, $10$, $100$ and so on. }

%%%%%%%%%%%%%%%%%%%%%%%%%%%%%%%%% 
\subsection{Repeated actions}
\label{sec:repeated_actions}

We next study models in which agents act repeatedly. We first examine rational settings, and then turn to imperfect rationality.

\subsubsection{Rational models}
\label{subsec:rational-repeated}

In simplest form, in repeated action models each agent takes an action in each one of infinitely many periods. Private signals may or may not be received repeatedly. In general, these are significant  deviations from the SBM. 
In the special case of observation structures where there are no cycles (where a sequence of observations starting from an  agent can never lead back to the same  agent), 
the logic of information cascades can be extended in a straightforward way. 

For example, suppose that we modify the SBM so that the order of moves is first $I_1$, then $(I_1,I_2)$, then $(I_1, I_2, I_3)$, and so on forever in an expanding sequence of sequences. We continue to assume that at every move by agent $I_m$, agent $I_m$ observes all agents $I_k$, $k < m$, and only these agents.  Then $I_1$ always sticks with $I_1$'s original choice. Because of this fact, so does $I_2$, and by induction the same is true for all agents.  So there are still information cascades and a failure of asymptotic learning. 

In this example, there is again an undue observational focus on a limited set of agents---early ones.   The lesson this illustrates is that bad aggregation in the SBM does not rely upon the fact that agents only move once. The key is  having a bad observation structure, such as one that is far from egalitarian.  

Under highly connected observation structures (such as one where  there is a path of observation between any pair of agents), the concepts of herd behavior and information cascades need to be generalized to apply  formally. Following the literature, we consider more general notions of agreement that capture phenomena that are similar to herding. Here {\em agreement} is the situation in which all agents eventually agree about what action is optimal (though they need not have identical beliefs). Likewise, we consider notions of \emph{learning} (or the failure thereof) that capture repeated action analogues of information cascades. Indeed, herding in sequential models is a form of agreement, since it occurs precisely when all agents (except a finite number, out of infinitely many) agree on the optimal action. In an analysis covering a wide class of social learning models, \cite{mossel2020social} find that  agreement and asymptotic learning are closely related, and provide conditions under which each of these phenomena occur.

An incorrect information cascade  implies that information is lost and that agents do not learn the true state. In repeated action models---in lieu of incorrect information cascades---we ask  whether agents learn the state, and we study the mechanisms that drive failures to do so. These mechanisms often resemble cascades in that some private information is permanently  lost, even though agents do not completely disregard their own signals in the early periods.

Information is lost for a reason that is similar to models of cascades in settings in which each agent acts only once. A point is reached at which an agent's signal is incrementally informative relative to social information, but owing to coarseness of the action space, the agent's action does not adjust to reflect this incremental information. So when social information is sufficiently precise,   actions do not reveal all the private information  to others.

The study of agreement in social learning goes back at least to  \citeasnoun{aumann:76}. This seminal paper showed that two agents who have a common prior, who receive a  signal regarding a binary state, and who have common knowledge of their posteriors, must have equal posteriors; agents cannot ``agree to disagree.''

Aumann's model does not consider the dynamics of how agents arrive at common knowledge of posteriors. Several subsequent models study how agreement is reached via social learning. \citeasnoun{geanakoplos:polemarchakis:1982} consider two  agents who observe private signals regarding a private state, and then repeatedly  tell each other their posteriors. The authors show that the agents reach common knowledge of posteriors, and hence their posteriors will be identical.

\citeasnoun{parikh:krasucki:1990} extend this conclusion to a network setting. They consider a finite set of agents connected by a network. In each period, each agent $I_n$ learns the posteriors of the members of her neighborhood $N_n$. Under the assumption that the network is {\em strongly connected},\footnote{The network is strongly connected if there is a chain of neighbors connecting every pair of agents: formally, for each two agents $I_n,I_m$ there is a sequence of agents $I_{n_1},I_{n_2},\ldots,I_{n_{k+1}}$ such that $I_n=I_{n_1}$, $I_m=I_{n_k}$, and $I_{n_{\ell+1}} \in N_{n_\ell}$ for $\ell=1,\ldots,k$.} Parikh and Krasucki show that the agents  reach common knowledge of posteriors, and hence agree. 

This result was  extended by \citeasnoun{gale:kariv:03} to agents who face a common, general decision problem at each period: they must choose an action, and receive a stage utility that depends on their action and the unknown state.\footnote{A \emph{stage utility} is a payoff received in a particular period of a dynamic game.} At each period, they observe their neighbors' actions. The agents are assumed to be myopic, so that at each period they choose an action that maximizes their stage utility. Neighboring agents do not necessarily eventually agree on actions, even at the limit. But any disagreement is due to indifference: if $I_n$'s neighbor $I_m \in N_n$ takes an action $b$ infinitely often, then $I_n$ must, at the limit, be indifferent between $a$ and $b$. This result was extended by \citeasnoun{rosenberg:solan:vieille:2009} to forward-looking agents who maximize their discounted expected utility, rather than myopically choosing an action in each period.

The mechanism at work here is the \emph{imitation principle} (also known as the improvement principle; see \citeasnoun{golub/sadler:16}), which asserts that in the long run an agent will be able to do at least as well as an agent that she observes, i.e., a neighbor.  Agent $I_n$ is always free to imitate $I_m$, i.e., choose at each time period the action that $I_m$ chose in the previous one. Since $I_n$ myopically maximizes her expected stage utility, her expected stage utility at time $t$ is at least as high as $I_m$'s expected stage utility at time $t-1$. Since the network is strongly connected, it follows that at the limit $t\to\infty$ all agents have the same expected stage utilities. 

Now suppose that with positive probability agent $I_n$ thought that agent $I_m$'s action were strictly suboptimal, at the limit $t \to \infty$. Then $I_m$ would have lower expected utility, in contradiction to the imitation principle. So even if  $I_n$  disagrees with $I_m$, $I_n$  must believe that $I_m$'s action generates equal utility to the action that $I_n$ chooses. I.e., $I_n$ is indifferent between these actions.

While neighboring agents that disagree must be indifferent, non-neighbors can disagree without indifference. Consider a network of four agents $I_1,I_2,I_3,I_4$, who are connected along a chain, with each observing both the agent's predecessor and successor (if there is one). It is possible for $I_1$ and $I_2$ to converge to action $L$, while $I_3$ and $I_4$ converge to $H$. In this case $I_2$ and $I_3$ must be indifferent between $L$ and $H$, but it is possible for $I_1$ and $I_4$ to not be indifferent. This happens in the case of binary signals and actions, when $I_1$ and $I_2$ receive an $\ell$ signal, $I_3$ and $I_4$ receive an $h$ signal, and the agents' tie breaking rule is to stick to their previous period action. In this case the agents' actions all immediately converge. After seeing that agent $I_3$ does not change her action, $I_2$ concludes that $I_4$ got an $h$, and thus $I_2$ becomes indifferent. Likewise, $I_3$ becomes indifferent. But $I_1$ does not know that $I_2$ is indifferent: from $I_2$'s point of view, it may well be possible that $I_3$ is also taking action $L$. Thus $I_1$ is not indifferent, and neither is $I_4$, and yet they disagree.

This raises the question of whether the possibility of indifference is robust. A partial answer is that when the action set is rich enough, such indifference is impossible, and the conclusion is again that agents must converge to the same action. \citeasnoun{mossel:sly:tamuz:2012} show that when private signals induce non-atomic beliefs (i.e., no belief occurs with positive probability)  then again such indifference is impossible, and hence all agents must converge to the same action.

Even when the agents do converge to the same action, it may be an incorrect one, so that the agents do not learn the state. In a quasi-Bayesian setting,\footnote{An important difference between the  \citeasnoun{bala:goyal:1998} model and the others discussed here is that in their model agents have strategic experimentation incentives  that determine endogenously the private signals that they observe.}  \citeasnoun{bala:goyal:1998} show that the learning outcome depends on the network geometry. An important example of a strongly connected network in which agents may  not  converge to the correct action is the {\it royal family}, in which all agents directly observe a small group, but that small group does not directly observe all others.

A similar phenomenon occurs in a setting with forward-looking, Bayesian agents who each receive a signal at period 0, and thereafter choose in each period a binary action with the objective of  matching a binary state (\citeasnoun{mossel:sly:tamuz:2015}). In a network with a royal family, incorrect signals received by the royal family can cause the entire population to eventually adopt the incorrect action. When this happens, the early period actions of agents in the population are still dependent upon their own private signals, but this information does not propagate through the network. The outcome is closely related to information cascades in models with a single action, in that  social information  can {\em eventually}  cause the agents to disregard their own private signals. 

Conversely, \cite{mossel:sly:tamuz:2015} show that in infinite networks that are  egalitarian, even with bounded signals, agents  all converge to the correct action. 
A network is said to be {\em egalitarian} if there are integers $d$ and $L$ such that (i) each agent observes at most $d$ others and (ii) if agent $I_n$ observes $I_m$, then there is a path of length at most $L$ from $I_m$ back to $I_n$. 
The first condition excludes agents who obtain large amounts of social information. The second excludes royal families, who are observed by many of their ``subjects'' but do not reciprocate by observing their subjects directly, or even indirectly through short paths. Another interpretation of the second condition is that it requires the network to be approximately undirected, where an undirected network is one in which which the second condition holds for $L=1$. 

Thus,  both \citeasnoun{bala:goyal:1998} and \citeasnoun{mossel:sly:tamuz:2015} conclude that  networks in which a subset of agents plays too important a role can hamper the flow of information and lead to failures of learning, and that asymptotic learning occurs in networks in which all agents in the network   play a similar role.\footnote{Interestingly, for undirected networks and myopic agents, \citeasnoun{mossel:sly:tamuz:2012} show that  agents always converge to the correct action, for any network topology---even non-egalitarian networks in which agents can observe arbitrarily many others.} 
The mechanism underlying the failure of asymptotic learning in non-egalitarian networks is similar to the cause of cascades in the SBM: many agents choose actions that are  heavily influenced by social information, and thus do not reveal their private signals. In the SBM, this social information comes from the early agents. In a setting with non-egalitarian networks and repeated actions, it comes from agents who become socially well-informed  because of their central locations in the network.  

When agents choose actions from a  set  that is rich enough to reveal their beliefs, information is aggregated regardless of the network structure (\citeasnoun{mueller2014does}). As a straightforward special case, when the SBM is modified to allow  a rich set of actions that reveal beliefs (see \S \ref{sec:ActionSpace}), information is perfectly aggregated, in the sense that agents converge to the same belief that they would hold if all private signals were to become public. 

Similarly, in a  network setting,
\citeasnoun{demarzo/vayanos/zwiebel:03} offer as a benchmark case a model in which agents  in a connected social network  are interested in a state $\theta \in \mathbb{R}$ for which all have a uniform (improper) prior. Each agent initially observes a Gaussian signal with expectation equal to $\theta$. 
In each period each agent observes her neighbors' posterior expectations of $\theta$, and updates her posterior using Bayes' Law. Each agent's posterior is   Gaussian, and thus is completely characterized by the agent's posterior expectation; the variance depends on the network structure and does not depend upon signal realizations. 
In this model the agents converge to the same belief they would have if they were to share their private signals, so that information is perfectly aggregated.\footnote{The main result of \citeasnoun{demarzo/vayanos/zwiebel:03}  describes the effects of persuasion bias, wherein  agents update beliefs under the mistaken premise  that other agents do not observe others.}

\citeasnoun{frongillo2011social} and \citeasnoun{dasaratha/golub/hak:19} consider a similar setting, but where the state exogenously changes with time, agents receive a new signal at each date, and signals have heterogeneous precisions. Since the state changes, the agents cannot learn it exactly. The efficiency of information aggregation depends on the social network structure and the signal structure. Information is better aggregated when there is large heterogeneity in signal precisions, because this helps agents filter out stale information that is entangled with information that is more relevant for the current state. 

%%%%%%%%%%%%%%%%%%%%%%%%%%%%%%%%% 
\subsubsection{Heuristic models}
\label{sec:heuristic_network_models}

A highly interdisciplinary literature studies heuristic models of social learning with repeated actions on networks.\footnote{We considered heuristic models under sequential (non-repeated) actions in \S \ref{sec:observability}.} Indeed, since repeated action models in networks tax the rationality assumption, in such settings heuristic decision making is often the more interesting case.

In mathematics and physics, models such as the ``voter model'' (\citeasnoun{holley:liggett:1975}) and the ``Deffuant model'' (\citeasnoun{deffuant2000mixing}) use tools from statistical mechanics to study interacting agents that follow rules that are simple enough that the agents resemble interacting particles. 
We refer the reader to other surveys and books that cover this field, such as \citeasnoun{castellano2009statistical} or \citeasnoun{boccaletti2006complex}.
In sociology, a quantitative literature  studies opinion exchanges on networks, partially motivated by the question of how to measure network centrality (see, e.g., \citeasnoun{katz1953new} and \citeasnoun{bonacich1987power}).

In a model that has been influential in economics as well as other fields, \citeasnoun{degroot:1974} considers a finite number of agents in a social network. Each agent is endowed with an initial subjective prior regarding a finite state. In each period each agent reveals her belief to her  network neighbors, and then updates her own belief to a weighted sum of her neighbors' beliefs. These weights $p_{nm}$ are exogenous and do not change with time. 

DeGroot finds that under mild conditions this process converges, and that  all agents converge to the same belief. This limit belief is equal to a weighted average of the initial beliefs; the weights in this average are given by the first eigenvector of the matrix $p_{nm}$. 

\citeasnoun{golub:jackson:10} use DeGroot's framework to study the aggregation of information. In their model, the state is any $\mu \in [0,1]$. The agents each start with a prior chosen from some distribution with expectation $\mu$, and follow DeGroot's update rule. Each agent uses equal weights for all the agents in her neighborhood. Golub and Jackson show that over time the agents' beliefs will converge arbitrarily close to $\mu$ (for sufficiently large networks) as long as no agent has too high a degree. The relevant  condition is that  the ratio between the maximum degree and the sum of degrees be sufficiently small, which can  be viewed as another notion of egalitarianism. Even though the learning process is far from rational, this conclusion is aligned with the aforementioned results of \citeasnoun{bala:goyal:1998} and \citeasnoun{mossel:sly:tamuz:2015}  that egalitarianism promotes the aggregation of information.

%%%%%%%%%%%%%%%%%%%%%%%%%%%%%%%%%%%%%%%%%%%%%%%%%%%%%%%%%%%%%%%%%%%%%%%%%%%%%%%%%%%%%%%%%%%%%%%%%%%%%%%%%%%%%%%%%%%%%%%%%%%%%%%%% 

\ifdefined\ALLCOMPILE\else\end{document}\fi
 % 15
\ifdefined\ALLCOMPILE
\saycurrfile
\else
\usepackage{jelall}
\renewcommand{\citeasnoun}[1]{\textcolor{red}{#1}}

\begin{document}
\fi

\section{Applications and Extensions} 
\label{sec:ApplicationsExtensions}

So far we have focused mainly on variations on the basic cascades and social learning  setting that are general in scope, or involve changing relatively few assumptions. We now turn to several extensions that are tailored to specific applications.  
In these applications, information externalities offer new insights into  social outcomes in both market and non-market settings.
 
%%%%%%%%%%%%%%%%%%%%%%%%%%%%%%%%% 

\subsection{Team Decisions, Optimal Contracting and Organizational Design}

Teams often need to make joint decisions that aggregate the information of  members who can  learn by observing the actions of other members.
Incentive problems and psychological bias influence whether agents make use of their own private signals, and therefore whether cascades hinder the efficiency of information aggregation. Such problems can  be addressed through the design of a communication network and an incentive scheme for the team. 
We first discuss settings with rational agents. 

 As illustrated by the SBM, incorrect cascades often form in a linear sequential observation network. As illustrated in the ``fashion leader'' example of  \S\ref{sec:binary-model}, the problem can be especially severe when better informed agents decide first, as this can trigger cascades very early. This problem can be addressed by anti-seniority voting systems (if seniority is associated with precision). However, as shown by \citeasnoun{ottaviani/sorensen:01}, ordering agents inversely with precision is not always optimal.

% See also Last sentence of BHW 1992 on designing observation structure/quarantining early agents
 
\citeasnoun{khanna/slezak:00} study the choice of network structure---teams versus hierarchy---for a firm that seeks to aggregate the signals of multiple employees.\footnote{We consider more generally the effects of networks on asymptotic learning in \S \ref{sec:networks}.}  Their analysis confirms the intuition that it can be desirable to assign  agents to  make decisions independently, preventing  information cascades, and allowing  a later observer (or more generally, observers) to obtain better social information. This is similar to the discussion of sacrificial lambs of  in \S \ref{sec:observability}.  

In \citeasnoun{khanna/slezak:00}, agents can  acquire a private binary signal about the state, which determines the profitability of a binary action on the part of the firm, such as whether or not to adopt a project. The precision of an agent's private signal  is increasing with costly private effort. Ex ante identical agents make their investigation decisions  before any social observation. In equilibrium they all choose the same effort level. Each agent's action consists of a recommendation (e.g., to adopt or reject the project).
The compensation contract consists of a wage, a bonus if an agent's  recommendation turns out to be  correct ex post, and, if agents are organized into teams, a possible team bonus that is received if the team recommendation is correct.

In teams, agent recommendations are publicly announced sequentially.  Owing to social observation, in equilibrium agents ex ante tend to free ride in generating private signals. Furthermore, unless the compensation scheme is  heavily weighted toward individual accuracy, agents tend to cascade on the recommendations of earlier agents. The  compensation scheme may optimally accommodate such cascading.
These effects limit the social information forthcoming from teams to top management for choosing the firm's action. 

The alternative to teams is hierarchical organization, wherein each agent  reports a recommendation to the top manager without benefit of social observation.  Such hierarchies reduce free-riding in information acquisition, and prevent the formation of incorrect cascades. Hierarchies therefore dominate, assuming that  direct communication among agents can be cheaply suppressed.\footnote{In a similar spirit, \citeasnoun{sgroi:02} analyzes the benefits to a social planner or a firm of forcing some agents (such as a subset of customers) to make their decisions early. In Sgroi's model, promotional activity by firms encourages some customers to decide early, which can be useful for increasing sales.}

In a multi-level hierarchy, an alternative to  forcing early agents to make recommendations without observing each other is to artificially coarsen their recommendations. For example, the recommendations of 3 agents can be aggregated into  a single overall adopt/reject recommendation on a project. This makes it unclear to an observer whether the preponderance in favor of an action was strong (3 to 0) or weak (2 to 1). Such coarsening can cause a supervisor to whom the agents report to  use the supervisor's own private information instead of cascading. This makes the supervisor's recommendation to the top layer of the firm informative. In some cases coarsening tends to improve the  firm's decision (\citeasnoun{arya:glover:mittendorf:2006}).

\subsection{Stigma, Prestige, and Related Phenomena}

As mentioned in the introduction, cascades may apply to adverse inferences drawn by participants in labor markets and the market for kidney transplants. We now say more about each of these.

Job seekers are often advised to try to avoid having gaps in their resumes, which might be indicative that previous potential employers have rejected their applications. \citeasnoun{kubler:weizsacker:2003}  model such unemployment stigma in a setting in which  employers receive noisy signals about the quality of potential employees. 
The Kubler and Weizsacker  model differs from the SBM in that both  employers and workers can take costly unilateral actions that modify signal precision. Kubler and Weizsacker
identify  an equilibrium in     which only cascades of rejection, not acceptance occur.
There is indeed evidence, including field evidence, that  employers view gaps in resume  as indicating rejection by other employers (\citeasnoun{oberholzer-gee:08}, \citeasnoun{kroft/lange/notowidigdo:13}).
     
A mechanism very similar to stigma  in labor markets can result in refusal of kidneys by patients who need kidney transplants. 
Patients are sorted into a queue, in order of severity of their conditions. When a kidney becomes available, patients are offered it in sequence. Each patient is offered the  kidney only if all predecessors rejected it.   In the model of \citeasnoun{zhang:10}, patients have heterogeneous preferences as they differ in compatibility with an available organ and in  their urgency for a transplant. There also is a common component to value, since all patients prefer a high quality kidney. Each patient's physician provides a medical judgement about the suitability of an available kidney, i.e., a private signal. If a patient is offered a kidney after refusals by earlier patients in the queue, she infers low quality of the organ, causing possible  reject cascades. 

Using data from the United States Renal Data System, \citeasnoun{zhang:10} finds strong evidence of social learning. Patients draw negative quality inferences from earlier
refusals, and thus become more likely to refuse a kidney. This leads to poor kidney utilization despite a severe shortage of available organs. There is also strong evidence of social learning in other health related decisions, such as choice of health plans (\citeasnoun{sorensen:06}).

Research on cultural evolution has hypothesized that the human mind has evolved to confer prestige upon successful individuals (\citeasnoun{henrich/gil-white:01}). In this theory, individuals who defer to an admired individual benefit from being granted  access to the  information possessed by that individual about how to be successful in their shared ecological environment. If different observers  observe different payoff realizations, they will have different information about who is a good decision maker. So observers    can acquire useful information by  observing whom others  defer to.  The Henrich and Gil-White perspective on prestige naturally  suggests that there can be information cascades in conferring prestige. If so, prestige may be a very noisy indicator of decision ability. 

\subsection{Social Information Use by Animals}
%  MOVED TO SEPARATE SUBSECTION foraging, habitat selection
% (mating, navigation, predator-avoidance)

Zoologists  have applied  information cascades models to the acquisition of social information by animals for  decisions about mating, navigation, predator-avoidance, foraging, and habitat selection.\footnote{With regard to human habitat choice, \citeasnoun{epstein:10} provides a model somewhat similar to the SBM in which there are cascades in the choice of country of emigration.} 
Several empirical studies of imitation in animals  conclude that social learning and cascades are important mechanisms, as distinct, e.g.,  from payoff externalities. Imitation occurs despite the fact that payoff externalities are often negative, as imitation increases  competition with others for the same food sources, territories, or mates.    

\citeasnoun{dall:giraldeau:olsson:mcnamara:stephens:2005} discuss several proposed examples of animal information cascades. An extensive empirical literature documents copying in  mate choices in many species, including humans  (\citeasnoun{waynforth:07}, \citeasnoun{anderson/surbey:14}, \citeasnoun{witte/kniel/kureck:15}). \citeasnoun{gibson/hoglund:92} provide evidence consistent with information cascades in the choices of sage hens observing mating display groups called leks. Consistent with observation of others dominating own-signals, there is  evidence of reversal of choices upon observation of others in guppie mating (e.g., \citeasnoun{dugatkin/godin:92}). 

\citeasnoun{giraldeau/valone/templeton:02} apply the logic of information cascades to understand the conditions under which animals will optimally acquire information socially.  The authors  propose three possible  examples of information cascades in nature: false alarm flights, wherein  ``groups of animals take fright and
retreat quickly to protective cover, sometimes even in the
absence of any obvious source of danger''; night roost site selection, wherein many birds adopt a site ``simply because the
social information indicates the site is profitable''; and mate choice copying, as just discussed. The authors argue that the possibility of incorrect cascades makes it unprofitable to expend resources to observe the actions of others.

%%%%%%%%%%%%%%%%%%%%%%%%%%%%%%%%% 
\subsection{Sequential Voting}

In primary elections for nominating party candidates for U.S. presidential elections, states that go early in the process exert a disproportionate influence on the ultimate outcome. One of the principal reasons is that the beliefs of later voters are influenced by the choices of earlier voters in New Hampshire and Iowa  (see \cite{bartels:88}).

While this finding is intuitive, in a sequential setting in which voters with private information signals seek to elect the best candidate, \citeasnoun{dekel:piccioni:2000} show that   there exist equilibria in which voters are not influenced by predecessors' votes. Essentially, each voter behaves as if she is 
{\it pivotal}, that is, as if others' votes were split equally between the two candidates. So a voter optimally  votes in accordance with  her private signal, regardless of choices of predecessors. This results in  asymptotic learning.
 % . Voter behavior is the same as  it would be under simultaneous voting.

However,  \citeasnoun{ali/kartik:12}  show that there also exist   equilibria in sequential voting in which voters are influenced by predecessors, resulting in information cascades. We consider a  modification of the SBM  to illustrate their main ideas.\footnote{Other applications of the  \cite{ali/kartik:12} model include  public good allocations and charitable giving.} Voters (agents) have identical preferences over two candidates, $L$ and $H$. 
Voters agree that candidate $H$ is {\it best} if and only if the  state is $H$.  
Each voter observes a private signal and the votes of her predecessors.  

The departure from the SBM is in the payoff of agents. Suppose  that the election is decided by a simple majority vote and that the total number of voters, $2N-1$, is odd. Let $a=(a_1,a_2,\ldots,a_{2N-1})$ be the votes (actions) of all agents and $C_\theta(a)$ be the number of votes for candidate $\theta$. As each voter wants the best  candidate  to win, the utility of each voter is
$$u(a)=\begin{cases}
1 & \mbox{if  } C_L(a) \ge N \mbox{ and } \theta=L\\
1 & \mbox{if  } C_H(a) \ge N \mbox{ and } \theta=H\\
0 & \mbox{otherwise}.
\end{cases}$$
Unlike the SBM, every agent cares about decisions taken by others.\footnote{By assumption there is no conflict of interest nor conformity preference nor congestion effects in the model. So the model does not share the types of  positive or negative externalities considered in  \S~\ref{sec:externalities}.}
An issue of interest  is whether, despite payoff interactions, voters behave {\it sincerely}, i.e.,  vote for the candidate they think is best given their beliefs. Ali-Kartik find that there is a perfect Bayesian equilibrium with sincere behavior. In this equilibrium, voters follow their signals unless a candidate leads by two or more votes, in which case they vote for the leading candidate regardless of their signal based on the inference that the leading candidate is better. So except in very close elections, this behavior results in  a (possibly incorrect) cascade; there is no asymptotic learning. 

As mentioned at  the start of this section, when there is sequential voting, early  success can  promote later success owing to inferences drawn by later voters.  Consistent with the social learning effects in  \citeasnoun{ali/kartik:12},  \citeasnoun{knight/schiff:10} document such political momentum.
Knight and Schiff estimate that  voters in early states had far greater influence than voters in later states on the outcome of the 2004 U.S. presidential primaries, and that campaign advertising choices took into account such momentum effects. 

\citeasnoun{battaglini:05} modifies Dekel-Piccione's  voting model to allow for a third choice -- an option to not vote. Battaglini shows that if there is even a small cost of voting, then Dekel and Piccione's result does not hold: the set of perfect equilibrium outcomes in simultaneous voting and in sequential voting are disjoint. Information aggregation is worse in sequential voting than in simultaneous voting when $N$ is large. Furthermore, when voters  have a direct preference for voting for the winner in addition to  a  preference for voting for the right candidate, as \citeasnoun{callander:2007} shows,  social learning can lead to a bandwagon effect.

In the papers discussed  in the previous paragraph, voters are  motivated by the fact that they   can be pivotal for the outcome. As other research emphasizes, other factors and motivations can also be very important, such as the cost of voting, the desire to vote for a winner, and direct utility from voting (or not voting) as an expression of  political viewpoint. It will be valuable to develop and test  empirical implications of models of social learning and voting to disentangle the different possible motivations for voting.

%%%%%%%%%%%%%%%%%%%%%%%%%%%%%%%%% 
\subsection{Politics and Revolutions}

When citizens publicly protest or revolt against their government, the actions of early individuals convey information about the prevalence of dissatisfaction to other observers, and of the risks of government sanctions. So there can be positive feedback in submission  or resistance to the regime. 

In \citeasnoun{kuran:87}, agents have both private preferences between two alternatives, and publicly declared preferences.  There is a payoff/preference externality in the form of pressure toward conformity. This results in an equilibrium with high declared support for an oppressive regime that most agents privately oppose. Kuran applies this idea to the persistence of the caste system in India. \citeasnoun{kuran:89} emphasizes that preference falsification and positive feedback result in multiple equilibria, and that sudden change (a ``prairie fire'') can occur as exogenous parameters shift when a critical threshold is crossed  that causes a shift from one equilibrium to another. 

\citeasnoun{lohmann:1994} explicitly models belief updating  to analyze the  maintenance  and collapse of political regimes as  information cascades.  In Lohmann's model, agents have different preferred policies, and have quadratic disutility from deviation between the government's policy and the preferred ones. Each agent also has a  disutility component from protesting. Disutility  decreases in the number of other agents who are protesting. The existing regime collapses once enough information is revealed to show that more than a critical number   of individuals support an alternative regime over the current one. 

Lohmann  applies the model to the Leipzig protests against the communist regime in East Germany during 1989-91. In her analysis,  even a small protest, when larger than expected, can reveal very strong opposition to the regime, causing the size of the protest to grow explosively. By the same token, when a large protest is expected, if the protest falls modestly short of expectations, the movement can rapidly collapse. Based on evidence from five cycles of protests, Lohmann argues that a dynamic information cascades model helps explain the fall of communism in East Germany.

\citeasnoun{ellis/fender:11} combine features of Lohmann's model with those of the model of democratization, repression and regime change of \citeasnoun{acemoglu/robinson:06}. In \citeasnoun{acemoglu/robinson:06}, in order to stave off revolution, a repressive regime that is run by the rich can choose to redistribute wealth via taxation and democratize by extending the voting franchise to the poor. Some fraction of the wealth is destroyed in a revolution, which limits the incentive of the poor  to rebel. 

Ellis and Fender extend the Acemoglu and Robinson analysis to a setting in which the poor have private information about regime strength. In what we call the $L$ state, a revolution would destroy a higher fraction of the wealth to be appropriated by the poor than in the $H$ state, making it less attractive for the poor to revolt in state $L$. So the state of nature characterizes the strength of the regime.

In the case of interest, the rich set the tax rate such that the poor would like to revolt if they knew that the state was $H$ and not revolt if they knew the state was $L$. Each poor agent (hereafter, ``agent'') has a conditionally independent  binary private signal  about the state. 
If the poor are not granted the franchise, poor agents decide in sequence whether to rebel against the regime, with each agent's decision observable to later agents. By assumption, a revolution occurs only if  every agent rebels. At first, an agent with an $h$ signal will rebel, whereas an agent with an $\ell$ signal will not, i.e., vetoes the revolution.  
However, after a sufficiently long sequence of rebels (i.e., $h$ signals), even an agent with an $\ell$ signal rebels. So in the model there are information cascades of revolution.     

%%%%%%%%%%%%%%%%%%%%%%%%%%%%%%%%% 

\subsection{Legal Precedent and Information Cascades}

Dating at least as far back as Oliver Wendell Holmes, a common interpretation by legal scholars of respect for precedent ({\it stare decisis}) is that  judges acquire information from past decisions in similar cases.   In the model of \citeasnoun{talley:1999},  this takes the form of imitation across courts, wherein a judge follows the decision of earlier judges ``notwithstanding the facts of the case before her.'' The resulting loss of the private signals of later judges who  respect precedent is also emphasized by \citeasnoun{vermeule:12} pp.\ 75-7. However, Talley argues that the conditions for cascades to occur in this setting are highly restrictive, in part because judges can relay  information via written opinions. 

\citeasnoun{daughety/reinganum:99} offer a more general model of imitation across courts. In their setting, $n$ appellate courts at the same level observe private signals about a binary state. This state  indicates the single correct decision for a set of related cases to be considered by these courts. The courts make decisions in exogenous sequence.  If signals are bounded,  imitation across the appellate courts results in information cascades, often incorrect; even if signals are unbounded, with many courts there is a high probability of extensive imitation, resulting in poor information aggregation. The authors offer several possible  examples of actual precedential cascades (pp. 161-5).

Daughety and Reinganum emphasize that incorrect cascades can be very persistent, despite the finding of BHW that cascades are fragile, because of the rarity of  shocks that might dislodge a judicial cascade. As discussed in \S \ref{sec:binary-model},  informative public disclosures can dislodge cascades. However, the authors argue that in the U.S. judicial setting, such shocks take the form of cases being brought to the Supreme Court opinions despite harmonious decisions of lower courts. Such review is rare, because the Supreme Court needs to wait for cases to be appealed, and because  decisions that are harmonious across courts are rarely reviewed. 

\citeasnoun{sunstein:09}\nada{p. 205} argues that the adoption of legislation sometimes takes the form of information cascades across nations.  Sunstein \nada{p. 172} also argues that legal resolution of constitutional questions, such as the once-common, now rejected, view  that the U.S. Constitution permits   racial segregation, may be the product of information cascades. 

%%%%%%%%%%%%%%%%%%%%%%%%%%%%%%%%% 
\subsection{Group Adoption of Conventions}

Social conventions often differ across groups and are sometimes transmitted from one group to the other. 
When there is limited observation across groups of agents, distinct cascades can temporarily form. \citeasnoun{fisher:wooders:2017} investigates the contagion of behavior between groups via  an individual who belongs to both groups. In their model, two SBMs, labelled A and B, run in parallel. A single agent, $I_n$, $n>2$, is common to both groups A and B and observes the actions of all predecessors in both groups. Every other agent is in only one group. 

Suppose that $n$ is odd, so that the difference between the number of Adopts and Rejects by $I_n$'s predecessors is even. 
If the two groups are in opposite cascades, then $I_n$ follows her own signal, which breaks the cascade  that opposes $I_n$'s signal. 
If group A is in an Adopt cascade and group B is not in a cascade (i.e., the  Adopt/Reject difference among $I_n$'s predecessors in Group B is zero), then $I_n$ Adopts and subsequently both groups are in an Adopt cascade, because  $I_n$'s successors in group B realize that $I_n$ has superior social information. Finally, if neither group is in a cascade, then both groups cascade on $I_n$'s action. 

%%%%%%%%%%%%%%%%%%%%%%%%%%%%%%%%% 

\subsection{Belief Cascades}
\label{subsec:beliefcascades}

Several authors have suggested that the dynamics of public opinions, such as the rise and fall of ideologies and conspiracy theories, are information cascades. \citeasnoun[p.50]{sunstein:19a}\nada{Conformity, p50} argues that the sudden popularity of ``fake news'' reports in recent years derives from information cascades.
\citeasnoun{kuran/sunstein:99,kuran/sunstein:00} have also  suggested that information cascades of belief play a role in how the public responds to health and financial crises, and resulting demands for risk regulation. 
\citeasnoun[p.18]{sunstein:19b} \nada{How Change Happens, p. 18} describes the adoption of the belief that genetic modification of food is or is not a serious problem as an information cascade. He further argues that there are cascades in norms as well as in factual beliefs.

Clearly, as in the SBM, cascades are possible in the binary decision of whether or not to forward a news item on social media.  It is less obvious whether we should expect to see information cascades in publicly expressed \emph{beliefs}.

In the SBM, agent $I_n$'s belief $\hat{p}_n$ about the state is a continuous variable.  Suppose that we extend the SBM by having each agent directly and truthfully communicate a report of her belief (probability assessment) to the next agent.  Then each agent's private signal always influences the next agent's belief, so no agent is ever in a cascade of reported beliefs.
However, in alternative settings there can be cascades---including incorrect cascades---in reported beliefs. We focus on one possible source of belief cascades, based upon discreteness in the communication channel.  

People often communicate in coarse categories, such as binary partitions.  When asked about a model of car, people often say that they like or dislike it, as opposed to reporting intensities of liking on a continuum,  detailed reasons for their opinions,  or probability distributions over which model is the best.\footnote{As the well-known hedge fund manager and self-improvement author Ray Dalio put it, ``It is common for conversations to consist of people sharing their conclusions rather than exploring the reasoning that led to those conclusions.''} Similarly, when asked for a suggestion about a movie or restaurant, or when discussing political topics, people  often just name the preferred  option, or say that an option is ``hot'' versus  ``sucks;'' or ``cool'' versus ``bogus.'' Such digitized communication is inevitable owing to limited  time and cognitive resources of speakers and listeners.

We modify the model of \citeasnoun{ADLO:11}, as discussed in \S\ref{sec:networks}, to capture such limited bandwidth in the communication of beliefs. Suppose that the action of each agent $I_n$ is to report  a binary indicator of state, $a_n \in \{L, H \}$. We consider the special case of  \citeasnoun{ADLO:11} in which agents  make decisions in a commonly known deterministic sequence, and each agent $I_{m}$ observes the actions  of some subset of  the agent's predecessors.\footnote{Alternatively,  each agent might also recall and pass on the  history of all beliefs that  have been reported to the agent.  If signals are discrete, such a setting would effectively be equivalent to the model of BHW.}  

We  interpret the report $a_n$ as the {\it reported belief} about the state. We will view reported belief $L$  as a claim that the state is probably $L$, and reported belief $H$ that the  state is probably $H$.\footnote{A possible behavioral extension of  the model would be to have  the reports sometimes understood naively by receivers as indicating that the state is $L$ or $H$ for sure. If, with some probability, receivers make this mistake, information cascades can start very quickly.} This could take the form of expressing that ``Candidate $H$ will win the election" or ``Candidate $H$ will lose the election."  
Specifically, $I_n$ {\it reports} belief $L$ if her true belief $\hat{p}_n < 0.5$, reports $H$ if $\hat{p}_n > 0.5$, and follows some indifference rule, such as flipping a coin between reporting $L$ or $H$, if  $\hat{p}_i = 0.5$. This reporting rule can be endogenized if agents desire a reputation with receivers for making good reports.  

A {\it belief cascade} is defined as a situation in which, having received the reported beliefs of some set of   predecessors, an agent’s reported belief is independent of the agent’s private signal.  A belief cascade is also an information cascade, as defined earlier, since the reported belief $a_n$ can be interpreted as the agent’s action. 

In this model, if signals are bounded, then as in \citeasnoun{ADLO:11} as discussed in \S\ref{sec:networks}, there is no asymptotic learning; even in the limit reported and actual beliefs are often incorrect. Furthermore, if signals are discrete, then for reasons similar to the model of BHW,  there are incorrect belief cascades. For example, the special case of this setting in which signals are binary and each agent observes all predecessors is isomorphic to the SBM.

Whether belief cascades occur in practice is an empirical question. A possible application is to the spread of scientific claims. Since it is costly to read original source articles in detail, scholars often rely on descriptions of the original findings in later publications.  \citeasnoun{greenberg:09} provides 
bibliometric evidence from the medical literature about the spread via chains of citations of unfounded scientific claims.
 %that tentative or nuanced claims  were often converted into bald assertions of fact by citing papers, leading to chains  of  citation for unfounded claims.

In the modified SBM example above, belief cascades derive from communication bandwidth constraints. However, belief cascades may also derive from other possible mechanisms.  In models of message sending with payoff interactions, message senders sometimes  strategically report only coarsened versions of their beliefs (\citeasnoun{crawford/sobel:82}). This raises the possibility that sequential reporting might result in cascades.  

Alternatively, in behavioral models with dual cognitive processing, an agent consists of two selves (sometimes called the  ``planner'' and the ``doer'') who face distinct  decision problems. This can  correspond, for example, to the System~2 versus System~1 thinking distinction of \citeasnoun{kahneman:11}. Suppose that the planner  rationally updates and assesses probabilities, and, when $\hat{p}_i > 0.5$ instructs the doer that  the state is $H$, and,  when $\hat{p}_i < 0.5$ instructs the doer that  the state is $L$. Such simplified instructions may be needed owing to limited cognitive processing power of the doer, who may face problems of distraction and time constraint.  If people are typically in ``doer'' mode when engaged in casual conversation with others, then only the coarsened information is reported. This would again result in cascades in reported beliefs. Furthermore, these reported beliefs correspond to the genuine beliefs of people when in doer mode.    

\ifdefined\ALLCOMPILE\else\end{document}\fi
 % 24 = 9 + many times one
\ifdefined\ALLCOMPILE
\saycurrfile
\else
\usepackage{jelall}
\renewcommand{\citeasnoun}[1]{\textcolor{red}{#1}}

\begin{document}
\fi

%%%%%%%%%%%%%%%%%%%%%%%%%%%%%%%%%%%%%%%%%%%%%%%%%%%%%%%%%%%%%%%%
\section{Conclusion and Prospect}
\label{sec:Conclusion}
%%%%%%%%%%%%%%%%%%%%%%%%%%%%%%%%%%%%%%%%%%%%%%%%%%%%%%%%%%%%%%%%

Theoretical analysis of social learning and information cascades has offered a new understanding of two   economic and social phenomena. The first is that  individuals who in the aggregate possess extensive information often  converge upon the same action, and that this action is  often wrong. The second is that people are attached to  popular actions very loosely, resulting in fragility of mass outcomes, and fads. 

In the social learning explanation, owing to an   information externality, the information in the history of actions accumulates more slowly than would be socially optimal. In the case of information cascades, this aggregation is completely blocked. 
Since cascades aggregate information poorly, even a small information shock, or even the possibility of one, can cause  actions to shift abruptly. The basic cascades model therefore implies that in such circumstances there will be occasional, irregular bouts of sudden change.  
The system systematically evolves to a precarious position, instead of having  fragility arise only under rare environmental conditions. 

Information externalities in social learning, and incorrect cascades in particular, occur under reasonably mild assumptions. As we have discussed, these effects have been applied to  a wide range of phenomena. 
The  literature has  provided  a rich set of insights by allowing for the possibilities that agents can choose when to act or whether to acquire private signals, or that the action being selected is a trade in a  marketplace.  Recent work has opened new vistas by analyzing social learning when agents  take repeated actions,  are imperfectly rational, or observe others in  complex social networks. We  highlight several directions for further research. 

\subsection{The Speed of Learning}

A recent line of work has studied the speed and efficiency of social learning (see \S\ref{section:speed of convergence}). In a wide range of settings, information externalities delay learning relative to the socially optimal rate.
There has been relatively little study of the comparative statics of speed of learning. 
An interesting issue to explore  is whether a version of the principle of countervailing adjustment (as discussed in  \S \ref{sec:binary-model}) applies in settings without information cascades.  
In settings with information  cascades, this principle maintains  that, under appropriate conditions,  agents adjust to additional information  by falling into cascades sooner, which opposes the accumulation of social information.
Similarly, in non-cascades settings,  parameter shifts that might seem to accelerate learning do not necessarily do so. 
A shift in the model that has a direct effect, at a given point in time, of making  social information  more  accurate will tend to make  the relevant agent's actions will depend more weakly on her  own private  signal. 
This can oppose the direct increase in  social information induced by the exogenous shift.

The literature places a strong emphasis on asymptotic learning. This focus does not address welfare, which, with discounting, depends on the actions of the early agents, rather than eventual limit outcomes. A natural benchmark is a scenario in which  a social planner whose aim is to maximize total discounted utility dictates decision rules to all agents. The question of how closely actual outcomes can approach this benchmark  deserves further  exploration (see \citeasnoun{chamley:04a,smith/sorensen/tian:20}).

\subsection{Imperfect Rationality}

Another rich direction for further research is how   imperfect rationality affects social learning. 
Existing models have allowed for overconfidence, and for people misestimating how much others are observing predecessors.  
Models with repeated actions on networks in which agents act myopically   also effectively allow for  the possibility that agents neglect the indirect effects of an action on the information content of the subsequent actions of others (\citeasnoun{HMST:21}). 
 % Mira Frick and Ryota Iijima and Yuhta Ishii, (Econometrica 2020) does have repeated actions and myopia, but it is not really a networks paper. There is independent random binary matching of agents, in a continuum of agents.

As discussed in \S \ref{sec:observability}, an insight  from previous work is that biases that cause agents to be more aggressive in using their own signals tend to improve  social information aggregation, whereas biases in the opposite direction worsen it.
These findings suggest that it will be possible to derive conclusions about the relative success of individuals and groups that contain different distributions of psychological agent types (e.g., \citeasnoun{bernardo:welch:2001}). 

The distribution of types affects both whether  there is asymptotic learning, and, when there is not, how much social learning occurs. 
In evolutionary settings, the success of individual types depends on  a balance between the direct benefit of having a  psychological  trait versus the  benefits to being in a group in which others have that trait. 
Analysis of this tension can endogenize which  imperfectly rational types will survive  in the long run.

Existing models differ in their conclusions about whether even a  small amount of bias tends to be socially amplified to generate   large and persistent  effects on social outcomes, such as failures of asymptotic learning.  This topic merits further exploration. 

Another question for further study is how imperfectly rational social learning affects boom bust dynamics in the adoption of behaviors such as investments, or  mergers and acquisitions.  We have discussed rational models of booms and busts (\citeasnoun{chamley/gale:94}), including some within market settings (\citeasnoun{lee:1998}). However,  many observers have argued that imperfect rationality contribute to such phenomena. 
Such imperfectly rational behavioral shifts are captured by the models  of  \citeasnoun{bohren/hauser:19} and  \citeasnoun{hirshleifer/plotkin:21}.
It will be valuable to develop both approaches for applications  and empirical testing.
A further promising direction is to explore product or security market settings with  imperfectly rational social learning  (see \citeasnoun{eyster/rabin/vayanos:18}).

An especially rich direction is to analyze how social learning and information cascades influence verbal discourse. This is of special interest owing to growing concerns about the spread of misinformation and fake news. The model of belief cascades in \S \ref{subsec:beliefcascades} is a small initial step. A further direction will be to incorporate a wider set  of psychological biases that influence human communication. These may include neglect of selection in what others choose to assert, and confirmation bias. Such analysis  can provide insight into why in some domains discourse grows ever more sophisticated (as with advance in the scientific scholarly community), whereas in others discourse remains persistently naive (as with much political discussion on social media).

\subsection{Repeated Actions}

As discussed in \S\ref{sec:repeated_actions}, the literature studying repeated actions by Bayesian agents is nascent. 
Only initial steps have been taken in the study of the speed of learning in repeated action settings (see \citeasnoun{HMST:21}). Likewise, the question of which social networks promote more efficient information aggregation deserves further exploration (\citeasnoun{golub:jackson:10,mossel:sly:tamuz:2015})

In a repeated action setting with discounting agents, even when an agent's utility depends only on the state and her action, there are strategic informational incentives. An agent may choose an action to  influence others to convey more information through their  future actions. For example, consider  an agent whose private information opposes the agent's social information.  The agent might follow  the private signal, lowering the agent's current payoff, if doing so generates informational value deriving from the reactions of others. The topic of strategic informational incentives has remained almost completely unexplored (but see the ``mad king'' example in \citeasnoun{mossel:sly:tamuz:2015}).

Only a handful of papers consider social learning with a changing state (see \citeasnoun{moscarini/ottaviani/smith:98,frongillo2011social}); even fewer do so in a repeated action setting (\citeasnoun{dasaratha/golub/hak:19}). Many interesting questions remain unexplored. In a steady state, how well attuned are actions to state? How does this concordance depend on the private signal structure, or  on deviations from rationality? How quickly does the  preponderant action change after a change in state? Is the probability of change in the preponderant action less than or greater than the probability of a change in state (inertia or impulsiveness respectively; see \citeasnoun{hirshleifer/welch:02})?

\subsection{Information Design}

A fruitful direction for further exploration is the problem of information design in sequential social learning settings. One possible purpose is simply to improve information aggregation. We have seen that this purpose is sometimes well-served by assigning some agents to be  sacrificial lambs. Alternatively, a social planner or a seller can  design or influence the signal structure or the observation network for  other purposes, in a form of  sequential Bayesian persuasion (\citeasnoun{kamenica/gentzkow:11}.

For example, the prices set by a monopolist  influence the decisions of early buyers, which in turn affect the information and decisions of later buyers (\citeasnoun{welch:92}). 
Similarly, a principal can  design an organization's observation structure to motivate agents to work hard or to choose desired actions (\citeasnoun{bose:orosel:ottaviani:vesterlund:2008}, and \citeasnoun{khanna/slezak:00} discussed in \S \ref{sec:ApplicationsExtensions}).

In these applications, the private information available to agents is taken as given.     There are also models of social learning wherein the principal assigns  private signals to  agents  \citeasnoun{kremer:mansour:perry:14,che:horner:18}. 
Several questions remain unanswered. When is it beneficial for a monopolist to have superior information about the product? When do principals benefit from the ability to commit as to possible dependence of the observation structure  on  history?  
How does the design of private signal assignment or observation structure interact with other levers for influencing buyer beliefs, such as setting prices?\footnote{For example, a seller can send free samples of a product to influencers, effectively providing them with private signals about quality, or can 
pay a social media platform to inform users that certain of  their friends have bought the product.} 

\subsection{Empirical Testing}

Our focus has been on theoretical models of cascades and social learning, but we have tried to show that the main ideas in this wide-ranging and sometimes rather technical literature are both accessible and potentially testable. 
We have discussed a few notable empirical contributions  that use archival or experimental data  to identify social influence effects.
An   empirical question which has received little attention, and which can help distinguish alternative theories of social influence, is whether conformists or deviants from the predominant action make better decisions.

Several recent developments  have created exciting  new  opportunities for   empirical testing. 
First, social media and other electronic databases have greatly increased the  information available to empiricists about individuals' social observation and connectedness in relation to   behavior (\citeasnoun{BCKSW:18}). Second, such datasets, in some cases with the benefit of  textual analysis, provide possible proxies for the information signals available to different agents or the beliefs that such agents have formed.  
Third, the rise of field experimentation promises to augment the insights provided by laboratory experimentation (see, e.g., \cite{salganik/dodds/watts:06}).   
Finally, several recent theoretical developments  are sources of  new empirical implications. These developments include  recognizing that agents are embedded in social observation networks, and that they often are imperfectly rational in the beliefs that they form.  
For all these reasons, empirical directions are likely to play an increasingly important role in the further advance of this field.

\ifdefined\ALLCOMPILE\else\end{document}\fi

 % 8

\clearpage

\AtEndDocument{\ifdefined\ALLCOMPILE\else
\documentclass[12pt]{article}
\usepackage{jelall}
\renewcommand{\citeasnoun}[1]{\textcolor{red}{#1}}

\begin{document}
\fi

\appendix

\section{Online Appendix}

\subsection{Further discussion of Section \ref{sec:binary-model} on the Simple Binary Model (SBM)}
\label{onlinesub:binary}

%%%%%%%%%%%

In \S \ref{sec:binary-model}, \S \ref{subsec:other-model-misspec}, we discussed that overconfidence  can improve social learning in the model of \citeasnoun{bernardo/welch:01}. We now provide  algebraic details. 

We consider a setting as in the SBM, except that  \I2 is overconfident, in the sense that \I2 incorrectly assesses \I2's precision at $p' = p + \epsilon$, where  $\epsilon \approx 0$.  $I_2$'s true precision of $p$ is known to all except \I2.  Observe that if the first two actions are $HL$ or $LH$, then starting with \I3 the model is isomorphic to the base model with \I3 as the first agent. 
So the probability of a correct cascade has two terms:
$$
p^2 +  2p(1-p) \frac{p(p+1)}{2(1 - p + p^2)}.
$$
The first term is the probability of a correct cascade occurring starting with \I3, 
  $$
  0.5 \ \Pr{a_1 a_2 = HH|\theta = H} + 0.5 \ \Pr{a_1 a_2 = LL|\theta = L} = p^2.
  $$
  The second term is the probability that the first two actions are  opposed, $2p(1-p)$, multiplied by the probability of a correct cascade in the long run in the SBM, as given by equation (\ref{e1}) above.
  The difference between this expression and the  probability that in the long run a correct cascade occurs  in the base model (again as given in (\ref{e1}))  is
  $$
 \frac{p(2p-1)(1-p)}{2(1 - p + p^2)}  > 0,
  $$
  since $p > 0.5$.  So overconfidence increases average accuracy and welfare.

%%%%%%%%%%%%%%%%%%%%%%%%%%%%%%%%%%%%

\subsection{Further discussion of Section \ref{sec:LimitedObservability} on limited communication or observation of past actions}
\label{onlinesubsec:observability}

We discuss here imperfectly rational models that make assumptions about the agent's mapping from observed actions and payoffs into the agent's actions.

\subsubsection{Heuristic action rules based on observation of averages of past payoffs and of action frequency}
\label{subsubsec:HeuristicAverages}

If rational agents could observe the average payoffs experienced from possible actions undertaken by many in a large population, they would always choose the right action. 
However, it is possible that agents fall short of rationality. 
In the model of \citeasnoun{ellison/fudenberg:93}, there is a continuum of agents, where each period some given fraction of them has the opportunity to update their choices.
Agents observe the average payoff of the two actions in the population from the previous period, and their popularities. 
Agents use rules of thumb for updating their actions, which take the form of  specified probability of switching action based on these two variables. 
The paper explores conditions on the weighting of these considerations that promote or hinder long-run correct decisions.  

\subsubsection{Heuristic action rules based on  sampling of past actions and payoffs}
\label{subsubsec:HeuristicSampling}

In \citeasnoun{ellison/fudenberg:95}, there is a large number of agents who take simultaneous actions at each discrete period.  
Each period, some fraction of agents exogenously stick to their current action, and the remainder observes an unbiased sample of the latest actions and payoffs of $N$  and choose action based on this and based on their own latest payoff.
Agents follow the following heuristic decision rule. 
If all the agents in an agent's sample takes the same action, the agent follows that action.  
If both actions are selected by at least one agent in the sample, the agent chooses whichever action has higher average payoff based on  observed reports and the agent's own latest experienced payoff.  

When the sample size $N$ is small, learning from the experience of others causes the system to evolve to universal adoption of the correct action. 
In contrast, when the sample size is large, there is strong pressure toward diversity of behavior.\footnote{This is because if one action is very unpopular, with large $N$ many adherents  of the more popular action  ``hear about'' the unpopular one and potentially convert to it.} 
So the system never fixes on the correct action.
This analysis is valuable in illustrating how non-obvious interesting conclusions about efficiency derive from reasonably plausible heuristic assumptions.  
On the other hand, when there is a `split decision' in an agent's sample, it would be reasonable for an agent to take into account that a preponderance of 99 Adopts to 1 Reject (for example) might suggest almost conclusively that adopt is superior.\footnote{Ellison and Fudenberg consider a related effect which they call `popularity weighting.'} 
This would tend to oppose the diversity effect discussed above. 
 
%%%%%%%%%%%%%%%%%%%%%%%%%%%%%%%%%%%%%%%%%%%%%%%%
\subsubsection{Recent past signals observable}
\label{subsubsec:RecentSignals}

% COmpare with CAO/H EXAMPLE --NOTE THAT CHAMLEY BOOK ALSO HAS AN EXMAPLE AS A PROBLEM BASED ON CAO/H***** \citeasnoun{cao/hirshleifer:97a}

The  SBM and the BHW model are based on the premise that  agents do not directly communicate their private signals.  If the signals of agents are fully communicated to their followers, learning would be efficient, and there would be asymptotic learning. 

In many practical contexts,  more information is indeed transmitted than just action choices.  For example, people are often  free to talk about their private information, though owing to time and cognitive constraints,  such communication may be imperfect.  This raises the question of whether incorrect cascades still occur, and whether
there is asymptotic learning.

We consider  here the possibility of limited communication of private signals.  Every private signal is directly communicated, which in principle could bring about asymptotic  learning.  However, we consider a case in which  each signal is passed on to just one other agent, and provide an example in which incorrect  cascades still form and last forever.  

Consider the SBM of Section \ref{sec:binary-model}, except  that in addition to observing all past actions, each agent observes the private signal of the agent's immediate predecessor.  In contrast with the SBM, we  assume that  when indifferent, each agent always follows her own signal. This assumption is convenient but not crucial. 

After actions $HH$ or $LL$, $I_3$ infers that the first two signals were $hh$ or $\ell \ell$ respectively, and falls into a cascade which could easily be incorrect.  This cascade can be broken. To see this, suppose that the first two actions are $HH$, and that $I_3$ observes $\ell$. So $I_3$ chooses $H$.  If $I_4$ observes $\ell$, this combines with observation of $I_3$'s $\ell$ to make $I_4$ indifferent, so $I_4$ rejects, breaking the cascade.

Consider next the case in which $I_3$ observes $h$ as well.  Now, $I_4$ adopts, because $I_4$ knows that the first 3 signals were all $h$.  At this point the cascade of adoption continues forever. $I_5$ knows that the first two signals were $hh$, and that the third and fourth signals were not $\ell \ell$.  (I.e., they were either $\ell h$, $h h$, or $h \ell$.)  So the net evidence in favor of state $H$ is at least slightly stronger than two $h$ signals.  So even in the worst case in which $I_5$ directly observes two $\ell$ signals ($I_5$'s  and $I_4$'s signals), $I_5$ strictly prefers $H$.  Similar reasoning shows that all subsequent agents adopt.

The  general point  that inefficient cascades can form and last forever does not require this arbitrary tie-breaking convention.\footnote{The following modification illustrates the point in a setting with no ties. Suppose that agents differ very slightly in their precisions, and that agents act in inverse order of their precisions, from low to high.  Now agents are never indifferent. Agents who are close to indifference strictly prefer to follow their own signals, and the same analysis holds.}  Intuitively, the ability to observe the predecessor's signal is a double-edged blade.  The immediate effect is to increase an agent's information, which increases the probability that an agent decides correctly.  But to the extent that this is the case, it later makes the action history more informative. That eventually encourages later agents to fall into a cascade, which blocks asymptotic learning.

In particular, the action sequence becomes informative enough that agents fall into cascades despite their access to an extra signal from the predecessor.  The  actions history eventually  overwhelms an agent's information (even inclusive of observation of a predecessor's signal), at which point the accuracy of the social belief stops growing.

Another way to think of this is that  being able to observe predecessors' signals is somewhat like  being able to observe multiple private signals instead of one.  This does not fundamentally change the argument for why incorrect cascades form---that the accumulation of information implicit in past actions must eventually overwhelm the signal(s) of a single agent.

\ifdefined\ALLCOMPILE\else\end{document}\fi

}
% \AtEndDocument{\tableofcontents}

\end{document}